\documentclass[%
 reprint,
 amsmath,amssymb,
 aps,
longbibliography,
]{revtex4-1}

\usepackage{xcolor}
\usepackage{graphicx}
\usepackage{dcolumn}
\usepackage{bm}
\usepackage{hyperref}
\usepackage{url}

\begin{document}

\preprint{APS/123-QED}

\title{The Neutrino Fast Flavor Instability in Three Dimensions}

\author{Sherwood Richers}
\email{srichers@berkeley.edu}
\affiliation{Department of Physics, University of California Berkeley, CA 94720}
\author{Donald Willcox}%
\affiliation{Computational Research Division, Lawrence Berkeley National Lab, CA 94720}
\author{Nicole Ford}
\affiliation{Computational Research Division, Lawrence Berkeley National Lab, CA 94720}

\begin{abstract}
Neutrino flavor instabilities have the potential to shuffle neutrinos between electron, mu, and tau flavor states, modifying the core-collapse supernova mechanism and the heavy elements ejected from neutron star mergers. Analytic methods indicate the presence of so-called fast flavor transformation instabilities, and numerical simulations can be used to probe the nonlinear evolution of the neutrinos. Simulations of the fast flavor instability to date have been performed assuming imposed symmetries. We perform simulations of the fast flavor instability that include all three spatial dimensions and all relevant momentum dimensions in order to probe the validity of these approximations. If the fastest growing mode has a wavenumber along a direction of imposed symmetry, the instability can be suppressed. The late-time equilibrium distribution of flavor, however, seems to be little affected by the number of spatial dimensions. This is a promising hint that the results of lower-dimensionality simulations to date have predictions that are robust against their the number of spatial dimensions, though simulations of a wider variety of neutrino distributions need to be carried out to support this claim more generally.
\end{abstract}

\maketitle


\section{Introduction}
Neutrinos are produced in immense numbers in core-collapse supernovae and neutron star mergers, but the neutrino's elusive nature and behavior currently limits our understanding of these explosive astrophysical phenomena. Core-collapse supernovae, produced when a massive star collapses after exhausting its nuclear fuel, rely on neutrinos to carry energy from the collapsed core to the infalling material below the shock front, and do so on a short enough timescale that they are able to propel the shock through the star (see e.g., \cite{janka_explosion_2012,branch_supernova_2017} for reviews). When two neutron stars or a neutron star and a black hole merge, the neutrinos emitted from the resulting hot accretion disk can enhance outflows that form heavy elements (see \cite{radice_dynamics_2020} for a recent review). In both cases, the matter ejected pollutes the surrounding environment with metals that later form more metal-rich stars, planets, and life. Furthermore, although mergers are the prime candidate source of some of the heaviest elements in the universe, neutrinos play a deciding role in determining whether the ejecta is sufficiently neutron-rich to efficiently form these elements (e.g., \cite{lippuner_signatures_2017,kullmann_dynamical_2021}).

Since we cannot directly see the interiors of supernovae and mergers and have only seen neutrinos from a single supernova \cite{alexeyev_detection_1988,bionta_observation_1987,hirata_observation_1987}, we resort to simulation to evaluate whether fundamental physics as it is understood today is capable of reproducing what we see in nature. Understanding supernova and merger dynamics without allowing neutrino flavor change is still an extremely active and productive field, as some of the arguments that neutrino flavor oscillations are not deeply important for the dynamics of supernovae have been rather convincing (e.g., \cite{duan_self-induced_2011,dasgupta_role_2012}). Promisingly, there is increasingly strong evidence that such simulations may in fact be capable of yielding explosions with energies trending toward those observed in nature (e.g., \cite{oconnor_global_2018,just_core-collapse_2018,burrows_overarching_2020,sandoval_three_2021}). Neutron star merger simulations are also becoming increasingly sophisticated tools for predicting the amount and composition of ejecta from compact object mergers, their electromagnetic and gravitational wave signals, and the type of compact central object left behind (e.g., \cite{mosta_magnetar_2020,nedora_mapping_2020,foucart_estimating_2021,zhu_fully_2021,metzger_neutrino-_2021,li_neutrino_2021,just_neutrino_2021}). These simulations have matured to include effects of general relativity, inviscid hydrodynamics, magnetic fields (or viscosity mimicking the action of magnetic fields), a dense nuclear equation of state, nuclear reactions, and transport of energy by neutrinos.

However, understanding supernovae and mergers may not be a simple matter of general relativistic neutrino radiation magnetohydrodynamics; we must also include the quantum nature of neutrinos in this already long list of relevant physics. There are three known species (flavors) of neutrinos, and a given neutrino is generally in a quantum superposition of all three states. Terrestrial experiments and observations of solar neutrinos show that this quantum state changes as the neutrino propagates according to the Schr\"odinger equation (e.g., \cite{ahmad_measurement_2001}, among many other detections since). Contributing to the Hamiltonian that drives this flavor change are the neutrino masses, the interactions of neutrinos with the background matter, and the interaction of neutrinos with other neutrinos (see, e.g., \cite{duan_collective_2010}). The latter of these is special in that it leads to a fascinating and poorly-understood nonlinear evolution of the flavor whenever neutrino-neutrino interactions are significant. Since different neutrino flavors have significantly different rates of interaction with matter, flavor-changing neutrinos can complicate our already limited understanding of the dynamics of supernovae and neutron star mergers (e.g., \cite{malkus_symmetric_2016,wu_imprints_2017,george_fast_2020,li_neutrino_2021,xiong_potential_2020}).

Recent developments have suggested the exciting possibility that neutrinos are unstable to flavor change deep in the engine of a supernova or neutron star merger \cite{sawyer_speed-up_2005,sawyer_neutrino_2016}. In fact, these instabilities are likely ubiquitous in both systems \cite{wu_fast_2017,morinaga_fast_2020}. The so-called fast flavor instability simply requires that, at a given location, there be an overabundance of neutrinos in some directions and an overabundance of antineutrinos moving in other directions \cite{morinaga_fast_2021}. Searches for these conditions in data from numerical simulations of both events that neglect flavor transforming physics have suggested that there are unstable conditions outside the supernova shock, inside the supernova shock, inside the turbulence supernova core, and around the accretion disk from a neutron star merger \cite{xiong_potential_2020,morinaga_fast_2020,tamborra_flavor-dependent_2017,abbar_occurrence_2019,m_d_azari_linear_2019,m_delfan_azari_fast_2020,abbar_fast_2020,nagakura_fast-pairwise_2019,glas_fast_2020,capozzi_fast_2021,abbar_characteristics_2021,wu_fast_2017,george_fast_2020,wu_imprints_2017}.

A first-principles global simulation of neutrino quantum kinetics in a supernova or a merger event remains intractable. However, it is still valuable to see how large an effect neutrino flavor transformation could have based on some computationally feasible prescription so as to help bound the realm of possibility. \citet{wu_imprints_2017} post-process simulation snapshots and tracer particle trajectories from a simulation of a NS-BH merger disk, modifying the neutrino interaction rates in locations with unstable neutrino distributions and assuming complete flavor mixing. The result was that the flavor conversion resulted in a more neutron-rich neutrino-driven wind and the formation of more heavy elements, though a similar analysis of a NS-NS merger disk showed the opposite \cite{george_fast_2020}. \citet{li_neutrino_2021} use the assumption of complete flavor mixing for unstable neutrino distributions and dynamically couple it to the hydrodynamics in simulations of a neutron star merger disk (a scenario similar to \cite{wu_imprints_2017}). This, too, resulted in a more neutron-rich outflow than when neutrinos are assumed not to change flavor. \citet{xiong_potential_2020} imposed multiple flavor transformation prescriptions on parametric models of the neutrino-driven wind following a CCSN. The flavor transformation resulted in more mass lost and a more proton-rich ejecta. Although a more robust parametric way to describe neutrino flavor transformation is needed, these results already show that neutrino flavor transformation can have a significant impact on galactic chemical evolution, supernova neutrino signals, and the properties of compact object mergers inferred from observation.

To understand the effects of the fast flavor instability, recent efforts have focused on directly simulating neutrino quantum kinetics. To make the problem computationally tractable, these simulations are performed using some combination of a limited physical domain, imposed symmetries, fewer neutrino flavors, and finite resolution. \citet{dasgupta_temporal_2015} and \citet{capozzi_collisional_2019} assume a two-beam model, where neutrinos can only move radially inward or radially outward. \citet{abbar_flavor_2015} and \citet{abbar_fast_2019} use a "line model" as a toy geometry that permits one free spatial and one free angular dimension. \citet{padilla-gay_multi-dimensional_2021} use a line-model-like symmetry, but also allow for temporal variations. \cite{padilla-gay_fast_2021,shalgar_change_2021,shalgar_three_2021,xiong_stationary_2021} allow asymmetry in only one momentum dimension, assuming homogeneity in the spatial dimensions. In \cite{shalgar_symmetry_2021} they extend this to two momentum dimensions. \citet{mirizzi_self-induced_2015} use a small number of neutrino directions, assume spatial translational symmetry in one direction and periodic symmetry in the other, and allow neutrinos to move in both dimensions parallel to an emitting neutrino plane. \citet{bhattacharyya_late-time_2020} and \citet{wu_collective_2021} assume translational symmetry in two spatial directions and azimuthal symmetry in neutrino momentum around the "free" spatial direction, and later extend the method to consider two spatial dimensions and two momentum dimensions \cite{bhattacharyya_fast_2021}. \citet{richers_particle--cell_2021} assume translational symmetry in two dimensions and periodic boundaries in the third, and allow both neutrino direction dimensions.

In this work, we attempt to take some of the guesswork out of interpreting whether simulation results are valid when symmetries are imposed. We simulate all dimensions (three spatial, two direction dimensions) relevant for the fast flavor instability where the neutrino potential dominates other potentials, imposing only periodic boundary conditions in all directions. In Sec.~\ref{sec:methods} we review the particle-in-cell method for neutrino quantum kinetics and describe enhancements made to the code since we published the one-dimensional results in \cite{richers_particle--cell_2021}. We describe the three-dimensional nature of the instability growth, instability saturation, and eventual quasi-equilibrium in Section~\ref{sec:results}. Finally, we conclude with Sec.~\ref{sec:conclusions} and describe the major results from this work that can be applied to global simulations of neutron star mergers and supernovae. We discuss the most important aspects of numerical convergence in the main text, but provide additional convergence details in App.~\ref{app:convergence}.

\section{Methods}
\label{sec:methods}
We simulate neutrino flavor transformation using the particle-in-cell code Emu \cite{richers_particle--cell_2021,noauthor_emu_2021}. In this method the neutrino field is represented by a large number of individual computational particles distributed throughout the domain. Each computational particle carries  several quantities: the number of neutrinos $N$ and antineutrinos $\bar{N}$ that the particle represents, the $3\times3$ density matrices for neutrinos ($\rho$) and antineutrinos ($\bar{\rho}$), the four-position $x^\alpha$ of the particle, and the four-momentum $p^\alpha$ of each neutrino or antineutrino in the particle. The particles move at the speed of light and the density matrices evolve following the Schr\"odinger equation as a collisionless approximation to the quantum kinetic equations:
\begin{equation}
  \begin{aligned}
    \frac{dx^\alpha}{dt} &= c \frac{p^\alpha}{p^0} \\
    \frac{dN}{dt}&=\frac{d\bar{N}}{dt} = \frac{dp^\alpha}{dt} = 0\\
    \frac{d\rho}{dt}&= \frac{-i}{\hbar}\left[\mathcal{H},\rho \right] \\
    \frac{d\bar{\rho}}{dt}&= \frac{-i}{\hbar}\left[\bar{\mathcal{H}},\bar{\rho} \right] \\
  \end{aligned}
\end{equation}
The potentials $\mathcal{H}$ and $\bar{\mathcal{H}}$ contain contributions from the neutrino masses, interactions with the surrounding matter, and interactions with other neutrinos as detailed in \cite{richers_particle--cell_2021}, although here we set the matter term to 0. All neutrinos in these simulations have an energy of $10\,\mathrm{MeV}$. We somewhat arbitrarily assign the neutrino parameters as $m_1=m_2=0$, $m_3=0.049\,\mathrm{eV}$, $\theta_{12}=10^{-6\circ}$, $\theta_{13}=48.3^\circ$, $\theta_{23}=8.61^\circ$, $\alpha_1=0$, $\alpha_2=0$, and $\delta_\mathrm{CP}=222^\circ$. The neutrino contribution to the potential for a neutrino moving in direction $\mathbf{\Omega}$ is
\begin{equation}
    \mathcal{H}_{\mathrm{neutrino},ab} = \sqrt{2} G_F \left[ (n_{ab} -\bar{n}_{ab}^* )- (\mathbf{f}_{ab}-\bar{\mathbf{f}}^*_{ab})\cdot \mathbf{\Omega} \right]\,\,,
\end{equation}
where $a$ and $b$ are flavor indices. Each particle contributes to the number density $n$ and number flux $\mathbf{f}$ in nearby grid cells according to a third-order shape function. $n$ and $\mathbf{f}$ are then interpolated at third order from the background grid. All of these steps are embedded in a fourth order Runge Kutta time integration scheme.

Although the original version of Emu was capable of three dimensional simulations, we change the time-integration procedure to be optimized for a larger number of particles per cell. Previously, particles were transferred between MPI blocks at each sub-step of the RK integration. We now instead add an additional layer of ghost zones to the outside of the domain (which do still get updated at every RK substep). This increases the number of grid cells that must be communicated between MPI ranks, but decreases the frequency with which particles need to be communicated by allowing the particles to avoid transferring to a new block until the end of the full time step.

For our spatial grid we use Cartesian coordinates and uniform grid spacing. The particles are initialized in the center of each cell and are distributed approximately uniformily in direction such that each particle represents the same amount of solid angle as described in \cite{richers_particle--cell_2021}. We place 64 particles in the $\hat{x}-\hat{y}$ plane in our production simulations, which results in a total of 1506 particles per cell.

\subsection{Initial Conditions}
\begin{figure}
    \centering
    \includegraphics[width=0.75\linewidth]{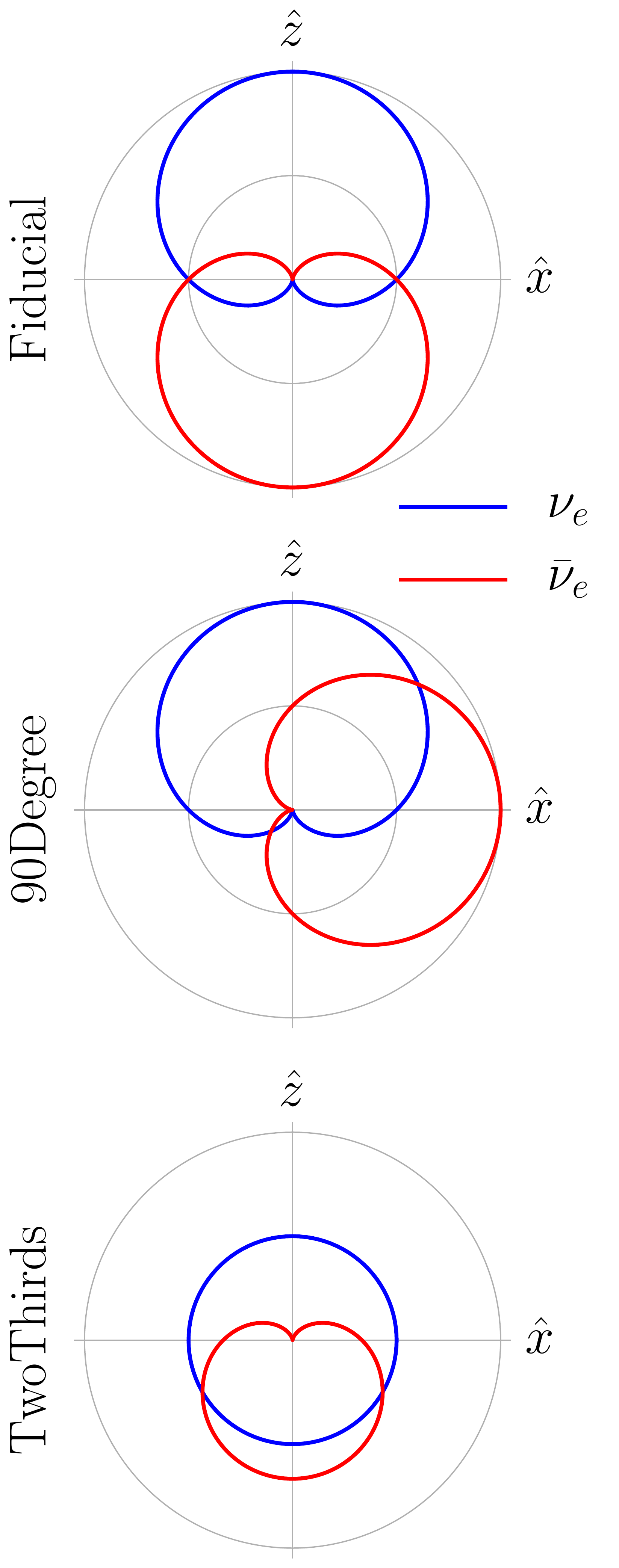}
    \caption{Graphical representation of the three initial conditions considered in this work. The radial extent of the curves represents the differential number density of that species of neutrino in the corresponding direction. The real initial conditions are distributed in all three directions and we only show the part in the $\hat{x}-\hat{z}$ plane; the full distribution can be visualized by rotating each curve around its axis of symmetry.}
    \label{fig:initial_conditions}
\end{figure}

\label{sec:initial_conditions}
\begin{table*}
    \centering
    \begin{ruledtabular}
    \begin{tabular}{ccccccc}
        Name & $\mathcal{N}_x\times \mathcal{N}_y\times \mathcal{N}_z$ & $L_x\times L_y \times L_z$ & $n_{\{\nu_e,\bar{\nu}_e,\nu_\mu\}}$ & $(\mathbf{f}/n)_{\{\nu_e,\bar{\nu}_e,\nu_\mu\}}$\\
         & & (cm) & ($10^{32}$ cm$^{-3}$) &  &  \\ \hline
        Fiducial\_3D & 128$\times$128$\times$\phantom{0}128 & \phantom{0}8$\times$\phantom{0}8$\times$\phantom{00}8 & & &\\
        Fiducial\_2D & \phantom{00}1$\times$128$\times$\phantom{0}128& \phantom{0}8$\times$\phantom{0}8$\times$\phantom{00}8& \{4.89,4.89,0\} & \{1/3$\hat{z}$, -1/3$\hat{z}$, --\}\\
        Fiducial\_1D & \phantom{00}1$\times$\phantom{00}1$\times$1024& \phantom{0}8$\times$\phantom{0}8$\times$\phantom{0}64& & &\vspace{0.1in}\\
        90Degree\_3D & 128$\times$128$\times$\phantom{0}128 & \phantom{0}8$\times$\phantom{0}8$\times$\phantom{00}8 & & &\\
        90Degree\_2D & \phantom{00}1$\times$128$\times$\phantom{0}128 & \phantom{0}8$\times$\phantom{0}8$\times$\phantom{00}8 & \{4.89,4.89,0\} & \{1/3$\hat{z}$, 1/3$\hat{x}$, --\} \\
        90Degree\_1D & \phantom{00}1$\times$\phantom{00}1$\times$1024 & \phantom{0}8$\times$\phantom{0}8$\times$\phantom{0}64 &  & \vspace{0.1in}\\\
        TwoThirds\_3D & 128$\times$128$\times$\phantom{0}128 & 32$\times$32$\times$\phantom{0}32 & & &\\
        TwoThirds\_2D & \phantom{00}1$\times$128$\times$\phantom{0}128& 32$\times$32$\times$\phantom{0}32& \{4.89,3.26,0\} & \{0, -1/3$\hat{z}$, --\} \\
        TwoThirds\_1D & \phantom{00}1$\times$\phantom{00}1$\times$1024&32$\times$32$\times$256 & & & \\
    \end{tabular}
    \end{ruledtabular}
    \caption{List of simulation parameters. $\mathcal{N}_x$, $\mathcal{N}_y$, and $\mathcal{N}_z$ are the number of grid cells in each direction. All boundary conditions are periodic, so a grid size of 1 implies imposed homogeneity in the corresponding direction. Similarly, $L_x$, $L_y$, and $L_z$ denote the physical extent of the domain in each direction, although the domain size in the direction of unit grid size carries no actual meaning. The 1D simulations have a larger domain size to be in line with the convergence criteria established in \cite{richers_particle--cell_2021}, and the multidimensional simulations are based on the convergence criteria established in App.~\ref{app:convergence}. All simulations were done with an angular resolution of 64 particles in the $\hat{x}-\hat{y}$ plane, corresponding to 1506 particles per cell. $n_{\nu_e}$, $n_{\bar{\nu}_e}$, and $n_{\nu_x}$ denote the initial number density of electron neutrinos, electron antineutrinos, and heavy lepton neutrinos, respectively. These are all three-flavor calculations, and both mu and tau neutrino and antineutrino densities are set to 0. The initial flux factor and flux direction of each neutrino species is listed in the final column. Graphical representations of the Fiducial, 90Degree, and TwoThirds initial conditions are shown in Fig.~\ref{fig:initial_conditions}.}
    \label{tab:simulations}
\end{table*}

We focus on three different physical conditions to elucidate the role of symmetries in the outcome of the fast flavor instability. In all cases, the simulations begin with only electron neutrinos and antineutrinos, and no heavy lepton neutrino content. The Fiducial case is the same as that in \cite{richers_particle--cell_2021}, is described in the top grouping of Table~\ref{tab:simulations}, and is depicted in the top panel of Figure~\ref{fig:initial_conditions}. There are an equal number of electron neutrinos and antineutrinos, and the density of each depends linearly on the cosine of the angle from the $z$-axis:
\begin{equation}
\begin{aligned}
    \frac{dn_{\nu_e}}{d\Omega} &= \frac{n_{\nu_e}}{4\pi} (1 + \mathbf{\Omega} \cdot \hat{z})\\
    \frac{dn_{\bar{\nu}_e}}{d\Omega} &= \frac{n_{\bar{\nu}_e}}{4\pi} (1 - \mathbf{\Omega} \cdot \hat{z})\,\,,    
    \end{aligned}
\end{equation}
where $d\Omega$ is the solid angle differential and $\mathbf{\Omega}$ is again the direction vector. The 90Degree simulations have an identical initial electron neutrino distribution, but the electron antineutrinos are instead distributed proportional to the cosine of the angle from the $\hat{x}$ direction:
\begin{equation}
\begin{aligned}
    \frac{dn_{\nu_e}}{d\Omega} &= \frac{n_{\nu_e}}{4\pi} (1 + \mathbf{\Omega} \cdot \hat{z})\\
    \frac{dn_{\bar{\nu}_e}}{d\Omega} &= \frac{n_{\bar{\nu}_e}}{4\pi} (1 + \mathbf{\Omega} \cdot \hat{x})\,\,.
    \end{aligned}
\end{equation}
This creates a distribution that is symmetric in the number of neutrinos and antineutrinos, but there is no axis of symmetry in momentum space. Finally, the TwoThirds simulations have an initially isotropic distribution of electron neutrinos with the same integrated number density as the other two cases. The electron antineutrinos have the same linear dependence as in the Fiducial simulations, but with an overall density scaled by $2/3$:
\begin{equation}
\begin{aligned}
    \frac{dn_{\nu_e}}{d\Omega} &= \frac{n_{\nu_e}}{4\pi}\\
    \frac{dn_{\bar{\nu}_e}}{d\Omega} &= \frac{n_{\bar{\nu}_e}}{4\pi} (1 - \mathbf{\Omega} \cdot \hat{z})\,\,.  
    \end{aligned}
\end{equation}
This makes for an asymmetry between the numbers of neutrinos and antineutrinos, and makes the depth and size of the ELN crossing significantly smaller than in the other cases. Although none of these calculations are directly representative of realistic astrophysical conditions, they elucidate which features are likely accurately captured by simulations with artificially imposed symmetries.

In practice, this means that the weight of a computational particle (i.e., the number of physical neutrinos it represents) is proportional to the angular distribution described above, and we initialize the quantum density matrix for each particle such that $\rho_{ee}=\bar{\rho}_{ee}=1$ and all other components are zero. We then place a uniform random number in the range $(-10^{-6},10^{-6})$ in the real and imaginary components of each off-diagonal matrix element and normalize the diagonal elements to ensure a unit flavor vector length. This choice of random initial conditions prevents us from evaluating exact convergence, especially once the system saturates and becomes chaotic. Changing the number of particles necessarily changes the initial conditions in a random way. However, in the same manner, this does also prevent the results of the simulation from depending on the precise functional form of the initial conditions, as all modes are present at some amplitude in the randomized initial conditions. It was also pointed out by \cite{wu_collective_2021} that the final equilibrium distributions depend increasingly little on the precise form of the initial perturbations, a point that we reinforce in this work.

The spatial resolution of 1024 cells in the 1D simulations follows the resolution requirements established in \cite{richers_particle--cell_2021}. However, we found that the results of multi-dimensional simulations are both less sensitive to spatial resolution and more sensitive to angular resolution (see Appendix~\ref{app:convergence}). For this reason, all of our production simulations for this work have four times the number of particles per cell, but the multidimensional simulations are only 128 cells on a side. We find similar results down to a domain size of $8\,\mathrm{cm}$ on a side for the multi-dimensional Fiducial and 90Degree simulations. The TwoThirds simulation set has a smaller antineutrino density and crossing strength, and therefore the dynamics are considerably slower and have longer characteristic length scales. Because of this, we extend the domain to $32\,\mathrm{cm}$ on a side for the multi-dimensional TwoThirds simulations and $256\,\mathrm{cm}$ for the 1D TwoThirds simulation.

\section{Results}
\label{sec:results}

\begin{figure*}[!ht]
    \centering
    \includegraphics[width=.32\linewidth]{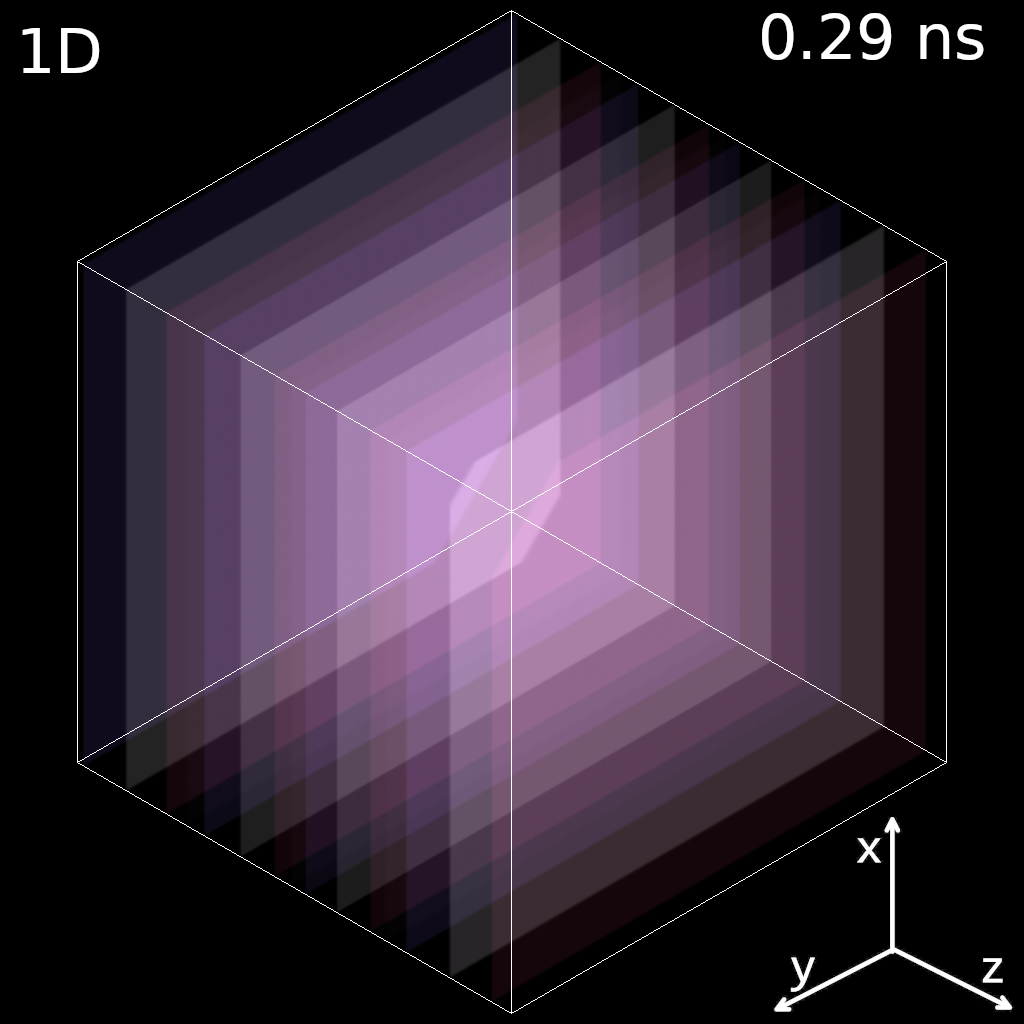}
    \includegraphics[width=.32\linewidth]{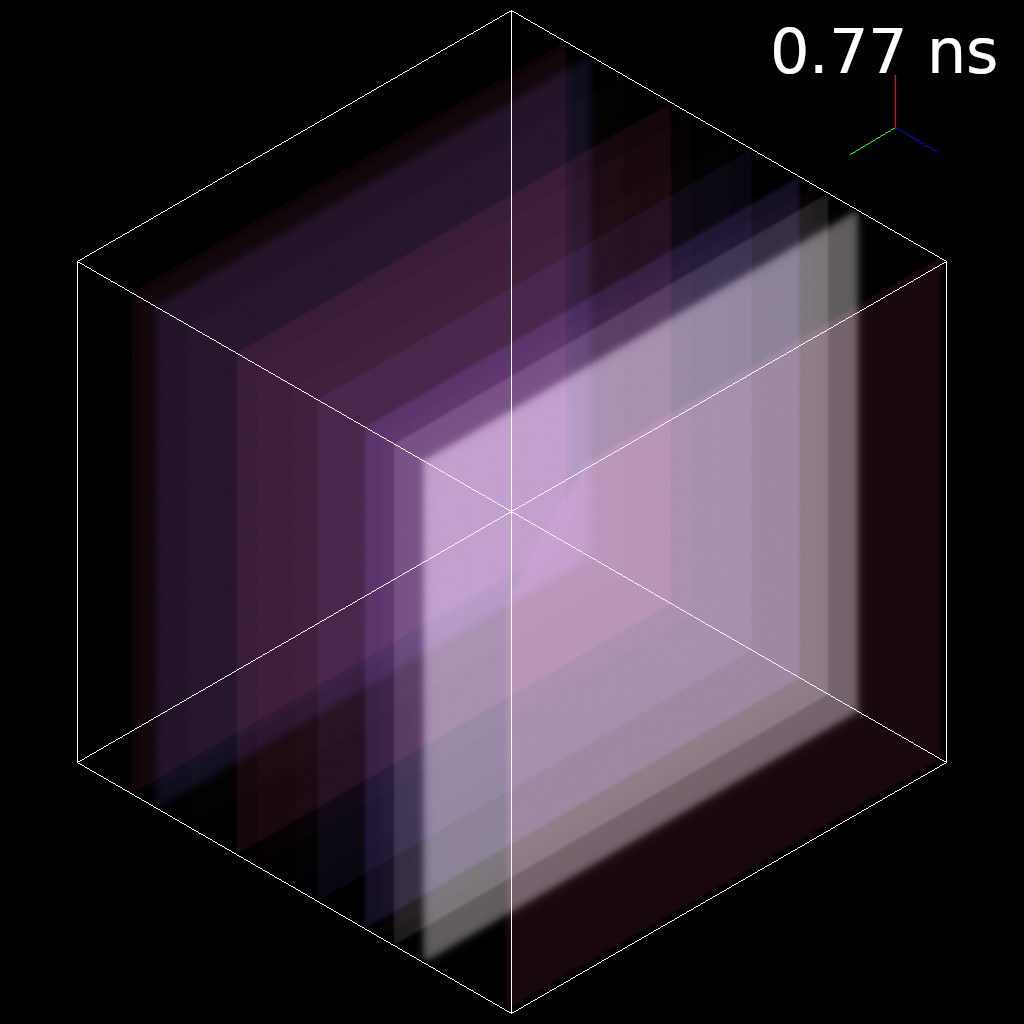}     \includegraphics[width=.32\linewidth]{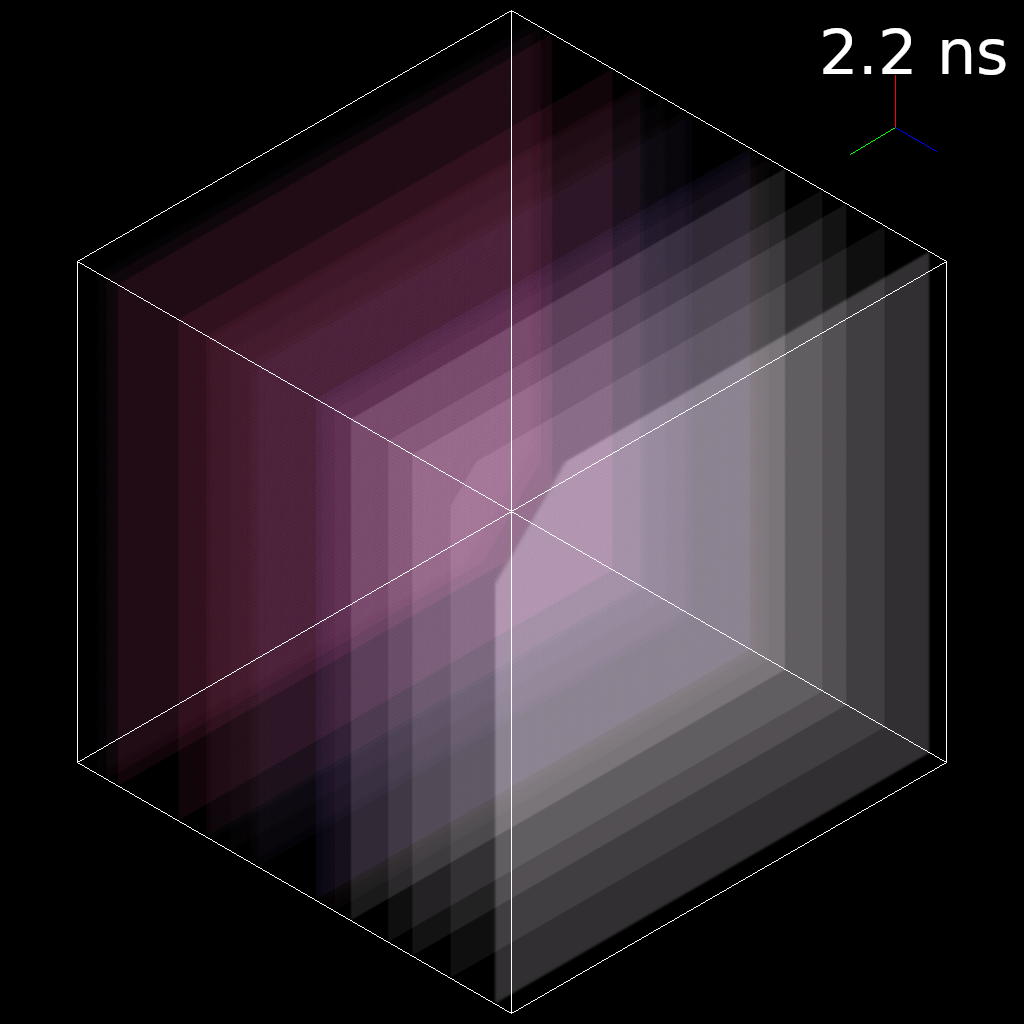}\\     \includegraphics[width=.32\linewidth]{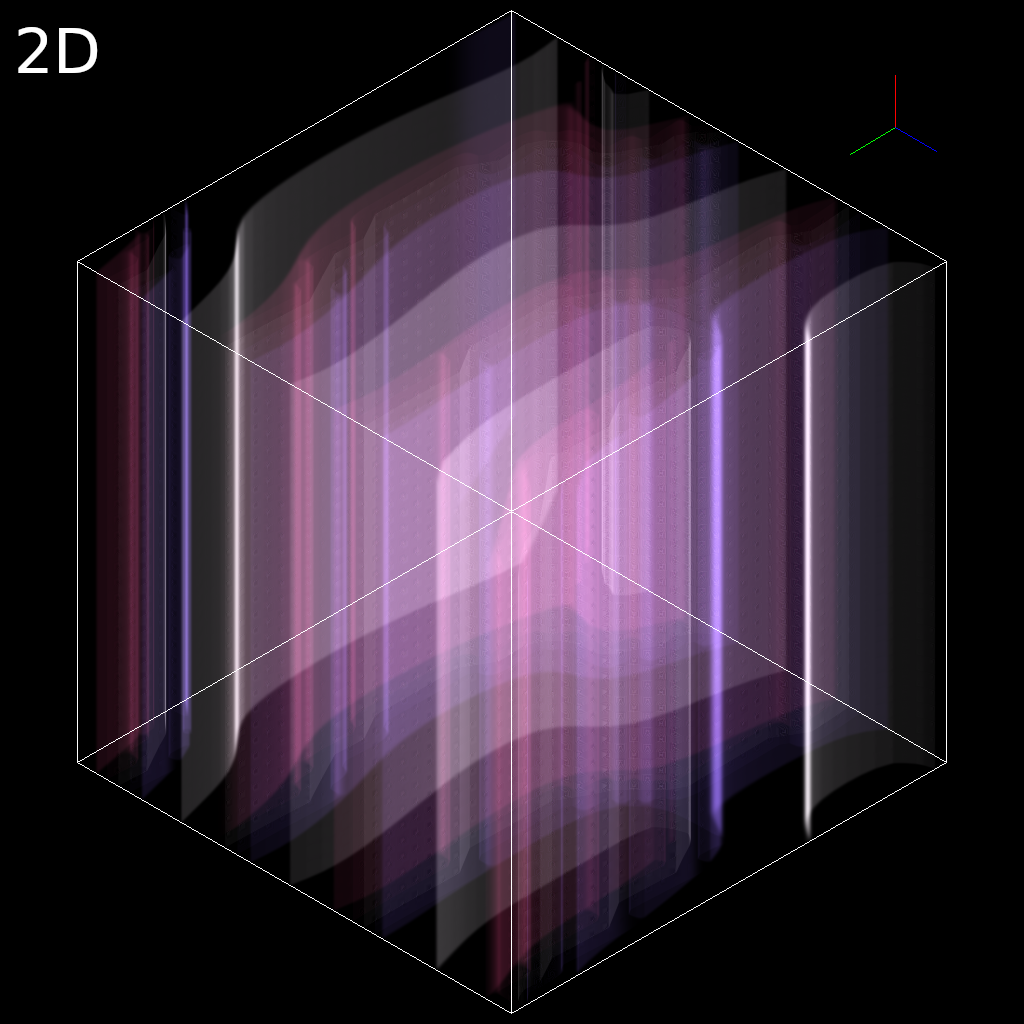}
    \includegraphics[width=.32\linewidth]{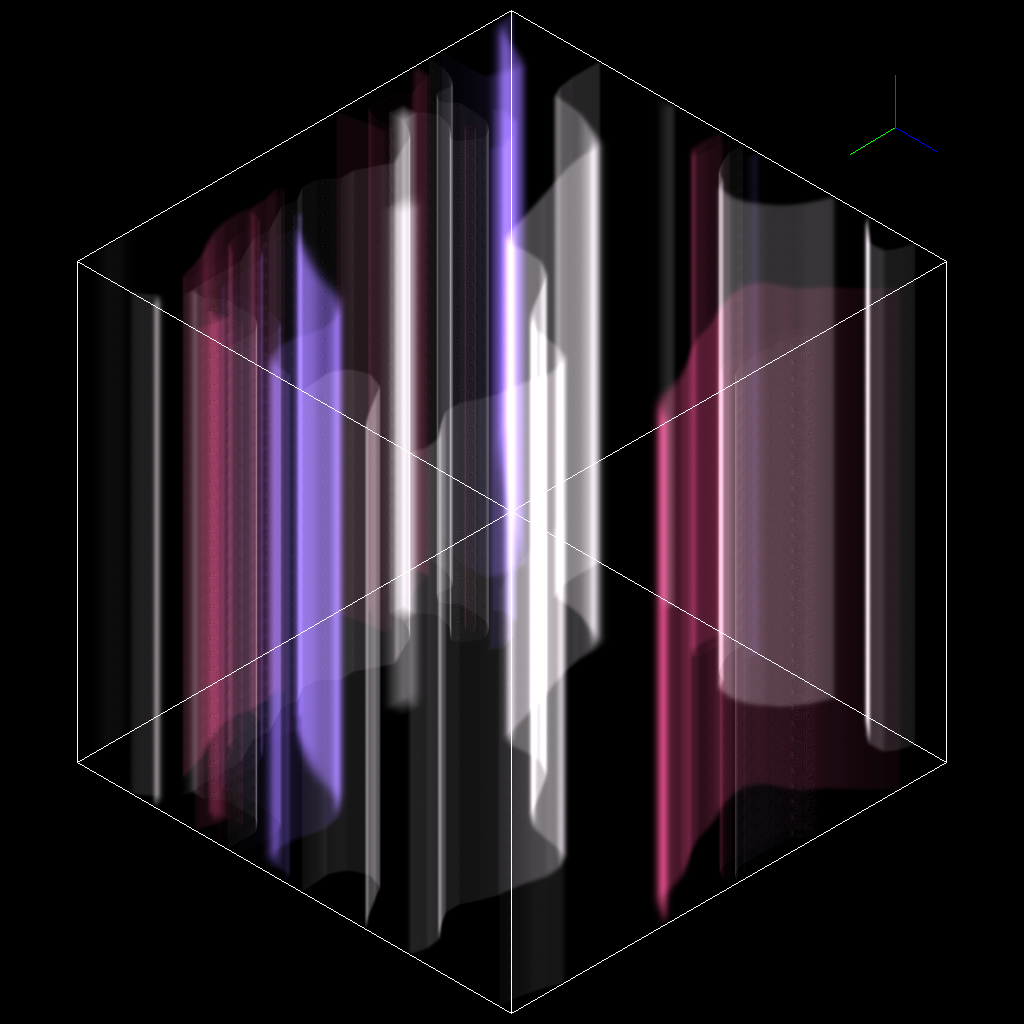}     \includegraphics[width=.32\linewidth]{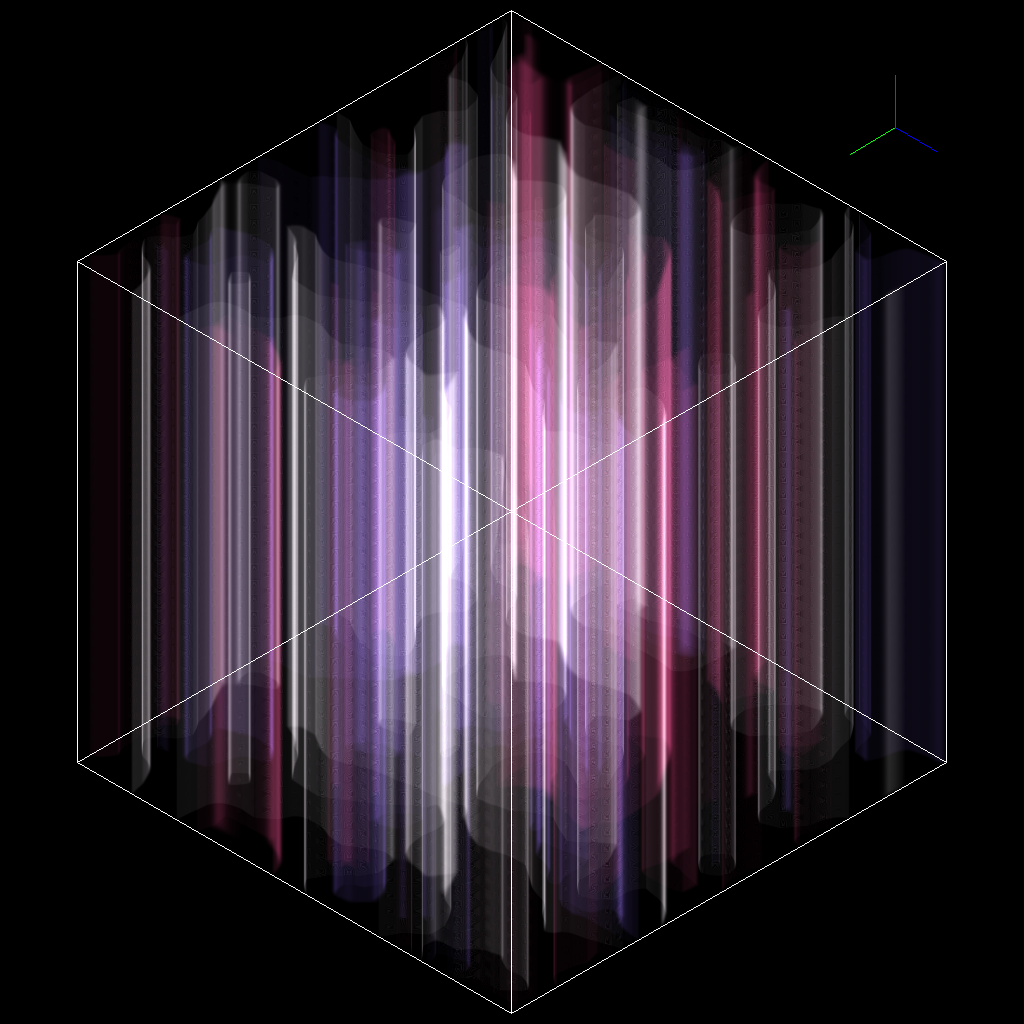}\\     \includegraphics[width=.32\linewidth]{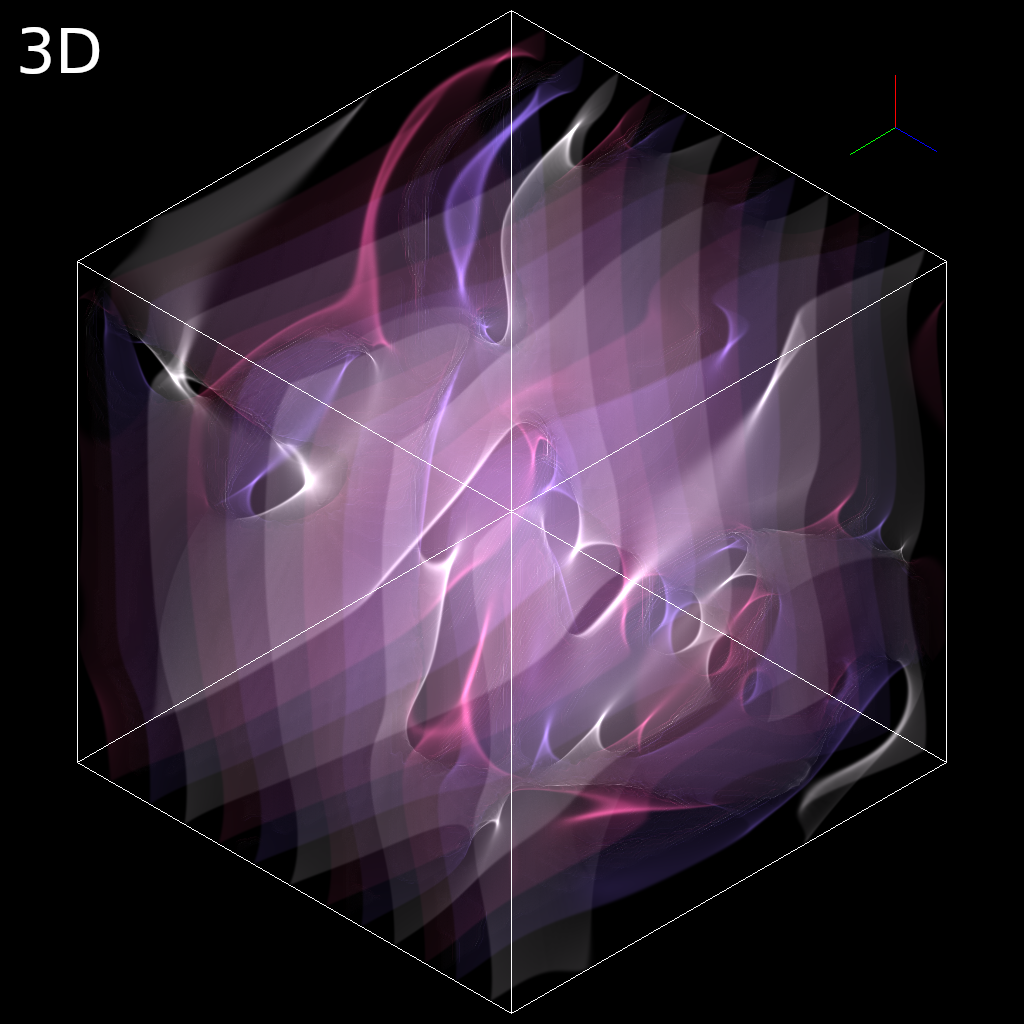}
    \includegraphics[width=.32\linewidth]{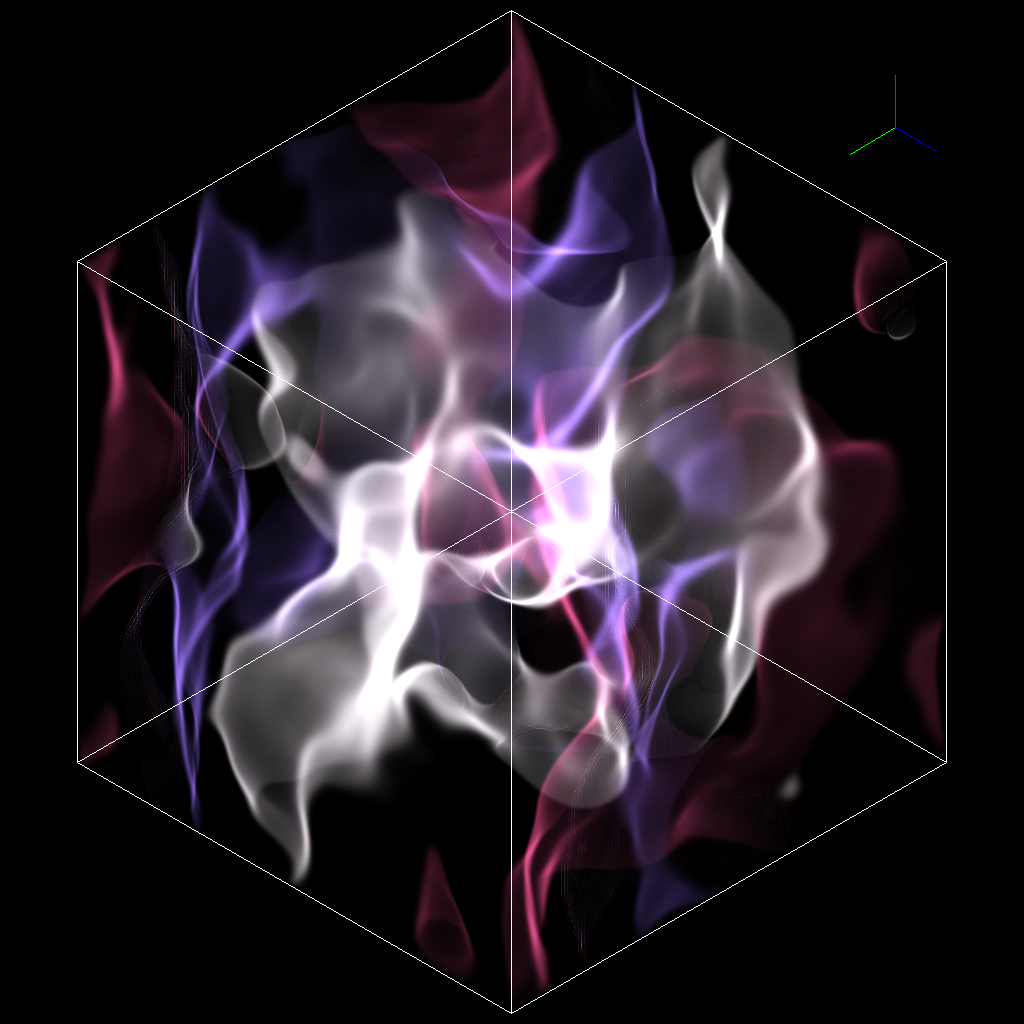}     \includegraphics[width=.32\linewidth]{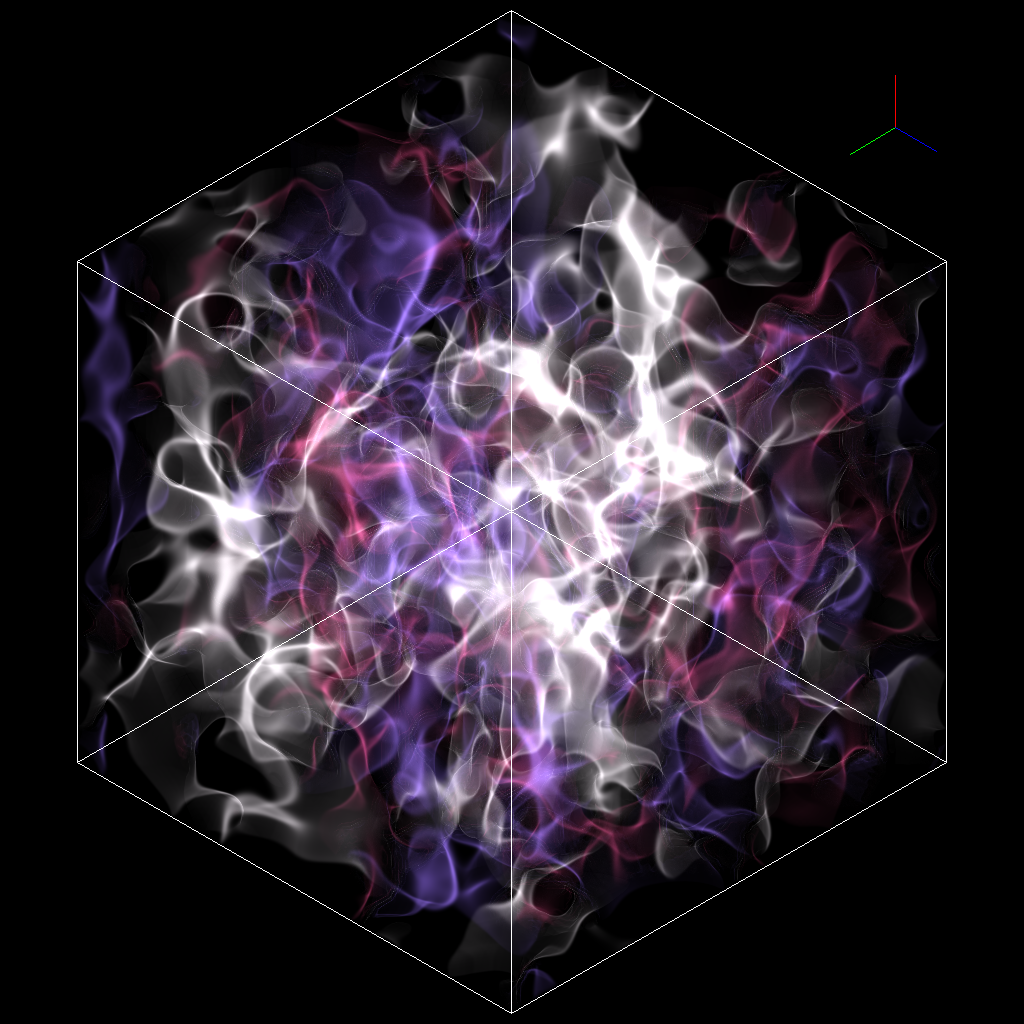}\\
    \caption{Volume rendering of contours of constant phase of $n_{e\mu}$ for the Fiducial\_1D (top row), Fiducial\_2D (center row), and Fiducial\_3D (bottom row) simulations. Phases of $-2\pi/3$, $0$, and $2\pi/3$ are shown in blue, white, and red, respectively. The left column shows the results at $t=0.29\,\mathrm{ns}$ during the linear growth phase of the fast flavor instability, the center column shows the results at $t=0.77\,\mathrm{ns}$ after the instability saturates, and the right column shows the results at $t=2.2\,\mathrm{ns}$ as the distribution is building power on small scales. The phase of $n_{e\mu}$ demonstrates wavefronts of the fastest growing unstable mode. The Fiducial\_1D data is copied into the $x$ and $y$ dimensions and the Fiducial\_2D data is copied into the $x$ direction for visualization purposes. Although there is significant multidimensional structure, the 3D results are qualitatively and quantitatively similar to the 1D and 2D results.}
    \label{fig:fiducial_volume_rendering_growth}
\end{figure*}

We present a series of simulations described in Section~\ref{sec:initial_conditions} designed to elucidate the differences between simulations with and without imposed spatial symmetries. All three distributions are unstable to the fast flavor instability, so their evolution is initially characterized by a linear growth phase, where unstable modes grow exponentially. Following this, the amplitude of the growing modes approach their maximal values, the evolution equations stop being well approximated by their linearized form, and power moves from the fastest growing mode to many other modes. Eventually, the distribution settles into a quasi-equilibrium with increasingly small fluctuations around average values. We will go through each of these three phases in detail and show that while artificially imposed spatial and directional symmetries can suppress instability by limiting which modes can be expressed, the final abundances of each flavor are the same between 1D, 2D, and 3D simulations for each of our three initial neutrino distributions. There are also some more subtle differences in the evolution of the Fourier and angular power spectra, but interpreting these details requires significant care in the context of numerical limitations.

\subsection{Overall Appearance}

We found that the complex phase of $n_{e\mu}$ is particularly useful in visualizing the complex matrix-valued solution, especially during the growth phase, as it clearly shows wavefronts of the flavor-transforming modes. In Fig.~\ref{fig:fiducial_volume_rendering_growth} we render surfaces of constant complex phase of this quantity at three points in time (each column corresponds to a snapshot time) and for 1D (top row), 2D (middle row), and 3D (bottom row) simulations. The Fiducial\_1D simulation (top row) only has a finite computational extent in the $\hat{z}$ direction, but is visualized by extruding the data in the $\hat{x}$ and $\hat{y}$ directions. Similarly, the Fiducial\_2D simulation (middle row) is simulated with finite computational extent in the $\hat{y}$ and $\hat{z}$ directions, but visualized by extruding in the $\hat{x}$ direction.

\subsubsection{Linear Growth}
During the growth phase of the Fiducial\_1D simulation (top left panel), we see evenly-spaced flat sheets of constant phase, indicating that the phase is varying in the $\hat{z}$ direction with a wavelength corresponding to the fastest growing mode (this was demonstrated in greater detail in \cite{richers_particle--cell_2021}). Due to the symmetry between the $\pm \hat{z}$ directions of the Fiducial initial conditions, the real part of the frequency of the fastest growing mode is zero and this is a standing wave; even as the mode amplitude grows, the surfaces of constant phase do not move until the instability saturates. 

The same structure exists in the multidimensional simulations, though the planes are distorted and interconnected. The center-left panel shows the equivalent data from the Fiducial\_2D simulation.  In the bottom left panel is the data from the Fiducial\_3D simulation. The wavelength of the fastest growing mode in the $\hat{z}$ direction in the multidimensional simulations matches that expected from the 1D analysis in \cite{richers_particle--cell_2021} even though the one-dimensional dispersion analysis does not account for the existence of variations in the $x$ and $y$ directions.  Animations of these visualizations show that the kinks and holes in the phase contours of the multidimensional simulations are also essentially unchanging until the instability saturates. The exact form of these features is non-unique and randomly determined by the perturbations to the initial conditions. In summary, shortly after the start of the simulation, the distribution settles into a particular multidimensional eigenmode that grows with the same characteristic growth rate and wavelength as the purely planar solution in the 1D simulation.

There are also small transient effects due to our choice of neutrino mixing parameters. The data plotted in the left column of Fig.~\ref{fig:fiducial_volume_rendering_growth} is entirely unpolluted by neutrino mass effects, since we choose $m_1=m_2=0$. However, $n_{e\tau}$ (not shown) quickly overcomes its random initial phases and establishes a constant phase throughout the domain instead of the planar structure seen in $n_{e\mu}$. This is a result of the $1\leftrightarrow 3$ vacuum mixing that grows linearly regardless of the amplitude of the initial perturbations. The fastest growing mode, however, still grows on top of the vacuum oscillations. Within $0.2\,\mathrm{ns}$ the fastest growing mode overtakes the vacuum oscillations and creates a phase pattern just like in $n_{e\mu}$. On the other hand, $n_{\mu\tau}$ (also not shown) has a phase distribution that is negative the phase $n_{e\mu}$ for the first $0.2\,\mathrm{ns}$. This reflects the fact that some of $n_{e\mu}$ induced by the fast flavor instability is subsequently being pushed into $n_{\mu\tau}$ by the vacuum potential. After $t=0.2\,\mathrm{ns}$, the phase of $n_{\mu\tau}$ also transforms to an altogether different distribution, instead varying on length scales comparable to $8\,\mathrm{cm}$ size of the domain.

The 90Degree\_3D simulation (not shown) proceeds quite similarly, except that the wavevector of the fastest growing neutrino mode points in the direction of $(\hat{z}-\hat{x})/\sqrt{2}$ (and in the opposite direction for antineutrinos). The phase pattern is again stationary, but with a longer wavelength due to the smaller magnitude of the self-interaction potential resultant from neutrino and antineutrino distributions with a smaller angle between them. In the 1D and 2D simulations, the $\hat{x}$ direction is assumed to be homogeneous, so the fastest growing is instead one with a wavevector in the $\hat{z}$ direction. The TwoThirds simulation instead has a fastest growing mode with nonzero real frequency. The phase pattern (also with a wavevector parallel to $\hat{z}$) of $n_{e\mu}$ and $n_{e\tau}$ drift along $\hat{z}$ with time. The effects of the vacuum potential are identical in these simulations as in the Fiducial simulation described above.

\subsubsection{Saturation}
The magnitude of off-diagonal components of a particle's density matrix cannot be larger than $\mathrm{Tr}(\rho)/2$, preventing infinite exponential growth. When the particle quantum states approach this limit in the Fiducial simulations at $t\approx0.3\,\mathrm{ns}$, the evolution equations begin to manifest their nonlinearity and the fastest growing modes cease to approximate the solution. The bottom-center panel of Fig.~\ref{fig:fiducial_volume_rendering_growth} shows the state of the Fiducial\_3D simulation at $t=0.77\,\mathrm{ns}$, well after the saturation of the instability has dismantled the fastest growing mode. The main features have a size scale larger than the wavelength of the fastest growing mode, but with no preferred direction. We will describe this more quantitatively in Sec.~\ref{sec:power_spectrum}. Quasi-planar structures occasionally spontaneously form in the phase of $n_{e\mu}$ with a random orientation, but are again destroyed within a few tenths of a nanosecond. The Fiducial\_2D (middle center panel) simulation qualitatively (and we will later see, also quantitatively) reproduces this analogously in two dimensions. Although the Fiducial\_1D (top center panel) simulation is still restricted to one spatial dimension, this chaotic, nonlinear behavor is also consistent with multi-dimensional simulations, though it is more obvious in the curves plotted in \cite{richers_particle--cell_2021} than in the volume renderings in Fig.~\ref{fig:fiducial_volume_rendering_growth}.

After $t\approx1\,\mathrm{ns}$, the distribution develops significantly more structure at smaller scales (e.g., bottom right panel of Fig.~\ref{fig:fiducial_volume_rendering_growth}). Although we are able to demonstrate convergence of the shape of this distribution (see App.~\ref{app:convergence}), the time at which this high-frequency structure arises depends on the angular resolution. We will try to argue in Sec.~\ref{sec:power_spectrum} and Sec.~\ref{sec:angular_structure} that this is a result of the interplay between modes of small spatial and angular scales, and that adding more particles delays the onset of these high-frequency fluctuations by reducing the initial amplitude of the angular modes. However, a full understanding of the origin of this feature requires further investigation. The same process occurs in the 1D (top right) and 2D (center right) simulations.

The same comments can be made about the 90\_Degree simulations, except that the smaller potentials lead to a later time of saturation at $t\approx0.5\,\mathrm{ns}$ and the onset of high-frequency structures at $t\approx1.5\,\mathrm{ns}$. The TwoThirds simulations, in turn, saturate at $t\approx1.5\,\mathrm{ns}$ and do not develop a strong high-frequency component in $n_{e\mu}$ over the $5\,\mathrm{ns}$ of the simulation. The fastest growing mode in the TwoThirds simulations is however not completely destroyed and continues to dominate the solution of $n_{e\mu}$ and $n_{e\tau}$, though with some random fluctuations on top of it. This is a reflection of the fact that the amount of possible flavor transformation is much smaller, so the post-saturation distribution remains much more similar to the pre-saturation distribution than in the Fiducial or 90Degree simulations.

\subsection{Average Flavor Evolution}
\label{sec:average_flavor}

Perhaps the most important metric of flavor transformation is the expectation value of the amount of each flavor of neutrino. We represent this by averaging the density matrices of particles as
\begin{equation}
    \langle \rho \rangle = \frac{\sum_p N \rho}{\sum_p N}\,\,,
\end{equation}
where the sum is over computational particles. As described in Sec.~\ref{sec:methods}, each particle has a unit-trace density matrix $\rho$ representing the flavor state of each neutrino and a scalar $N$ indicating the number of physical neutrinos that particle represents (and similarly for antineutrino quantities). In these simulations, all neutrinos and antineutrinos begin in electron flavor states, so $\langle \rho_{ee} \rangle$ is equivalent to the survival probability.

\begin{figure}[!ht]
    \centering
    \includegraphics[width=\linewidth]{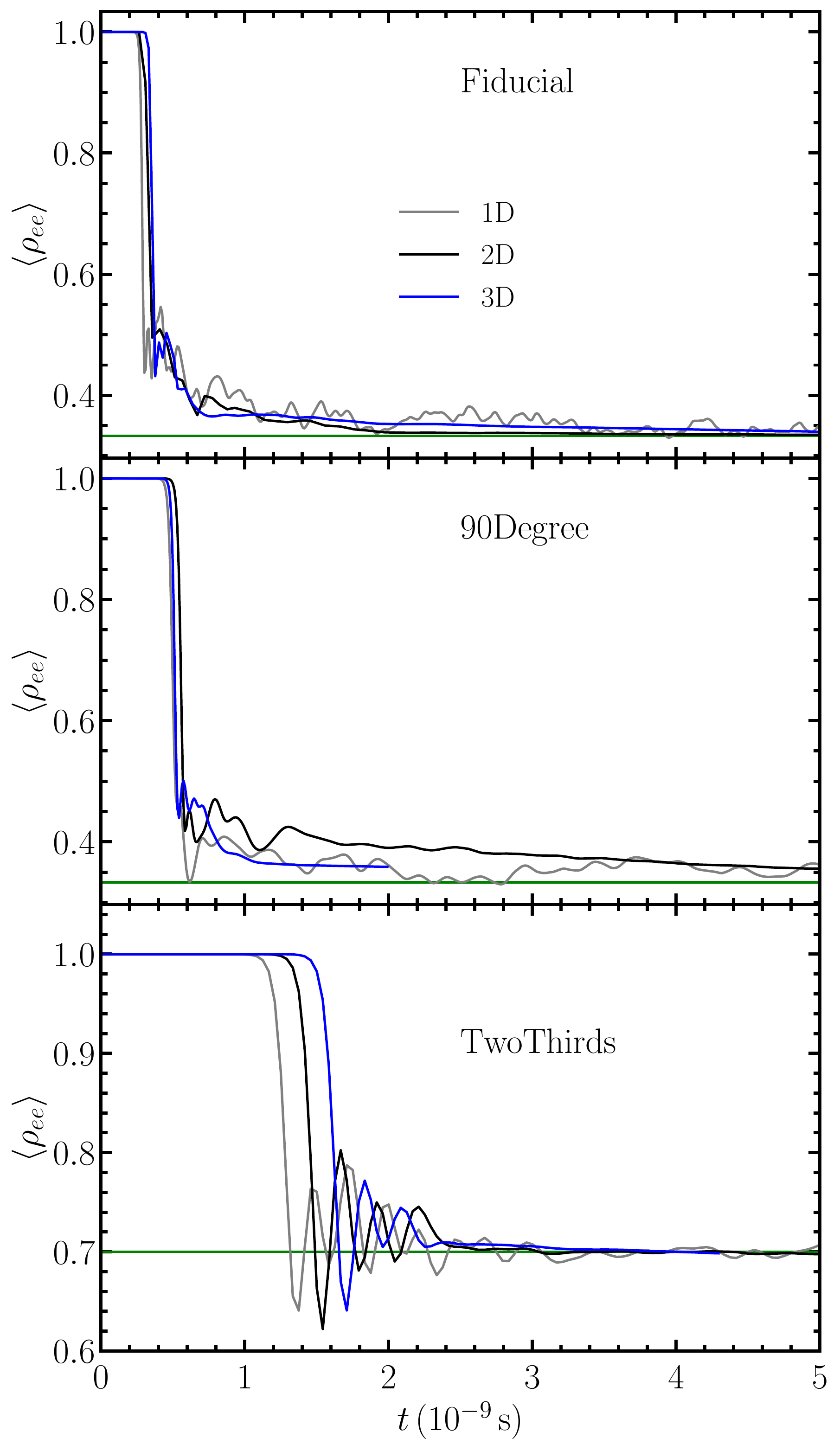}
    \caption{Time evolution of fraction of electron flavor neutrinos for the Fiducial (top), 90Degree (center), and TwoThirds (bottom) simulations. The 1D, 2D, and 3D simulation data are plotted in gray, black, and blue, respectively. The green line marks the equilibrium fraction of electron neutrinos ($1/3$ for Fiducial and 90Degree, 0.7 for TwoThirds). Equilibrium abundances (at late times) of each neutrino species seems to be independent of simulation dimensionality. The instability growth rates show weak dimensionality dependence in the 90Degree simulations, indicated by a slight offset in the time of the drop. The offsets in the drop for the TwoThirds simulation reflects initial mode amplitude differences in 1D, 2D, and 3D.}
    \label{fig:avgfee}
\end{figure}
\begin{figure}
    \centering
    \includegraphics[width=\linewidth]{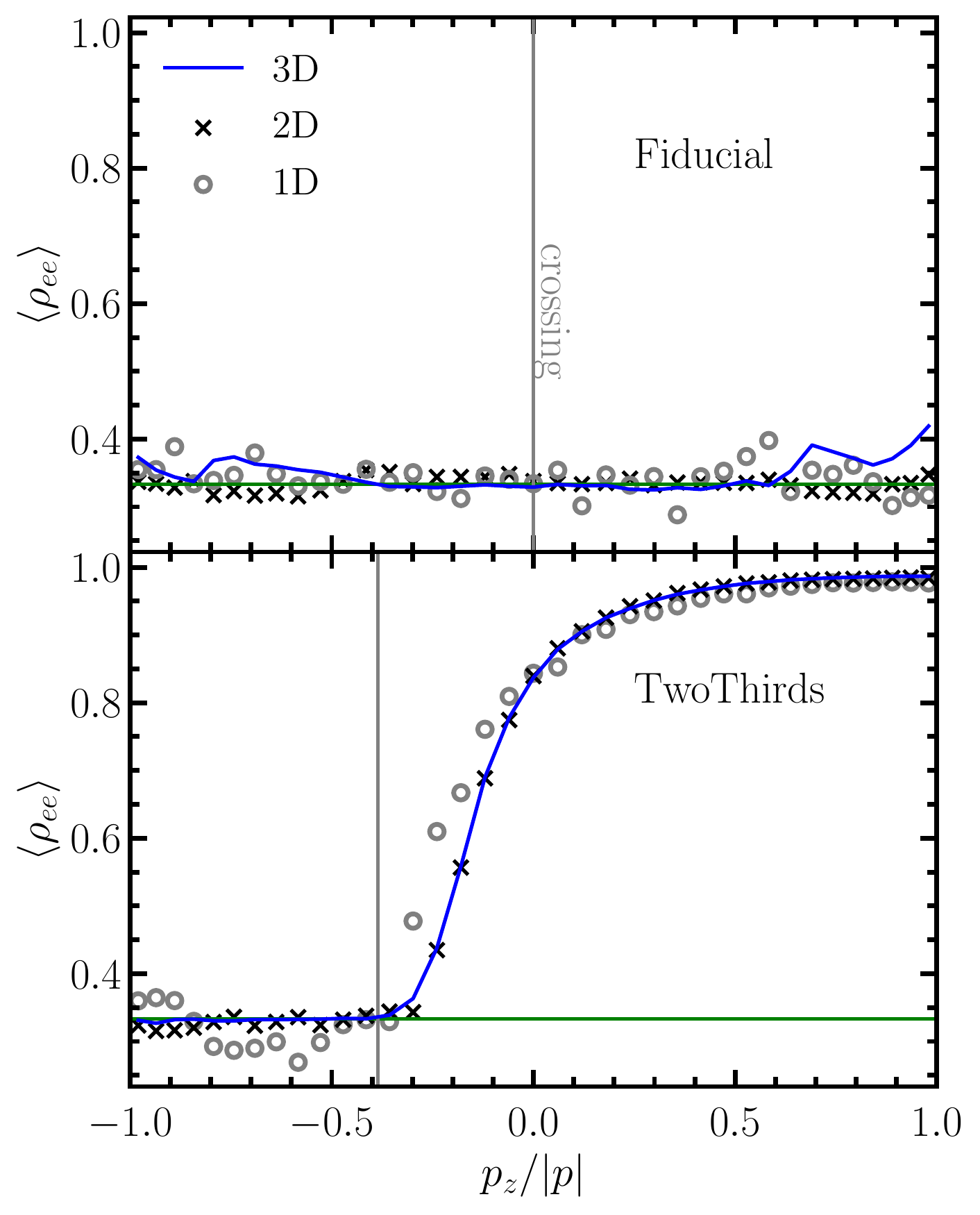}
    \caption{Probability that a neutrino (top) or antineutrino (bottom) starting in the electron flavor ends up in the electron (left), muon (center), or tauon (right) flavor state as a function of polar angle. Blue lines, black crosses, and gray circles are from 3D, 2D, and 1D simulations, respectively. The vertical gray line shows the angle of the crossing; to the left of this line there are more antineutrinos than neutrinos. The green horizontal line is at $\langle\rho_{ee}\rangle=1/3$, marking complete flavor mixing. Flavor content equilibrates between the flavors in the antineutrino-dominated region but no flavor transformation occurs far from the crossing.}
    \label{fig:averaged_flavor_evolution}
\end{figure}

The evolution of $\langle\rho_{ee}\rangle$ plotted in the top panel of Fig.~\ref{fig:averaged_flavor_evolution} further shows the dimension-independence of the solution for the Fiducial simulations. The values start on the left side of the plot at 1 because all neutrinos begin in electron neutrino states (modulo small perturbations). The amplitude of these perturbations grows exponentially until $t\approx0.3\,\mathrm{ns}$, where the amplitudes have grown to order unity, characterized by a steep drop in $\langle \rho_{ee}\rangle$. This happens at a very similar time for 1D, 2D, and 3D simulations, as the fastest growing mode can manifest in all three with the same growth rate. Progressing farther to the right on the plot, the averaged survival probabilities of the 1D, 2D, and 3D simulations all approach the equilibrium value of $1/3$ (horizontal green line indicating complete flavor mixing) at the same rate, albeit with random fluctuations. The nonlinearity of the equations after the saturation of the instability cause the precise solution to be chaotic, so this randomness is expected. The simulations of lower dimensionality have larger fluctuations, though this is simply a result of the fact that there are fewer grid cells to average over in the domain.

The slight spread in the time of the drop for the Fiducial simulations is due to differences in the initial amplitude of the fastest growing mode and not due to differences in the instability growth rate (all three clock in at $6.3\times10^{10}\,\mathrm{s}^{-1}$). Higher-dimensional simulations have more grid cells, which implies a larger number of possible modes, but the contribution to each of those modes from the random perturbations is smaller. The TwoThirds simulation (bottom panel of Fig.~\ref{fig:avgfee}) is much more sensitive to this effect, as the saturation of the instability in the 3D simulation is about $0.2\,\mathrm{ns}$ later than in the 1D simulations, even though the growth rate in all three TwoThirds simulations is measured at $1.3\times10^{10}\,\mathrm{s}^{-1}$. Following the saturation of the instability in the TwoThirds simulations, about 30\% of the neutrinos and 45\% of the antineutrinos have converted to a heavy lepton flavor. We will get to the 90Degree simulations shortly.

The inability of the TwoThirds simulations to undergo complete flavor transformation is understandable intuitively. Since the exponential growth from the fast flavor instability is entirely a result of the self-interaction terms (assuming of course that other contributions to the Hamiltonian are small), we can rely on the symmetries in that portion of the Hamiltonian. Specifically, the self-interaction term has no preferred flavor, owing to the fact that it arises only from neutral current interactions. The antineutrino Hamiltonian is also trivially related to the neutrino Hamiltonian ($\bar{H}_\mathrm{neutrino}=-H_\mathrm{neutrino}^*$). As a result, the system must conserve the total flavor charge $q_a=n_a - \bar{n}_a$ for each flavor $a$ and the evolution of the antineutrinos mirrors that of neutrinos. In the TwoThirds simulations, there is an excess of electron neutrinos over antineutrinos in the $+\hat{z}$ direction. If neutrinos and antineutrinos moving in this direction begin transforming flavor to a heavy lepton flavor, the total number of electron neutrinos would decrease faster than the number of electron antineutrinos, which on its own would result in a change in $q_a$. This transformation must be complemented by flavor transformation on the other side of the crossing, where electron antineutrinos are more abundant than electron neutrinos, in a way that prevents $q_a$ from changing. This is another way of stating that the fast flavor instability is fundamentally a multi-direction phenomena. However, it is not yet clear to us how to predict the final abundances without carrying out a simulation.

The 1D and 2D 90Degree simulations do, however, differ slightly in their measured growth rate ($3.6\times10^{10}\,\mathrm{s}^{-1}$) from the 3D simulation ($4.3\times10^{10}\,\mathrm{s}^{-1}$) because the former are unable to support the true fastest growing mode. The electron antineutrino distribution is pointed in the $\hat{x}$ direction, but the 90Degree\_2D simulation only allows inhomogeneity in the $\hat{y}$ and $\hat{z}$ directions and the 90Degree\_1D simulation allows inhomogeneity only along $\hat{z}$. That is, the antineutrino distribution is pointing out of the plane of the 2D computational domain, and the fastest growing mode has a wavevector in a direction between the two distributions. As part of our numerical tests (App.~\ref{app:convergence}), we did try simulations with a grid instead in the $\hat{x}-\hat{z}$ plane, which resulted in a growth rate that matched the 3D simulation. Despite the differences in growth rate, the 1D, 2D, and 3D simulations all approach the same equilibrium value of $\langle\rho_{ee}\rangle = 1/3$.

The Fiducial and 90Degree simulations of all dimensionalities approach an equilibrium characterized by complete flavor mixing. This is reflected in the tendency of $\langle\rho_{ee}\rangle$ toward $1/3$ in Figure~\ref{fig:avgfee}. We also show the azimuthally-integrated equilibrium distributions for the Fiducial simulations in the top panel of Figure~\ref{fig:averaged_flavor_evolution}. The gray vertical line shows the polar angle where the crossing in the original distribution was, and the blue curve, black marks, and gray marks respectively show the 3D, 2D, and 1D averages at that polar angle at $t\approx5\,\mathrm{ns}$. At this point, there still are some fluctuations, but it is clear that all polar angles show even flavor mixing. The asymmetry of the 90Degree simulations would require a mollweide projection color plot to show the spatially-averaged survival probability for each direction, since there is no reason to expect azimuthal symmetry around any particular axis. We do not show the resulting plot because it is exceedingly boring - the survival probabilities also all lie near 1/3 for all dimensionalities.

The TwoThirds simulation (bottom panel) is more interesting. The location of the crossing in the original distribution is closer to the $-\hat{z}$ direction (also apparent in Figure~\ref{fig:initial_conditions}). The result is a distribution that has a much smaller region where there is an excess of antineutrinos over neutrinos, and in that region the difference between the differential neutrino and antineutrino densities are smaller. We see that in simulations of all three dimensionalities, the this region "inside" the crossing completely mixes, and the region outside the crossing compensates in a way that preserves the total neutrino-antineutrino asymmetry. This independently corroborates similar results found in \cite{wu_collective_2021}. However, an explanation of the precise functional form of the compensating flavor transformation to the right of the gray line still eludes us. The 1D results seem to have significant scatter from the 2D and 3D results, but this appears to be because the 1D simulations have more difficulty relaxing. If the simulation is run for a longer period of time, (not shown to preserve the common time snapshot), they points do fluctuate around the multidimensional results.

\subsection{Power Spectrum}
\label{sec:power_spectrum}

\begin{figure*}[!ht]
    \centering
    \includegraphics[width=\linewidth]{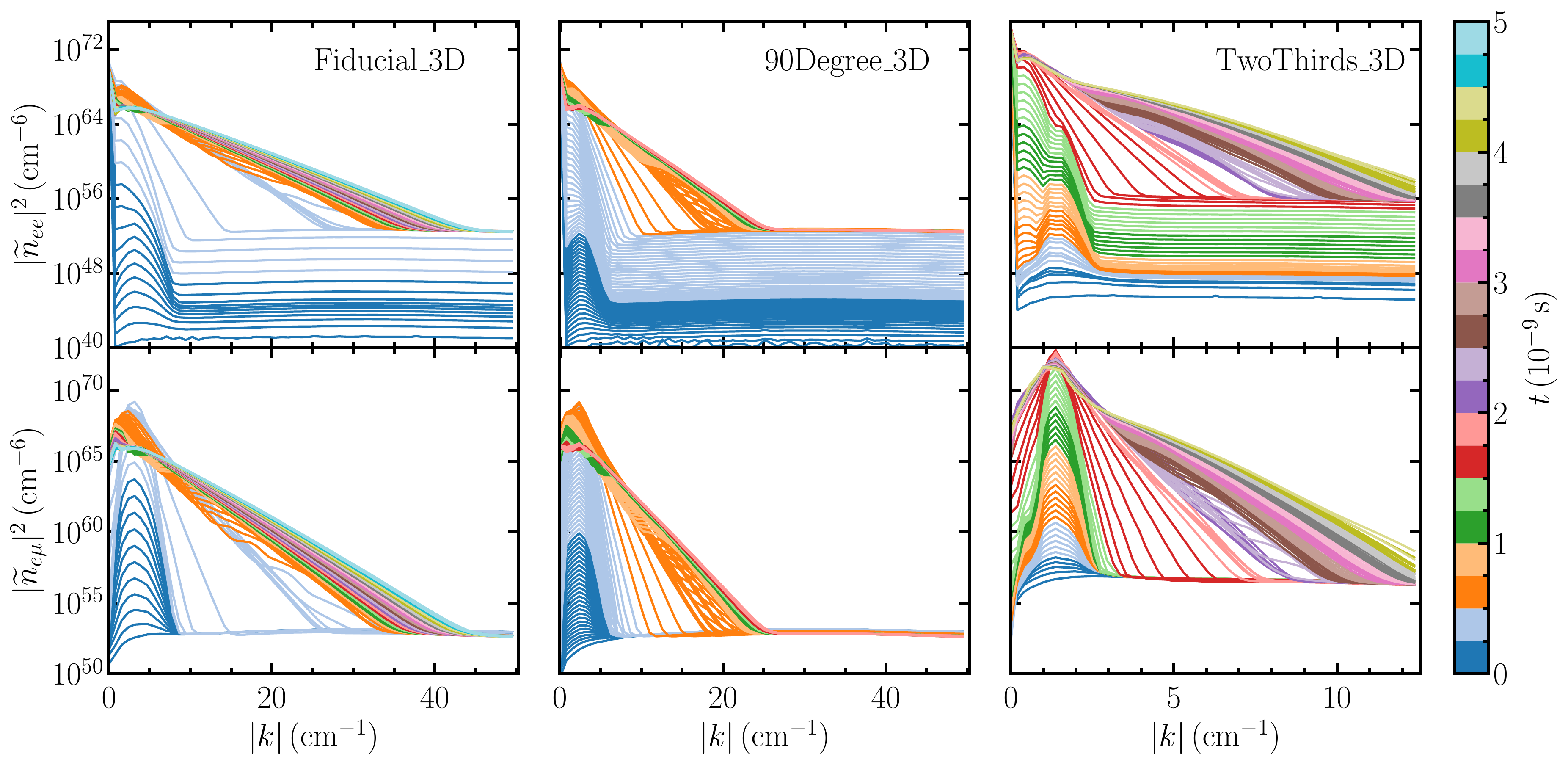}
    \caption{Time evolution of the Fourier power spectra as defined in Eq.~\ref{eq:spectrum_shell} for the Fiducial\_3D (left column), 90Degree\_3D (center column) and TwoThirds\_3D (right column) simulations. The color indicates the time of the snapshot for which the power spectrum was evaluated. Early times are characterized by a rising bump with a peak at the fastest growing wavenumber, and late times show an exponential distribution centered at $k=0$. The slow spread of the exponential tail is a numerical artifact (see App.~\ref{app:convergence}).}
    \label{fig:fft}
\end{figure*}
We now turn our attention to the evolution of the Fourier power spectrum of the neutrino distribution. In particular, we assume the convention that
\begin{equation}
    \widetilde{n}_{ab}(\mathbf{k}) = \sum_{l=0}^{n_x-1} \sum_{m=0}^{n_y-1} \sum_{n=0}^{n_z-1}  e^{-i\mathbf{k}\cdot\mathbf{x}_{ijk}} n_{ab}(\mathbf{x}_{lmn})\,\,,
\end{equation}
where we use $l$, $m$, and $n$ to denote the index of the $x$, $y$, and $z$ position of the grid cell and $a$ and $b$ to denote the flavor index. The above is evaluated as a discrete Fast Fourier Transform using Scipy \cite{virtanen_scipy_2020}, and we use the resulting grid of wavenumbers $k$ from 0 to $\pi/L_z$ with spacing $\Delta k = \pi/\Delta z$ (where $\Delta z=L_z/n_z$ is the spatial grid size). We integrate power in shells of magnitude of $\mathbf{k}$ according to
\begin{equation}
    |\widetilde{n}_{ab}|^2 (k) = \sum_{k_\mathrm{min} < |\mathbf{k}| \le k_\mathrm{max}} |\widetilde{n}_{ab}(\mathbf{k})|^2\,\,.
    \label{eq:spectrum_shell}
\end{equation}
$k_\mathrm{min}=k-\Delta k/2$ and $k_\mathrm{max}=k+\Delta k/2$ define the inner and outer radius of the spherical shell (in $\mathbf{k}$ space) over which power is integrated for each value of $k$.

Returning to the Fiducial\_3D simulation, the spectra of $n_{ee}$ and $n_{e\mu}$ are shown in the left column of Fig.~\ref{fig:fft}. The color denotes the time at which the spectrum was calculated; all of the dark blue curves are from $t\in [0,0.25)\,\mathrm{ns}$ and hence all represent points in time in the linear growth phase (which ends at $t\approx0.3\,\mathrm{ns}$). The results are essentially identical to those in the 1D simulations of \cite{richers_particle--cell_2021} during this phase. The fastest growing mode, characterized by a wavelength of $\lambda=2.2\,\mathrm{cm}$, is visible as an exponentially growing bump in $|\widetilde{n}_{e\mu}|^2$. $|\widetilde{n}_{ee}|^2$ also grows sympathetically, since an increase in $|n_{e\mu}|$ requires a decrease in $n_{ee}$ (the quantum state "rotates" away from the pure electron flavor state). Meanwhile, the noise floor ($k\gtrsim1.5\,\mathrm{cm}^{-1}$) increases linearly (indicated by horizontal lines that become increasingly dense) at a scale that is not visible on the bottom panel, and slows down toward $|\widetilde{n}_{ee}|\approx10^{44}\,\mathrm{cm}^{-6}$. This reflects the behavior of the spectrum of $n_{\mu\tau}$ (not shown), which grows because the vacuum part of the Hamiltonian rotates perturbations into $n_{e\mu}$. The subsequent exponential growth of the flat part of the spectrum (indicated by evenly spaced light blue horizontal lines) is likely a numerical artifact that tracks the exponential growth of the physical instability peak. The increasingly large amplitude of the physical instability excites artificial vibrations throughout the domain, but at amplitudes more than ten orders of magnitude smaller than the peak itself. Higher-resolution simulations have a lower noise floor, and the floor stops growing exactly when the amplitude of the peak ceases to be able to grow. 

When the instability saturates (light blue curves), power rapidly spreads to larger wavenumbers and immediately establishes an exponential distribution that intersects the noise floor at $k\approx 25\,\mathrm{cm}^{-1}$ (for the resolution described in Tab.~\ref{tab:simulations}). This (still light blue) tail remains static for around a tenth of a nanosecond before the high-$k$ part of the spectrum suddenly kicks out to intersect the noise floor at $k\approx33\,\mathrm{cm}^{-1}$ (rightmost light blue curves). As time progresses (orange curves), this feature travels up the slope.

The multidimensional simulation spectra differ significantly from the 1D simulation spectra after saturation. First, the much larger number of cells results in a significantly smoother spectrum due to the summation in Eq.~\ref{eq:spectrum_shell}. Second, the multidimensional simulations are both much more sensitive to the fidelity of the simulation (especially the angular resolution; see Appendix~\ref{app:convergence}), and converge more quickly with resolution. Because of this, the long-term spread of the exponential tail (orange and beyond in the left column of Fig.~\ref{fig:fft}) seems to be a numerical artifact. As the angular resolution increases this spread slows, and including more spatial dimensions makes the difference between simulations of different resolutions more severe.

\begin{figure}
    \centering
    \includegraphics[width=\linewidth]{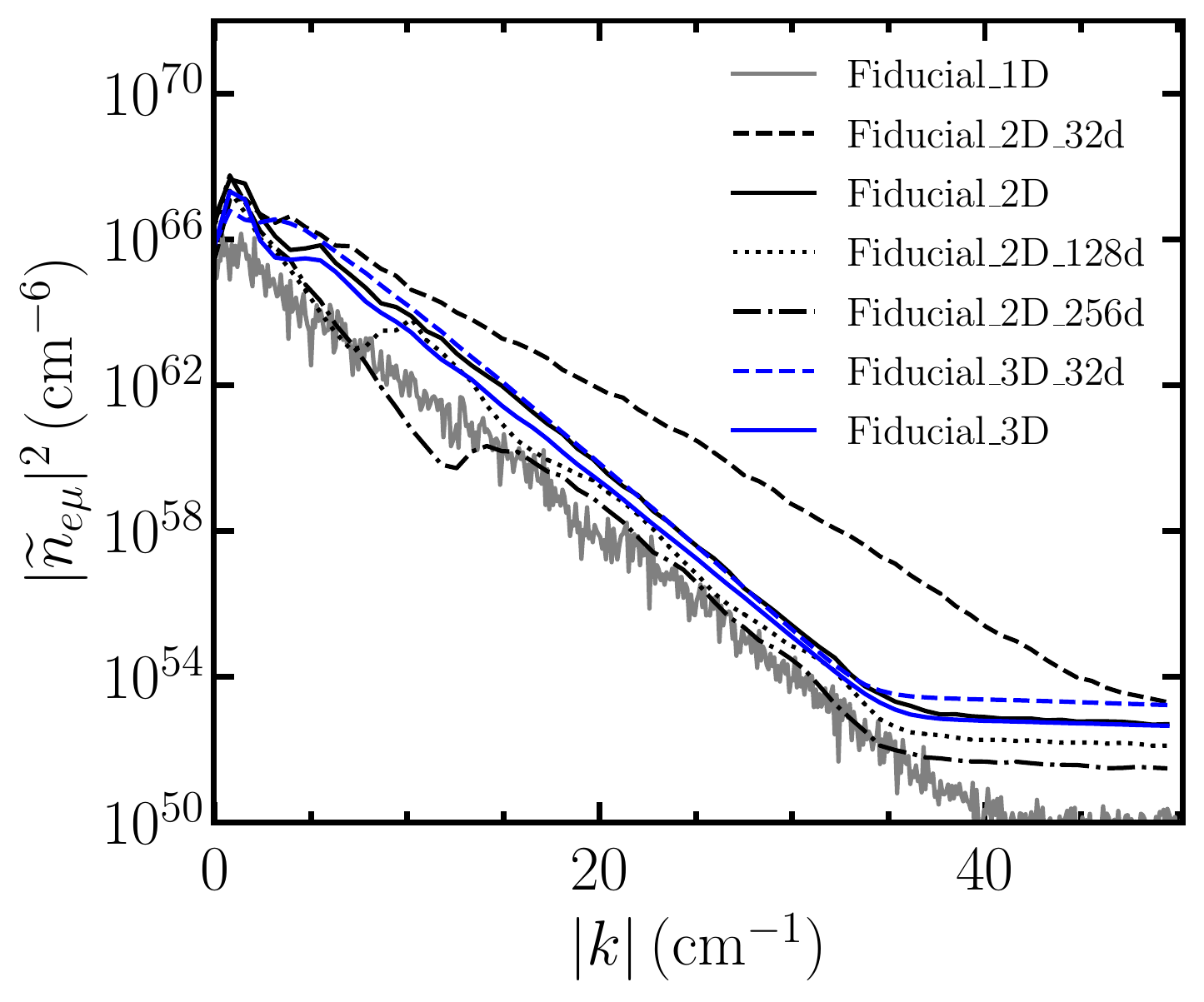}
    \caption{Snapshot of the power spectrum of $n_{e\mu}$ at $t=1\,\mathrm{ns}$ for the 1D, 2D, and 3D Fiducial simulations under variations in the number of particles per cell. 32d means 32 equatorial directions or 378 particles per cell, 128d means 6022 particles per cell, and 256 means 24088 particles per cell. Solid curves reproduce a single snapshot of the data shown in Fig.~\ref{fig:fft}. The 2D and 3D results at similar resolutions have very similar peak behavior, and the resolution dependence is similar for 2D and 3D simulations (though 1D simulations behave differently). The four black curves show that the slope of the exponential tail on either side of the upward-propagating feature is robust, but the time at which that feature is launched is increasingly late with increasing resolution.}
    \label{fig:fft_1ns_convergence}
\end{figure}
We will argue in Sec.~\ref{sec:angular_structure} that this feature is a result of high-$k$ modes interacting with high-$l$ angular modes, though we still lack a complete description. Even so, at this point we can provide some numerical evidence suggesting the credibility of the tail kick. The size, shape, and speed, and timing of this feature is consistent between the 2D and 3D simulations of the same resolution. This is indicated by the overlap of the solid black and blue curves in Fig.~\ref{fig:fft_1ns_convergence}, which shows both spectra at $t\approx1\,\mathrm{ns}$. The size, shape, and speed are consistent between simulations of different resolutions as indicaated by the series of black curves in Fig.~\ref{fig:fft_1ns_convergence}, which show the spectra from the Fiducial\_2D simulation, along with simulations at higher angular resolution (Fiducial\_2D\_128d and Fiducial\_2D\_256d). Note that the slope of the spectrum on either side of the bump is the same. The sole difference is the location of the bump, an indication that the feature was launched sooner in the lower-resolution simulations. Looking at the 1D data (gray), this feature is not at all apparent, though the (significantly noisier due to fewer cells to average over) spectrum appears to already match the slope in the multidimensional simulations from after the tail kick is launched. The lower resolution 2D and 3D simulations (dashed black and blue, respectively) differ greatly for $k\gtrsim 5\,\mathrm{cm}^{-1}$, even though they were quite similar for the standard resolution. This is because, since even at this early time, the artificial rapid spread of the exponential tail is significantly faster in the 2D simulation than in the 3D simulation (see App.~\ref{app:convergence}).

A similar process occurs in the 90Degree\_3D simulations (middle column of Fig.~\ref{fig:fft}). The fastest growing mode has a slightly longer wavelength (smaller $k$). This is a result of the smaller average angle between neutrinos and antineutrinos compared to the Fiducial simulation, which results in a smaller average strength of the self-interaction Hamiltonian. The instability saturates at $t\approx0.5\,\mathrm{ns}$ (dark orange curves), and a similar tail-kick travels up the exponential tail.

The TwoThirds simulation (right column of Fig.~\ref{fig:fft}) shows two separate modes growing, and in this case the azimuthally asymmetric mode dominates the symmetric mode. In the top right panel, the amplitude of the $k\approx0.4\,\mathrm{cm}^{-1}$ mode is initially imperceptible for $t\leq 0.5\,\mathrm{ns}$ (dark blue and light blue curves), but by $t=0.8\,\mathrm{ns}$ (light orange), it has overtaken the amplitude of the mode at $k\approx1.5\,\mathrm{cm}^{-1}$. That is to say that the mode causing the low-$k$ feature is initially smaller amplitude than the high-$k$ feature, but has a faster growth rate. However, the bottom right panel shows that the mode amplitude remains very small in $\widetilde{n}_{e\mu}$.

This is naturally explained by a rapidly-growing, azimuthally odd mode with a long wavelength. If the value of $\langle \rho_{e\mu}\rangle (\hat{x})=-\langle \rho_{e\mu}\rangle (-\hat{x})$, when $n_{e\mu}$ is calculated these two contributions cancel out. However, in both cases $\langle \rho_{ee}\rangle$ must decrease as the flavor vector rotates away from the electron flavor axis, so the mode invisible in $\widetilde{n}_{e\mu}$ is still represented in $\widetilde{n}_{ee}$. If we compute the cell-integrated transverse neutrino flux $f_{e\mu,x}=\sum N \rho_{e\mu} p_x/|p|$ (where the sum is over all particles in a grid cell), the factor $p_x/|p|$ itself is odd in direction, so $f_{e\mu,x}$ is not sensitive to isotropic modes and only shows direction-odd modes. Indeed, plots of the phase of this quantity (not shown) exhibit a wave with a wavelength of $16\,\mathrm{cm}$ and wavevector along $\hat{z}$, consistent with the low-$k$ peak.

The behavior of the exponential tail in the TwoThirds simulation is similar to that of the Fiducial and 90Degree calculations in that it also temporarily freezes ($t\approx1.8\,\mathrm{ns}$, salmon curves) before the high-$k$ part of the tail kicks out and sends a bump traveling up the slope. Finally, the main peak in the bottom right plot is slowly moving toward $k=0$ as in the other simulations. The movement of the peak is slower in the TwoThirds simulation because of the overall longer average timescale from the weaker self-interaction potential.

\subsection{Angular Structure}
\label{sec:angular_structure}
None of the simulations in this work restrict any angular dimensions; every simulation in Table~\ref{tab:simulations} has 1506 particles in each cell with momenta distributed evenly over the full $4\pi$ steradians of solid angle. Even so, restrictions in spatial dimensions can have some impact on the angular structure of the results, and restrictions in angular dimensions can prevent modes from growing even if all three spatial dimensions are included.

\begin{figure}[!ht]
    \includegraphics[width=.8\linewidth]{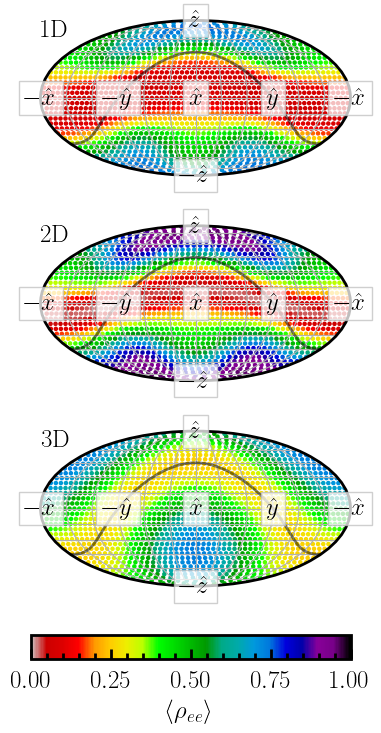}
    \caption{Mollweide projection of the electron neutrino abundance in the 90Degree simulation at $t=0.56\,\mathrm{ns}$, or shortly before the flavor instability saturates. The 1D (top) and 2D (center) show similar mode structure, though the mode in the 3D simulation (bottom) much more closely tracks the initial ELN crossing (gray curve).}
    \label{fig:mollweide_90deg}
\end{figure}
The imposed symmetries significantly affect the structure of the fastest growing mode in the 90Degree simulations, since the distribution itself does not have any axis of symmetry. We show in Fig~\ref{fig:mollweide_90deg} a mollweide projection of the spatially averaged value of $\rho_{ee}$ for each particle direction in the three 90Degree simulations at $t=0.56\,\mathrm{ns}$, shortly before the instability saturates. The precise colors should not be compared directly, as they can differ between panels due to slightly different simulation output times. Instead, the valuable information is in the shape of the regions that exhibit strong flavor transformation. In the 3D simulation (bottom panel), the fastest growing mode exhibits neutrinos in the directions near the original ELN crossing (gray curve) transforming flavor most quickly. In the 1D calculation (top panel), the band of maximal flavor transformation does tend slightly toward the ELN crossing, but appears much more tightly bound to the equatorial plane as a result of the influence of the imposed translational symmetry. The 2D calculation (center panel) is quite similar to the 1D calculation, since the imposed symmetry results in the majority of the antineutrino distribution being pointed in a direction ($\hat{x}$) that is assumed to be homogeneous. Indeed, repeating the 90Degree\_2D simulation with the imposed symmetry instead in the $\hat{y}$ direction (not shown) yields a growing mode with an angular structure matching that of the 3D simulation. This corroborates the argument in Sec.~\ref{sec:average_flavor} that because the fastest growing mode cannot exist in the 90Degree\_1D and 90Degree\_2D simulations, a different, more slowly growing mode compatible with the imposed symmetry takes the mantle during the growth of the instability.

\begin{figure}
    \centering
    \includegraphics[width=\linewidth]{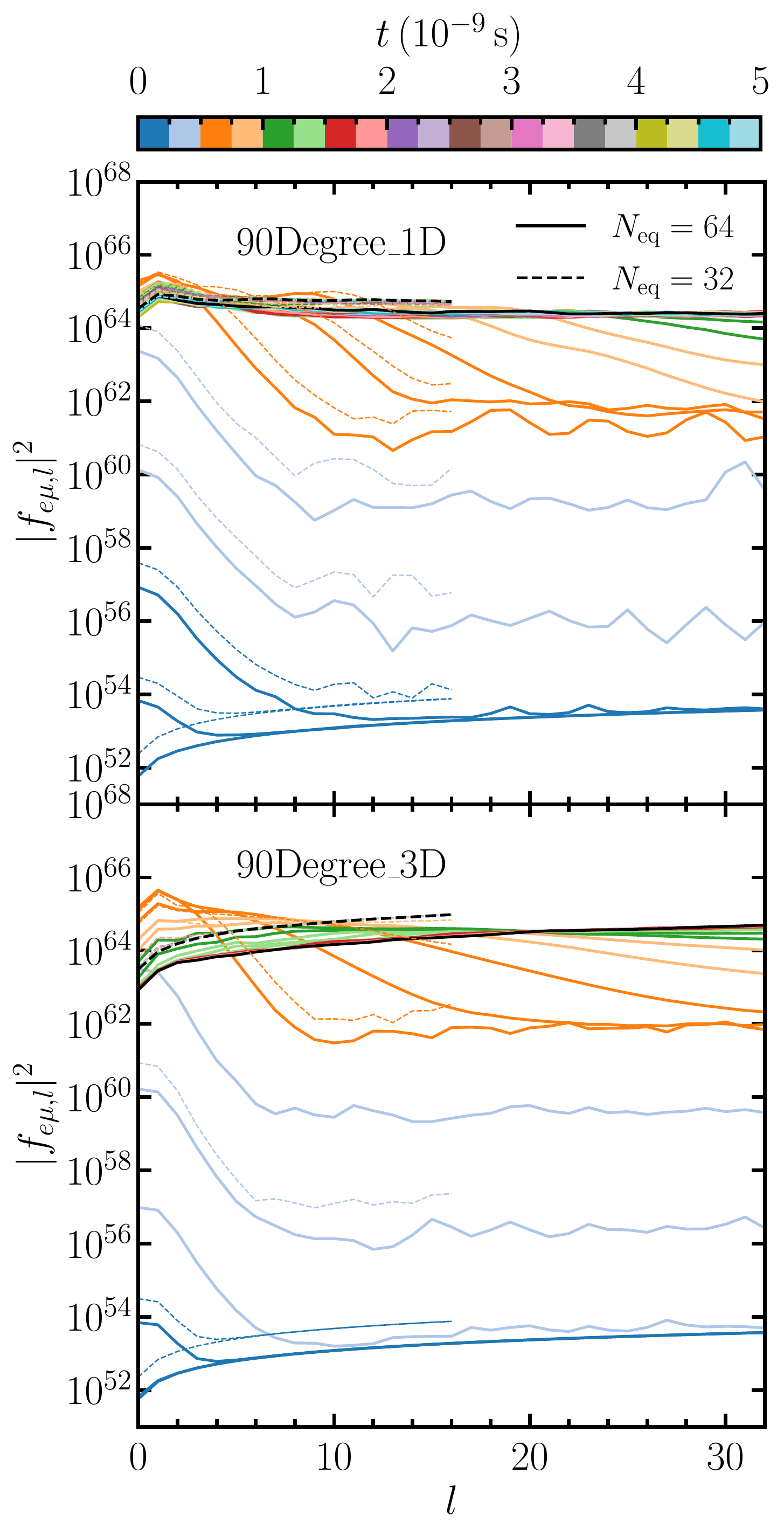}
    \caption{Evolution of the angular power spectrum of $f_{e\mu}$ as defined in Eq.~\ref{eq:angular_power} for the 1D (top panel) and 3D (bottom panel) 90Degree simulations. Solid lines show data from the standard simulations listed in Tab.~\ref{tab:simulations} with 1506 computational particles per cell, while dashed lines show data from otherwise identical simulations with 378 particles per cell. The standard simulations contain significant information up to $l=32$, while the low-resolution simulations only contain significant information up to $l=16$. The time of the snapshot used to make the curve is indicated by color in the same way as in Fig.~\ref{fig:fft}. The "tail kick" in Fig.~\ref{fig:fft} corresponds to the saturation of low-$l$ angular power (dark orange curves) and not to power cascading to smaller scales than the angular resolution can support (light orange curves).}
    \label{fig:angular_power_spectrum}
\end{figure}
We can describe the angular structure of the neutrino radiation field in terms of a spherical harmonic power spectrum. A continuous angular distribution function can be decomposed into spherical harmonics as
\begin{equation}
    f_{ab}(\cos\theta,\phi) =  \sum_{l=0}^{\infty}\sum_{m=-l}^l f_{ab,lm} Y_{lm}(\cos\theta,\phi)\,\,.
\end{equation}
In this section, $f_{ab}$ refers to the angular distribution function flavor matrix, which is related to the number density matrix via $\int d\Omega f_{ab}=n_{ab}$, rather than the number flux vector. We can approximately evaluate the coefficients $f_{ab,lm}$ as \cite{fornberg_spherical_2014}
\begin{equation}
\begin{aligned}
    f_{ab,lm} &=  \int d\Omega f_{ab}(\cos\theta,\phi) Y_{lm}^*(\cos\theta,\phi) \\
    &\approx \frac{1}{V} \sum_p N_p \rho_{p,ab} Y_{lm}^*(\cos\theta_p,\phi_p)
    \end{aligned}
\end{equation}
We have assumed that all particles occupy the same solid angle in direction, which they do approximately by construction. We then evaluate a one-dimensional angular power spectrum as
\begin{equation}
    |f_{ab,l}|^2 = \sum_{m=-l}^l |f_{ab,lm}|^2
    \label{eq:angular_power}
\end{equation}

Looking first at the power spectrum of $f_{e\mu}$ in the 90Degree\_3D simulation (bottom panel of Fig.~\ref{fig:angular_power_spectrum}), we see that the initial random perturbations have a power spectrum characterized by less power on smaller angular scales (lowest dark blue curve), as low-$l$ modes effectively average over a large number of incoherently random values. The fastest growing mode begins to emerge in the next two dark blue curves corresponding to $t<0.25\,\mathrm{ns}$, exhibited by an angular spectrum peaked at $l=0$ and that extends out to $l=10$. The mode then grows in amplitude with the same spectral shape for the next $0.25\,\mathrm{ns}$ (light blue curves), bringing up an artificial the high-$l$ plateau with it, similar to the exponential of the Fourier plateau discussed in Sec.~\ref{sec:power_spectrum}. It is once again important to note that the level of the artificial plateau is several orders of magnitude below the peak. After $t=0.5\,\mathrm{ns}$ (dark orange curves), the angular power spectrum stops growing and power cascades away from $l=0$ to smaller angular scales.

The time during which the angular spectrum is extending to higher $l$ (dark orange curves) corresponds to the time of the "tail kick" feature in Fig.~\ref{fig:fft} (see Sec.~\ref{sec:power_spectrum}). It is not until $t\approx0.75\,\mathrm{ns}$ (light orange curves) that this cascade hits $l=32$, the highest-$l$ mode representable by our 1506 particles per cell (64 particles in the $\hat{x}-\hat{y}$ plane in momentum space). By this point, the tail-kick feature is already halfway up the Fourier power spectrum slope in Fig.~\ref{fig:fft}. Hence, we believe the tail kick feature is a physical result of modes of high-$l$ interacting with modes of high-$k$, rather than a numerical artifact of our limited angular resolution.

We commented in Sec.~\ref{sec:power_spectrum} that the tail-kick occurs later for simulations with higher angular resolution (more particles per cell). In the top panel of Fig.~\ref{fig:angular_power_spectrum} we also see dashed lines that depict the spectra from otherwise equivalent simulations performed with fewer particles per cell. Since these low-resolution simulations have only 32 particles in the $\hat{x}-\hat{y}$ plane in momentum space, the highest-$l$ mode that is represented in the data is $l=16$. The initial noise level (bottom dashed blue curve) is a factor of 4 higher because each mode now represents an average over a factor of 4 fewer particles. This factor of 4 is maintained throughout the growth phase (dark blue and light blue, note that there are fewer dashed curves because the data output frequency was a factor of 2 lower), so it reaches its maximal amplitude sooner than the standard resolution simulation.

For $t\gtrsim0.75$ (light orange and later solid curves), power continues flowing from low $l$ to high $l$ modes. By the end of the simulation (solid black curve), the angular power spectrum again takes on the shape representative of incoherent random noise, much like the random initial conditions (lowest dark blue curve), but with a much larger amplitude. Once again, the power in the lower-resolution simulation (dashed black curve) is a factor of 4 larger because each mode averages over a factor of 4 fewer incoherent particles. This spread to small angular scales is consistent with that observed by \citet{johns_fast_2020}, though our method is not susceptible to the same kind of spurious oscillations associated with a boundary condition imposed at small angular scales. Since all of these results are distributed from a set of computational particles, there is no explicit angular boundary condition enforced. Rather, it just arises naturally (though no less artificially) as a limited number of particles are less able to provide the angular detail that a larger number of particles could.

Looking briefly at the top panel of Fig.~\ref{fig:angular_power_spectrum}, we see that the same line of reasoning applies to the 90Degree\_1D simulation. The exception is the form of the angular power spectrum at the end of the simulation (solid black curve). The shape is mostly flat, unlike the monotonically rising power spectrum indicative of incoherent random noise (like the bottom most dark blue curve). This is a result of the fact that the restricted spatial dimensions suppresses angular anisotropy, as was reported in \cite{richers_particle--cell_2021}.

The Fiducial and TwoThirds simulations have initial conditions that are initially symmetric around the $\hat{z}$ direction, but the evolution of the angular spectrum (not shown) is still very similar to the 90Degree simulation. In \cite{richers_particle--cell_2021}, we reported that in the Fiducial\_1D simulation, this azimuthal symmetry is very well preserved through the entirety of the simulation, even though the initial perturbations are random and not axisymmetric. When we relax the assumption of homogeneity in $\hat{x}$ and $\hat{y}$, the Fiducial\_2D and Fiducial\_3D simulations are similarly characterized by a fastest growing mode that is azimuthally symmetric around the $\hat{z}$ axis. However, the high degree of azimuthal symmetry begins to breaks down even before the instability saturates, and fluctuations begin to exist on increasingly smaller angular scales. The fastest growing mode in the TwoThirds simulations is azimuthally asymmetric (see Section~\ref{sec:power_spectrum}), but fluctuations once again move from large to small angular scales upon saturation of the instability. In the case of the TwoThirds simulations, it is clear that if axial symmetry were imposed, only the slower mode would be able to grow.

\section{Conclusions}
\label{sec:conclusions}
We perform the first simulations of the fast flavor instability in three spatial dimensions and two momentum dimensions (i.e. all dimensions relevant when the vacuum Hamiltonian is negligible). We simulate each of three idealized initial conditions in one, two, and three spatial dimensions in order to probe for effects of artificially-imposed symmetries. In all cases, the simulations are characterized by a linear growth phase where unstable modes grow exponentially, followed by saturation of the instability, where power cascades to smaller spatial (Fig.~\ref{fig:fiducial_volume_rendering_growth}) and angular (Fig.~\ref{fig:angular_power_spectrum}) scales. For our choice of initial conditions, the predictions of the abundance of each neutrino flavor after the fast flavor instability saturates are independent of the dimensionality of the simulation (Fig.~\ref{fig:avgfee} and \ref{fig:averaged_flavor_evolution}). This is a very welcome result that supports the validity of the many previous simulations performed with imposed symmetries, though our set of three initial conditions is far from exhaustive.

However, imposed symmetries are not without impact. The growth rate of the instability is mildly dependent on simulation dimensionality when the fastest growing mode has a wavevector with some component along a direction assumed to be homogeneous (Fig.~\ref{fig:fft}). When spatial symmetries are imposed, the angular structure of the fastest growing mode can also be significantly different (Fig.~\ref{fig:mollweide_90deg}). In addition, the fastest growing mode in our "TwoThirds" set of simulations is axially asymmetric. Imposing axial symmetry would preclude this mode from existing, and one would only see the more slowly-growing axially symmetric mode (right column of Fig.~\ref{fig:fft}).

Similar to the 1D calculations in \cite{richers_particle--cell_2021}, Fourier transforms of the neutrino distribution show exponentially growing perturbations in agreement with the dispersion relation. When the instability saturates, power quickly jumps to higher wavenumbers, establishing an exponential tail (Fig.~\ref{fig:fft}). Our multidimensional simulations show a slower growth of this tail and stronger dependence on the angular resolution. From these data, we expect that physical distributions would be characterized by a power distribution at late time centered at $k=0$ and with a static exponential tail. The much larger number of grid cells also make the power spectrum much clearer, such that we can identify a "tail-kick" feature in the spectrum, where shortly after saturation of the instability, the high-$k$ part of the spectrum suddenly grows, sending a bump up the slope of the power spectrum toward small wavenumbers. We believe this is related to the growth of modes at small angular scales interacting physically with modes of small spatial scales, though more work is required to confirm and explain this on a fundamental level.

The particle-in-cell method keeps some features of a multi-body quantum system in that the flavor vector length of each particle cannot change or average out, but only rotate. However, the largest physical shortcoming of this work is the lack of multi-body entanglement. There have been significant advances on this front recently (e.g. \cite{rrapaj_exact_2020,roggero_dynamical_2021,roggero_entanglement_2021,hall_simulation_2021}), and if entanglement proves to be be important in general in the limit of many particles, these results will likely need to be reconsidered.

In the future it will be important to simulate a larger variety of initial conditions to map out the steady-state flavor distribution induced by each. A confidence that the abundances of each neutrino flavor is independent of any numerical choices will make sub-grid modeling a much more attractive and feasible prospect. In addition, it will be interesting to simulate cases where the matter and vacuum potentials are relevant in order to determine if multidimensional effects are relevant more broadly. 

\section{Acknowledgements}
We are very grateful to Gail McLaughlin, Samuel Flynn, Evan Grohs, and James Kneller for many useful discussions. This material is based upon work supported by the National Science Foundation under Award No. 2001760. This research was supported by the Exascale Computing Project (17-SC-20-SC), a collaborative effort of the U.S. Department of Energy Office of Science and the National Nuclear Security Administration. This work was supported in part by the U.S. Department of Energy, Office of Science, Office of Workforce Development for Teachers and Scientists (WDTS) under the Science Undergraduate Laboratory Internship (SULI) program. This work used the Bridges-2 system, which is supported by NSF award number ACI-1445606, at the Pittsburgh Supercomputing Center (PSC). This research used the Cori supercomputer of the National Energy Research Scientific Computing Center (NERSC), a U.S. Department of Energy Office of Science User Facility located at Lawrence Berkeley National Laboratory, operated under Contract No. DE-AC02-05CH11231.

In this work we make use of {\tt SymPy} \cite{meurer_sympy_2017}, {\tt NumPy} \cite{walt_numpy_2011}, {\tt MatPlotLib} \cite{hunter_matplotlib_2007}, {\tt SciPy} \cite{virtanen_scipy_2020}, and {\tt yt} \cite{turk_yt_2011}.

\bibliography{references}

\begin{thebibliography}{68}%
\makeatletter
\providecommand \@ifxundefined [1]{%
 \@ifx{#1\undefined}
}%
\providecommand \@ifnum [1]{%
 \ifnum #1\expandafter \@firstoftwo
 \else \expandafter \@secondoftwo
 \fi
}%
\providecommand \@ifx [1]{%
 \ifx #1\expandafter \@firstoftwo
 \else \expandafter \@secondoftwo
 \fi
}%
\providecommand \natexlab [1]{#1}%
\providecommand \enquote  [1]{``#1''}%
\providecommand \bibnamefont  [1]{#1}%
\providecommand \bibfnamefont [1]{#1}%
\providecommand \citenamefont [1]{#1}%
\providecommand \href@noop [0]{\@secondoftwo}%
\providecommand \href [0]{\begingroup \@sanitize@url \@href}%
\providecommand \@href[1]{\@@startlink{#1}\@@href}%
\providecommand \@@href[1]{\endgroup#1\@@endlink}%
\providecommand \@sanitize@url [0]{\catcode `\\12\catcode `\$12\catcode
  `\&12\catcode `\#12\catcode `\^12\catcode `\_12\catcode `\%12\relax}%
\providecommand \@@startlink[1]{}%
\providecommand \@@endlink[0]{}%
\providecommand \url  [0]{\begingroup\@sanitize@url \@url }%
\providecommand \@url [1]{\endgroup\@href {#1}{\urlprefix }}%
\providecommand \urlprefix  [0]{URL }%
\providecommand \Eprint [0]{\href }%
\providecommand \doibase [0]{http://dx.doi.org/}%
\providecommand \selectlanguage [0]{\@gobble}%
\providecommand \bibinfo  [0]{\@secondoftwo}%
\providecommand \bibfield  [0]{\@secondoftwo}%
\providecommand \translation [1]{[#1]}%
\providecommand \BibitemOpen [0]{}%
\providecommand \bibitemStop [0]{}%
\providecommand \bibitemNoStop [0]{.\EOS\space}%
\providecommand \EOS [0]{\spacefactor3000\relax}%
\providecommand \BibitemShut  [1]{\csname bibitem#1\endcsname}%
\let\auto@bib@innerbib\@empty
\bibitem [{\citenamefont {Janka}(2012)}]{janka_explosion_2012}%
  \BibitemOpen
  \bibfield  {author} {\bibinfo {author} {\bibfnamefont {Hans-Thomas}\
  \bibnamefont {Janka}},\ }\bibfield  {title} {\enquote {\bibinfo {title}
  {Explosion {Mechanisms} of {Core}-{Collapse} {Supernovae}},}\ }\href
  {\doibase 10.1146/annurev-nucl-102711-094901} {\bibfield  {journal} {\bibinfo
   {journal} {Annual Review of Nuclear and Particle Science}\ }\textbf
  {\bibinfo {volume} {62}},\ \bibinfo {pages} {407--451} (\bibinfo {year}
  {2012})}\BibitemShut {NoStop}%
\bibitem [{\citenamefont {Branch}\ and\ \citenamefont
  {Wheeler}(2017)}]{branch_supernova_2017}%
  \BibitemOpen
  \bibfield  {author} {\bibinfo {author} {\bibfnamefont {David}\ \bibnamefont
  {Branch}}\ and\ \bibinfo {author} {\bibfnamefont {J.~Craig}\ \bibnamefont
  {Wheeler}},\ }\href {\doibase 10.1007/978-3-662-55054-0} {\emph {\bibinfo
  {title} {Supernova {Explosions}}}},\ Astronomy and {Astrophysics} {Library}\
  (\bibinfo  {publisher} {Springer Berlin Heidelberg},\ \bibinfo {address}
  {Berlin, Heidelberg},\ \bibinfo {year} {2017})\BibitemShut {NoStop}%
\bibitem [{\citenamefont {Radice}\ \emph {et~al.}(2020)\citenamefont {Radice},
  \citenamefont {Bernuzzi},\ and\ \citenamefont
  {Perego}}]{radice_dynamics_2020}%
  \BibitemOpen
  \bibfield  {author} {\bibinfo {author} {\bibfnamefont {David}\ \bibnamefont
  {Radice}}, \bibinfo {author} {\bibfnamefont {Sebastiano}\ \bibnamefont
  {Bernuzzi}}, \ and\ \bibinfo {author} {\bibfnamefont {Albino}\ \bibnamefont
  {Perego}},\ }\bibfield  {title} {\enquote {\bibinfo {title} {The {Dynamics}
  of {Binary} {Neutron} {Star} {Mergers} and {GW170817}},}\ }\href {\doibase
  10.1146/annurev-nucl-013120-114541} {\bibfield  {journal} {\bibinfo
  {journal} {Annual Review of Nuclear and Particle Science}\ }\textbf {\bibinfo
  {volume} {70}},\ \bibinfo {pages} {95--119} (\bibinfo {year}
  {2020})}\BibitemShut {NoStop}%
\bibitem [{\citenamefont {Lippuner}\ \emph {et~al.}(2017)\citenamefont
  {Lippuner}, \citenamefont {Fernández}, \citenamefont {Roberts},
  \citenamefont {Foucart}, \citenamefont {Kasen}, \citenamefont {Metzger},\
  and\ \citenamefont {Ott}}]{lippuner_signatures_2017}%
  \BibitemOpen
  \bibfield  {author} {\bibinfo {author} {\bibfnamefont {Jonas}\ \bibnamefont
  {Lippuner}}, \bibinfo {author} {\bibfnamefont {Rodrigo}\ \bibnamefont
  {Fernández}}, \bibinfo {author} {\bibfnamefont {Luke~F.}\ \bibnamefont
  {Roberts}}, \bibinfo {author} {\bibfnamefont {Francois}\ \bibnamefont
  {Foucart}}, \bibinfo {author} {\bibfnamefont {Daniel}\ \bibnamefont {Kasen}},
  \bibinfo {author} {\bibfnamefont {Brian~D.}\ \bibnamefont {Metzger}}, \ and\
  \bibinfo {author} {\bibfnamefont {Christian~D.}\ \bibnamefont {Ott}},\
  }\bibfield  {title} {\enquote {\bibinfo {title} {Signatures of hypermassive
  neutron star lifetimes on r-process nucleosynthesis in the disc ejecta from
  neutron star mergers},}\ }\href {\doibase 10.1093/mnras/stx1987} {\bibfield
  {journal} {\bibinfo  {journal} {Monthly Notices of the Royal Astronomical
  Society}\ }\textbf {\bibinfo {volume} {472}},\ \bibinfo {pages} {904--918}
  (\bibinfo {year} {2017})}\BibitemShut {NoStop}%
\bibitem [{\citenamefont {Kullmann}\ \emph {et~al.}(2021)\citenamefont
  {Kullmann}, \citenamefont {Goriely}, \citenamefont {Just}, \citenamefont
  {Ardevol-Pulpillo}, \citenamefont {Bauswein},\ and\ \citenamefont
  {Janka}}]{kullmann_dynamical_2021}%
  \BibitemOpen
  \bibfield  {author} {\bibinfo {author} {\bibfnamefont {I.}~\bibnamefont
  {Kullmann}}, \bibinfo {author} {\bibfnamefont {S.}~\bibnamefont {Goriely}},
  \bibinfo {author} {\bibfnamefont {O.}~\bibnamefont {Just}}, \bibinfo {author}
  {\bibfnamefont {R.}~\bibnamefont {Ardevol-Pulpillo}}, \bibinfo {author}
  {\bibfnamefont {A.}~\bibnamefont {Bauswein}}, \ and\ \bibinfo {author}
  {\bibfnamefont {H.-T.}\ \bibnamefont {Janka}},\ }\bibfield  {title} {\enquote
  {\bibinfo {title} {Dynamical ejecta of neutron star mergers with nucleonic
  weak processes {I}: {Nucleosynthesis}},}\ }\href
  {http://arxiv.org/abs/2109.02509} {\bibfield  {journal} {\bibinfo  {journal}
  {arXiv:2109.02509 [astro-ph]}\ } (\bibinfo {year} {2021})},\ \bibinfo {note}
  {arXiv: 2109.02509}\BibitemShut {NoStop}%
\bibitem [{\citenamefont {Alexeyev}\ \emph {et~al.}(1988)\citenamefont
  {Alexeyev}, \citenamefont {Alexeyeva}, \citenamefont {Krivosheina},\ and\
  \citenamefont {Volchenko}}]{alexeyev_detection_1988}%
  \BibitemOpen
  \bibfield  {author} {\bibinfo {author} {\bibfnamefont {E.~N.}\ \bibnamefont
  {Alexeyev}}, \bibinfo {author} {\bibfnamefont {L.~N.}\ \bibnamefont
  {Alexeyeva}}, \bibinfo {author} {\bibfnamefont {I.~V.}\ \bibnamefont
  {Krivosheina}}, \ and\ \bibinfo {author} {\bibfnamefont {V.~I.}\ \bibnamefont
  {Volchenko}},\ }\bibfield  {title} {\enquote {\bibinfo {title} {Detection of
  the neutrino signal from {SN} {1987A} in the {LMC} using the {INR} {Baksan}
  underground scintillation telescope},}\ }\href {\doibase
  10.1016/0370-2693(88)91651-6} {\bibfield  {journal} {\bibinfo  {journal}
  {Physics Letters B}\ }\textbf {\bibinfo {volume} {205}},\ \bibinfo {pages}
  {209--214} (\bibinfo {year} {1988})}\BibitemShut {NoStop}%
\bibitem [{\citenamefont {Bionta}\ \emph {et~al.}(1987)\citenamefont {Bionta},
  \citenamefont {Blewitt}, \citenamefont {Bratton}, \citenamefont {Casper},
  \citenamefont {Ciocio}, \citenamefont {Claus}, \citenamefont {Cortez},
  \citenamefont {Crouch}, \citenamefont {Dye}, \citenamefont {Errede},
  \citenamefont {Foster}, \citenamefont {Gajewski}, \citenamefont {Ganezer},
  \citenamefont {Goldhaber}, \citenamefont {Haines}, \citenamefont {Jones},
  \citenamefont {Kielczewska}, \citenamefont {Kropp}, \citenamefont {Learned},
  \citenamefont {LoSecco}, \citenamefont {Matthews}, \citenamefont {Miller},
  \citenamefont {Mudan}, \citenamefont {Park}, \citenamefont {Price},
  \citenamefont {Reines}, \citenamefont {Schultz}, \citenamefont {Seidel},
  \citenamefont {Shumard}, \citenamefont {Sinclair}, \citenamefont {Sobel},
  \citenamefont {Stone}, \citenamefont {Sulak}, \citenamefont {Svoboda},
  \citenamefont {Thornton}, \citenamefont {van~der Velde},\ and\ \citenamefont
  {Wuest}}]{bionta_observation_1987}%
  \BibitemOpen
  \bibfield  {author} {\bibinfo {author} {\bibfnamefont {R.~M.}\ \bibnamefont
  {Bionta}}, \bibinfo {author} {\bibfnamefont {G.}~\bibnamefont {Blewitt}},
  \bibinfo {author} {\bibfnamefont {C.~B.}\ \bibnamefont {Bratton}}, \bibinfo
  {author} {\bibfnamefont {D.}~\bibnamefont {Casper}}, \bibinfo {author}
  {\bibfnamefont {A.}~\bibnamefont {Ciocio}}, \bibinfo {author} {\bibfnamefont
  {R.}~\bibnamefont {Claus}}, \bibinfo {author} {\bibfnamefont
  {B.}~\bibnamefont {Cortez}}, \bibinfo {author} {\bibfnamefont
  {M.}~\bibnamefont {Crouch}}, \bibinfo {author} {\bibfnamefont {S.~T.}\
  \bibnamefont {Dye}}, \bibinfo {author} {\bibfnamefont {S.}~\bibnamefont
  {Errede}}, \bibinfo {author} {\bibfnamefont {G.~W.}\ \bibnamefont {Foster}},
  \bibinfo {author} {\bibfnamefont {W.}~\bibnamefont {Gajewski}}, \bibinfo
  {author} {\bibfnamefont {K.~S.}\ \bibnamefont {Ganezer}}, \bibinfo {author}
  {\bibfnamefont {M.}~\bibnamefont {Goldhaber}}, \bibinfo {author}
  {\bibfnamefont {T.~J.}\ \bibnamefont {Haines}}, \bibinfo {author}
  {\bibfnamefont {T.~W.}\ \bibnamefont {Jones}}, \bibinfo {author}
  {\bibfnamefont {D.}~\bibnamefont {Kielczewska}}, \bibinfo {author}
  {\bibfnamefont {W.~R.}\ \bibnamefont {Kropp}}, \bibinfo {author}
  {\bibfnamefont {J.~G.}\ \bibnamefont {Learned}}, \bibinfo {author}
  {\bibfnamefont {J.~M.}\ \bibnamefont {LoSecco}}, \bibinfo {author}
  {\bibfnamefont {J.}~\bibnamefont {Matthews}}, \bibinfo {author}
  {\bibfnamefont {R.}~\bibnamefont {Miller}}, \bibinfo {author} {\bibfnamefont
  {M.~S.}\ \bibnamefont {Mudan}}, \bibinfo {author} {\bibfnamefont {H.~S.}\
  \bibnamefont {Park}}, \bibinfo {author} {\bibfnamefont {L.~R.}\ \bibnamefont
  {Price}}, \bibinfo {author} {\bibfnamefont {F.}~\bibnamefont {Reines}},
  \bibinfo {author} {\bibfnamefont {J.}~\bibnamefont {Schultz}}, \bibinfo
  {author} {\bibfnamefont {S.}~\bibnamefont {Seidel}}, \bibinfo {author}
  {\bibfnamefont {E.}~\bibnamefont {Shumard}}, \bibinfo {author} {\bibfnamefont
  {D.}~\bibnamefont {Sinclair}}, \bibinfo {author} {\bibfnamefont {H.~W.}\
  \bibnamefont {Sobel}}, \bibinfo {author} {\bibfnamefont {J.~L.}\ \bibnamefont
  {Stone}}, \bibinfo {author} {\bibfnamefont {L.~R.}\ \bibnamefont {Sulak}},
  \bibinfo {author} {\bibfnamefont {R.}~\bibnamefont {Svoboda}}, \bibinfo
  {author} {\bibfnamefont {G.}~\bibnamefont {Thornton}}, \bibinfo {author}
  {\bibfnamefont {J.~C.}\ \bibnamefont {van~der Velde}}, \ and\ \bibinfo
  {author} {\bibfnamefont {C.}~\bibnamefont {Wuest}},\ }\bibfield  {title}
  {\enquote {\bibinfo {title} {Observation of a neutrino burst in coincidence
  with supernova {1987A} in the {Large} {Magellanic} {Cloud}},}\ }\href
  {\doibase 10.1103/PhysRevLett.58.1494} {\bibfield  {journal} {\bibinfo
  {journal} {Physical Review Letters}\ }\textbf {\bibinfo {volume} {58}},\
  \bibinfo {pages} {1494--1496} (\bibinfo {year} {1987})},\ \bibinfo {note}
  {publisher: American Physical Society}\BibitemShut {NoStop}%
\bibitem [{\citenamefont {Hirata}\ \emph {et~al.}(1987)\citenamefont {Hirata},
  \citenamefont {Kajita}, \citenamefont {Koshiba}, \citenamefont {Nakahata},
  \citenamefont {Oyama}, \citenamefont {Sato}, \citenamefont {Suzuki},
  \citenamefont {Takita}, \citenamefont {Totsuka}, \citenamefont {Kifune},
  \citenamefont {Suda}, \citenamefont {Takahashi}, \citenamefont {Tanimori},
  \citenamefont {Miyano}, \citenamefont {Yamada}, \citenamefont {Beier},
  \citenamefont {Feldscher}, \citenamefont {Kim}, \citenamefont {Mann},
  \citenamefont {Newcomer}, \citenamefont {Van}, \citenamefont {Zhang},\ and\
  \citenamefont {Cortez}}]{hirata_observation_1987}%
  \BibitemOpen
  \bibfield  {author} {\bibinfo {author} {\bibfnamefont {K.}~\bibnamefont
  {Hirata}}, \bibinfo {author} {\bibfnamefont {T.}~\bibnamefont {Kajita}},
  \bibinfo {author} {\bibfnamefont {M.}~\bibnamefont {Koshiba}}, \bibinfo
  {author} {\bibfnamefont {M.}~\bibnamefont {Nakahata}}, \bibinfo {author}
  {\bibfnamefont {Y.}~\bibnamefont {Oyama}}, \bibinfo {author} {\bibfnamefont
  {N.}~\bibnamefont {Sato}}, \bibinfo {author} {\bibfnamefont {A.}~\bibnamefont
  {Suzuki}}, \bibinfo {author} {\bibfnamefont {M.}~\bibnamefont {Takita}},
  \bibinfo {author} {\bibfnamefont {Y.}~\bibnamefont {Totsuka}}, \bibinfo
  {author} {\bibfnamefont {T.}~\bibnamefont {Kifune}}, \bibinfo {author}
  {\bibfnamefont {T.}~\bibnamefont {Suda}}, \bibinfo {author} {\bibfnamefont
  {K.}~\bibnamefont {Takahashi}}, \bibinfo {author} {\bibfnamefont
  {T.}~\bibnamefont {Tanimori}}, \bibinfo {author} {\bibfnamefont
  {K.}~\bibnamefont {Miyano}}, \bibinfo {author} {\bibfnamefont
  {M.}~\bibnamefont {Yamada}}, \bibinfo {author} {\bibfnamefont {E.~W.}\
  \bibnamefont {Beier}}, \bibinfo {author} {\bibfnamefont {L.~R.}\ \bibnamefont
  {Feldscher}}, \bibinfo {author} {\bibfnamefont {S.~B.}\ \bibnamefont {Kim}},
  \bibinfo {author} {\bibfnamefont {A.~K.}\ \bibnamefont {Mann}}, \bibinfo
  {author} {\bibfnamefont {F.~M.}\ \bibnamefont {Newcomer}}, \bibinfo {author}
  {\bibfnamefont {R.}~\bibnamefont {Van}}, \bibinfo {author} {\bibfnamefont
  {W.}~\bibnamefont {Zhang}}, \ and\ \bibinfo {author} {\bibfnamefont {B.~G.}\
  \bibnamefont {Cortez}},\ }\bibfield  {title} {\enquote {\bibinfo {title}
  {Observation of a neutrino burst from the supernova {SN1987A}},}\ }\href
  {\doibase 10.1103/PhysRevLett.58.1490} {\bibfield  {journal} {\bibinfo
  {journal} {Physical Review Letters}\ }\textbf {\bibinfo {volume} {58}},\
  \bibinfo {pages} {1490--1493} (\bibinfo {year} {1987})}\BibitemShut {NoStop}%
\bibitem [{\citenamefont {Duan}\ and\ \citenamefont
  {Friedland}(2011)}]{duan_self-induced_2011}%
  \BibitemOpen
  \bibfield  {author} {\bibinfo {author} {\bibfnamefont {Huaiyu}\ \bibnamefont
  {Duan}}\ and\ \bibinfo {author} {\bibfnamefont {Alexander}\ \bibnamefont
  {Friedland}},\ }\bibfield  {title} {\enquote {\bibinfo {title}
  {Self-{Induced} {Suppression} of {Collective} {Neutrino} {Oscillations} in a
  {Supernova}},}\ }\href {\doibase 10.1103/PhysRevLett.106.091101} {\bibfield
  {journal} {\bibinfo  {journal} {Physical Review Letters}\ }\textbf {\bibinfo
  {volume} {106}},\ \bibinfo {pages} {091101} (\bibinfo {year}
  {2011})}\BibitemShut {NoStop}%
\bibitem [{\citenamefont {Dasgupta}\ \emph {et~al.}(2012)\citenamefont
  {Dasgupta}, \citenamefont {O'Connor},\ and\ \citenamefont
  {Ott}}]{dasgupta_role_2012}%
  \BibitemOpen
  \bibfield  {author} {\bibinfo {author} {\bibfnamefont {Basudeb}\ \bibnamefont
  {Dasgupta}}, \bibinfo {author} {\bibfnamefont {Evan~P.}\ \bibnamefont
  {O'Connor}}, \ and\ \bibinfo {author} {\bibfnamefont {Christian~D.}\
  \bibnamefont {Ott}},\ }\bibfield  {title} {\enquote {\bibinfo {title} {Role
  of collective neutrino flavor oscillations in core-collapse supernova shock
  revival},}\ }\href {\doibase 10.1103/PhysRevD.85.065008} {\bibfield
  {journal} {\bibinfo  {journal} {Physical Review D}\ }\textbf {\bibinfo
  {volume} {85}},\ \bibinfo {pages} {065008} (\bibinfo {year}
  {2012})}\BibitemShut {NoStop}%
\bibitem [{\citenamefont {O'Connor}\ \emph {et~al.}(2018)\citenamefont
  {O'Connor}, \citenamefont {Bollig}, \citenamefont {Burrows}, \citenamefont
  {Couch}, \citenamefont {Fischer}, \citenamefont {Janka}, \citenamefont
  {Kotake}, \citenamefont {Lentz}, \citenamefont {Liebendörfer}, \citenamefont
  {Messer}, \citenamefont {Mezzacappa}, \citenamefont {Takiwaki},\ and\
  \citenamefont {Vartanyan}}]{oconnor_global_2018}%
  \BibitemOpen
  \bibfield  {author} {\bibinfo {author} {\bibfnamefont {Evan}\ \bibnamefont
  {O'Connor}}, \bibinfo {author} {\bibfnamefont {Robert}\ \bibnamefont
  {Bollig}}, \bibinfo {author} {\bibfnamefont {Adam}\ \bibnamefont {Burrows}},
  \bibinfo {author} {\bibfnamefont {Sean}\ \bibnamefont {Couch}}, \bibinfo
  {author} {\bibfnamefont {Tobias}\ \bibnamefont {Fischer}}, \bibinfo {author}
  {\bibfnamefont {Hans-Thomas}\ \bibnamefont {Janka}}, \bibinfo {author}
  {\bibfnamefont {Kei}\ \bibnamefont {Kotake}}, \bibinfo {author}
  {\bibfnamefont {Eric~J.}\ \bibnamefont {Lentz}}, \bibinfo {author}
  {\bibfnamefont {Matthias}\ \bibnamefont {Liebendörfer}}, \bibinfo {author}
  {\bibfnamefont {O.~E.~Bronson}\ \bibnamefont {Messer}}, \bibinfo {author}
  {\bibfnamefont {Anthony}\ \bibnamefont {Mezzacappa}}, \bibinfo {author}
  {\bibfnamefont {Tomoya}\ \bibnamefont {Takiwaki}}, \ and\ \bibinfo {author}
  {\bibfnamefont {David}\ \bibnamefont {Vartanyan}},\ }\bibfield  {title}
  {\enquote {\bibinfo {title} {Global comparison of core-collapse supernova
  simulations in spherical symmetry},}\ }\href {\doibase
  10.1088/1361-6471/aadeae} {\bibfield  {journal} {\bibinfo  {journal} {Journal
  of Physics G Nuclear Physics}\ }\textbf {\bibinfo {volume} {45}},\ \bibinfo
  {pages} {104001} (\bibinfo {year} {2018})}\BibitemShut {NoStop}%
\bibitem [{\citenamefont {Just}\ \emph {et~al.}(2018)\citenamefont {Just},
  \citenamefont {Bollig}, \citenamefont {Janka}, \citenamefont {Obergaulinger},
  \citenamefont {Glas},\ and\ \citenamefont
  {Nagataki}}]{just_core-collapse_2018}%
  \BibitemOpen
  \bibfield  {author} {\bibinfo {author} {\bibfnamefont {O}~\bibnamefont
  {Just}}, \bibinfo {author} {\bibfnamefont {R}~\bibnamefont {Bollig}},
  \bibinfo {author} {\bibfnamefont {H-Th}\ \bibnamefont {Janka}}, \bibinfo
  {author} {\bibfnamefont {M}~\bibnamefont {Obergaulinger}}, \bibinfo {author}
  {\bibfnamefont {R}~\bibnamefont {Glas}}, \ and\ \bibinfo {author}
  {\bibfnamefont {S}~\bibnamefont {Nagataki}},\ }\bibfield  {title} {\enquote
  {\bibinfo {title} {Core-collapse supernova simulations in one and two
  dimensions: comparison of codes and approximations},}\ }\href {\doibase
  10.1093/mnras/sty2578} {\bibfield  {journal} {\bibinfo  {journal} {Monthly
  Notices of the Royal Astronomical Society}\ }\textbf {\bibinfo {volume}
  {481}},\ \bibinfo {pages} {4786--4814} (\bibinfo {year} {2018})}\BibitemShut
  {NoStop}%
\bibitem [{\citenamefont {Burrows}\ \emph {et~al.}(2020)\citenamefont
  {Burrows}, \citenamefont {Radice}, \citenamefont {Vartanyan}, \citenamefont
  {Nagakura}, \citenamefont {Skinner},\ and\ \citenamefont
  {Dolence}}]{burrows_overarching_2020}%
  \BibitemOpen
  \bibfield  {author} {\bibinfo {author} {\bibfnamefont {Adam}\ \bibnamefont
  {Burrows}}, \bibinfo {author} {\bibfnamefont {David}\ \bibnamefont {Radice}},
  \bibinfo {author} {\bibfnamefont {David}\ \bibnamefont {Vartanyan}}, \bibinfo
  {author} {\bibfnamefont {Hiroki}\ \bibnamefont {Nagakura}}, \bibinfo {author}
  {\bibfnamefont {M~Aaron}\ \bibnamefont {Skinner}}, \ and\ \bibinfo {author}
  {\bibfnamefont {Joshua~C}\ \bibnamefont {Dolence}},\ }\bibfield  {title}
  {\enquote {\bibinfo {title} {The overarching framework of core-collapse
  supernova explosions as revealed by {3D} fornax simulations},}\ }\href
  {\doibase 10.1093/mnras/stz3223} {\bibfield  {journal} {\bibinfo  {journal}
  {Monthly Notices of the Royal Astronomical Society}\ }\textbf {\bibinfo
  {volume} {491}},\ \bibinfo {pages} {2715--2735} (\bibinfo {year}
  {2020})}\BibitemShut {NoStop}%
\bibitem [{\citenamefont {Sandoval}\ \emph {et~al.}(2021)\citenamefont
  {Sandoval}, \citenamefont {Hix}, \citenamefont {Messer}, \citenamefont
  {Lentz},\ and\ \citenamefont {Harris}}]{sandoval_three_2021}%
  \BibitemOpen
  \bibfield  {author} {\bibinfo {author} {\bibfnamefont {Michael~A.}\
  \bibnamefont {Sandoval}}, \bibinfo {author} {\bibfnamefont {W.~Raphael}\
  \bibnamefont {Hix}}, \bibinfo {author} {\bibfnamefont {O.~E.~Bronson}\
  \bibnamefont {Messer}}, \bibinfo {author} {\bibfnamefont {Eric~J.}\
  \bibnamefont {Lentz}}, \ and\ \bibinfo {author} {\bibfnamefont {J.~Austin}\
  \bibnamefont {Harris}},\ }\bibfield  {title} {\enquote {\bibinfo {title}
  {Three {Dimensional} {Core}-{Collapse} {Supernova} {Simulations} with 160
  {Isotopic} {Species} {Evolved} to {Shock} {Breakout}},}\ }\href
  {http://arxiv.org/abs/2106.01389} {\bibfield  {journal} {\bibinfo  {journal}
  {arXiv:2106.01389 [astro-ph]}\ } (\bibinfo {year} {2021})}\BibitemShut
  {NoStop}%
\bibitem [{\citenamefont {Mösta}\ \emph {et~al.}(2020)\citenamefont {Mösta},
  \citenamefont {Radice}, \citenamefont {Haas}, \citenamefont {Schnetter},\
  and\ \citenamefont {Bernuzzi}}]{mosta_magnetar_2020}%
  \BibitemOpen
  \bibfield  {author} {\bibinfo {author} {\bibfnamefont {Philipp}\ \bibnamefont
  {Mösta}}, \bibinfo {author} {\bibfnamefont {David}\ \bibnamefont {Radice}},
  \bibinfo {author} {\bibfnamefont {Roland}\ \bibnamefont {Haas}}, \bibinfo
  {author} {\bibfnamefont {Erik}\ \bibnamefont {Schnetter}}, \ and\ \bibinfo
  {author} {\bibfnamefont {Sebastiano}\ \bibnamefont {Bernuzzi}},\ }\bibfield
  {title} {\enquote {\bibinfo {title} {A magnetar engine for short {GRBs} and
  kilonovae},}\ }\href {\doibase 10.3847/2041-8213/abb6ef} {\bibfield
  {journal} {\bibinfo  {journal} {The Astrophysical Journal}\ }\textbf
  {\bibinfo {volume} {901}},\ \bibinfo {pages} {L37} (\bibinfo {year}
  {2020})},\ \bibinfo {note} {arXiv: 2003.06043}\BibitemShut {NoStop}%
\bibitem [{\citenamefont {Nedora}\ \emph {et~al.}(2020)\citenamefont {Nedora},
  \citenamefont {Schianchi}, \citenamefont {Bernuzzi}, \citenamefont {Radice},
  \citenamefont {Daszuta}, \citenamefont {Endrizzi}, \citenamefont {Perego},
  \citenamefont {Prakash},\ and\ \citenamefont {Zappa}}]{nedora_mapping_2020}%
  \BibitemOpen
  \bibfield  {author} {\bibinfo {author} {\bibfnamefont {Vsevolod}\
  \bibnamefont {Nedora}}, \bibinfo {author} {\bibfnamefont {Federico}\
  \bibnamefont {Schianchi}}, \bibinfo {author} {\bibfnamefont {Sebastiano}\
  \bibnamefont {Bernuzzi}}, \bibinfo {author} {\bibfnamefont {David}\
  \bibnamefont {Radice}}, \bibinfo {author} {\bibfnamefont {Boris}\
  \bibnamefont {Daszuta}}, \bibinfo {author} {\bibfnamefont {Andrea}\
  \bibnamefont {Endrizzi}}, \bibinfo {author} {\bibfnamefont {Albino}\
  \bibnamefont {Perego}}, \bibinfo {author} {\bibfnamefont {Aviral}\
  \bibnamefont {Prakash}}, \ and\ \bibinfo {author} {\bibfnamefont {Francesco}\
  \bibnamefont {Zappa}},\ }\bibfield  {title} {\enquote {\bibinfo {title}
  {Mapping dynamical ejecta and disk masses from numerical relativity
  simulations of neutron star mergers},}\ }\href
  {http://arxiv.org/abs/2011.11110} {\bibfield  {journal} {\bibinfo  {journal}
  {arXiv:2011.11110 [astro-ph, physics:gr-qc]}\ } (\bibinfo {year}
  {2020})}\BibitemShut {NoStop}%
\bibitem [{\citenamefont {Foucart}\ \emph {et~al.}(2021)\citenamefont
  {Foucart}, \citenamefont {Moesta}, \citenamefont {Ramirez}, \citenamefont
  {Wright}, \citenamefont {Darbha},\ and\ \citenamefont
  {Kasen}}]{foucart_estimating_2021}%
  \BibitemOpen
  \bibfield  {author} {\bibinfo {author} {\bibfnamefont {Francois}\
  \bibnamefont {Foucart}}, \bibinfo {author} {\bibfnamefont {Philipp}\
  \bibnamefont {Moesta}}, \bibinfo {author} {\bibfnamefont {Teresita}\
  \bibnamefont {Ramirez}}, \bibinfo {author} {\bibfnamefont {Alex~James}\
  \bibnamefont {Wright}}, \bibinfo {author} {\bibfnamefont {Siva}\ \bibnamefont
  {Darbha}}, \ and\ \bibinfo {author} {\bibfnamefont {Daniel}\ \bibnamefont
  {Kasen}},\ }\bibfield  {title} {\enquote {\bibinfo {title} {Estimating
  outflow masses and velocities in merger simulations: impact of r-process
  heating and neutrino cooling},}\ }\href {http://arxiv.org/abs/2109.00565}
  {\bibfield  {journal} {\bibinfo  {journal} {arXiv:2109.00565 [astro-ph,
  physics:gr-qc]}\ } (\bibinfo {year} {2021})}\BibitemShut {NoStop}%
\bibitem [{\citenamefont {Zhu}\ and\ \citenamefont
  {Rezzolla}(2021)}]{zhu_fully_2021}%
  \BibitemOpen
  \bibfield  {author} {\bibinfo {author} {\bibfnamefont {Zhenyu}\ \bibnamefont
  {Zhu}}\ and\ \bibinfo {author} {\bibfnamefont {Luciano}\ \bibnamefont
  {Rezzolla}},\ }\bibfield  {title} {\enquote {\bibinfo {title} {Fully
  general-relativistic simulations of isolated and binary strange quark
  stars},}\ }\href {http://arxiv.org/abs/2102.07721} {\bibfield  {journal}
  {\bibinfo  {journal} {arXiv:2102.07721 [astro-ph, physics:gr-qc]}\ }
  (\bibinfo {year} {2021})}\BibitemShut {NoStop}%
\bibitem [{\citenamefont {Metzger}\ and\ \citenamefont
  {Fernández}(2021)}]{metzger_neutrino-_2021}%
  \BibitemOpen
  \bibfield  {author} {\bibinfo {author} {\bibfnamefont {Brian~D.}\
  \bibnamefont {Metzger}}\ and\ \bibinfo {author} {\bibfnamefont {Rodrigo}\
  \bibnamefont {Fernández}},\ }\bibfield  {title} {\enquote {\bibinfo {title}
  {From {Neutrino}- to {Photon}-cooled in {Three} {Years}: {Can} {Fallback}
  {Accretion} {Explain} the {X}-{Ray} {Excess} in {GW170817}?}}\ }\href
  {\doibase 10/gmt587} {\bibfield  {journal} {\bibinfo  {journal} {The
  Astrophysical Journal Letters}\ }\textbf {\bibinfo {volume} {916}},\ \bibinfo
  {pages} {L3} (\bibinfo {year} {2021})}\BibitemShut {NoStop}%
\bibitem [{\citenamefont {Li}\ and\ \citenamefont
  {Siegel}(2021)}]{li_neutrino_2021}%
  \BibitemOpen
  \bibfield  {author} {\bibinfo {author} {\bibfnamefont {Xinyu}\ \bibnamefont
  {Li}}\ and\ \bibinfo {author} {\bibfnamefont {Daniel~M.}\ \bibnamefont
  {Siegel}},\ }\bibfield  {title} {\enquote {\bibinfo {title} {Neutrino {Fast}
  {Flavor} {Conversions} in {Neutron}-{Star} {Postmerger} {Accretion}
  {Disks}},}\ }\href {\doibase 10/gkzfd4} {\bibfield  {journal} {\bibinfo
  {journal} {Physical Review Letters}\ }\textbf {\bibinfo {volume} {126}},\
  \bibinfo {pages} {251101} (\bibinfo {year} {2021})},\ \bibinfo {note}
  {publisher: American Physical Society}\BibitemShut {NoStop}%
\bibitem [{\citenamefont {Just}\ \emph {et~al.}(2021)\citenamefont {Just},
  \citenamefont {Goriely}, \citenamefont {Janka}, \citenamefont {Nagataki},\
  and\ \citenamefont {Bauswein}}]{just_neutrino_2021}%
  \BibitemOpen
  \bibfield  {author} {\bibinfo {author} {\bibfnamefont {Oliver}\ \bibnamefont
  {Just}}, \bibinfo {author} {\bibfnamefont {Stephane}\ \bibnamefont
  {Goriely}}, \bibinfo {author} {\bibfnamefont {Hans-Thomas}\ \bibnamefont
  {Janka}}, \bibinfo {author} {\bibfnamefont {Shigehiro}\ \bibnamefont
  {Nagataki}}, \ and\ \bibinfo {author} {\bibfnamefont {Andreas}\ \bibnamefont
  {Bauswein}},\ }\bibfield  {title} {\enquote {\bibinfo {title} {Neutrino
  absorption and other physics dependencies in neutrino-cooled black-hole
  accretion disks},}\ }\href {http://arxiv.org/abs/2102.08387} {\bibfield
  {journal} {\bibinfo  {journal} {arXiv:2102.08387 [astro-ph]}\ } (\bibinfo
  {year} {2021})}\BibitemShut {NoStop}%
\bibitem [{\citenamefont {Ahmad}\ \emph {et~al.}(2001)\citenamefont {Ahmad},
  \citenamefont {Allen}, \citenamefont {Andersen}, \citenamefont {Anglin},
  \citenamefont {Bühler}, \citenamefont {Barton}, \citenamefont {Beier},
  \citenamefont {Bercovitch}, \citenamefont {Bigu}, \citenamefont {Biller},
  \citenamefont {Black}, \citenamefont {Blevis}, \citenamefont {Boardman},
  \citenamefont {Boger}, \citenamefont {Bonvin}, \citenamefont {Boulay},
  \citenamefont {Bowler}, \citenamefont {Bowles}, \citenamefont {Brice},
  \citenamefont {Browne}, \citenamefont {Bullard}, \citenamefont {Burritt},
  \citenamefont {Cameron}, \citenamefont {Cameron}, \citenamefont {Chan},
  \citenamefont {Chen}, \citenamefont {Chen}, \citenamefont {Chen},
  \citenamefont {Chon}, \citenamefont {Cleveland}, \citenamefont {Clifford},
  \citenamefont {Cowan}, \citenamefont {Cowen}, \citenamefont {Cox},
  \citenamefont {Dai}, \citenamefont {Dai}, \citenamefont {Dalnoki-Veress},
  \citenamefont {Davidson}, \citenamefont {Doe}, \citenamefont {Doucas},
  \citenamefont {Dragowsky}, \citenamefont {Duba}, \citenamefont {Duncan},
  \citenamefont {Dunmore}, \citenamefont {Earle}, \citenamefont {Elliott},
  \citenamefont {Evans}, \citenamefont {Ewan}, \citenamefont {Farine},
  \citenamefont {Fergani}, \citenamefont {Ferraris}, \citenamefont {Ford},
  \citenamefont {Fowler}, \citenamefont {Frame}, \citenamefont {Frank},
  \citenamefont {Frati}, \citenamefont {Germani}, \citenamefont {Gil},
  \citenamefont {Goldschmidt}, \citenamefont {Grant}, \citenamefont {Hahn},
  \citenamefont {Hallin}, \citenamefont {Hallman}, \citenamefont {Hamer},
  \citenamefont {Hamian}, \citenamefont {Haq}, \citenamefont {Hargrove},
  \citenamefont {Harvey}, \citenamefont {Hazama}, \citenamefont {Heaton},
  \citenamefont {Heeger}, \citenamefont {Heintzelman}, \citenamefont {Heise},
  \citenamefont {Helmer}, \citenamefont {Hepburn}, \citenamefont {Heron},
  \citenamefont {Hewett}, \citenamefont {Hime}, \citenamefont {Howe},
  \citenamefont {Hykawy}, \citenamefont {Isaac}, \citenamefont {Jagam},
  \citenamefont {Jelley}, \citenamefont {Jillings}, \citenamefont {Jonkmans},
  \citenamefont {Karn}, \citenamefont {Keener}, \citenamefont {Kirch},
  \citenamefont {Klein}, \citenamefont {Knox}, \citenamefont {Komar},
  \citenamefont {Kouzes}, \citenamefont {Kutter}, \citenamefont {Kyba},
  \citenamefont {Law}, \citenamefont {Lawson}, \citenamefont {Lay},
  \citenamefont {Lee}, \citenamefont {Lesko}, \citenamefont {Leslie},
  \citenamefont {Levine}, \citenamefont {Locke}, \citenamefont {Lowry},
  \citenamefont {Luoma}, \citenamefont {Lyon}, \citenamefont {Majerus},
  \citenamefont {Mak}, \citenamefont {Marino}, \citenamefont {McCauley},
  \citenamefont {McDonald}, \citenamefont {McDonald}, \citenamefont
  {McFarlane}, \citenamefont {McGregor}, \citenamefont {McLatchie},
  \citenamefont {Drees}, \citenamefont {Mes}, \citenamefont {Mifflin},
  \citenamefont {Miller}, \citenamefont {Milton}, \citenamefont {Moffat},
  \citenamefont {Moorhead}, \citenamefont {Nally}, \citenamefont {Neubauer},
  \citenamefont {Newcomer}, \citenamefont {Ng}, \citenamefont {Noble},
  \citenamefont {Norman}, \citenamefont {Novikov}, \citenamefont {O'Neill},
  \citenamefont {Okada}, \citenamefont {Ollerhead}, \citenamefont {Omori},
  \citenamefont {Orrell}, \citenamefont {Oser}, \citenamefont {Poon},
  \citenamefont {Radcliffe}, \citenamefont {Roberge}, \citenamefont
  {Robertson}, \citenamefont {Robertson}, \citenamefont {Rowley}, \citenamefont
  {Rusu}, \citenamefont {Saettler}, \citenamefont {Schaffer}, \citenamefont
  {Schuelke}, \citenamefont {Schwendener}, \citenamefont {Seifert},
  \citenamefont {Shatkay}, \citenamefont {Simpson}, \citenamefont {Sinclair},
  \citenamefont {Skensved}, \citenamefont {Smith}, \citenamefont {Smith},
  \citenamefont {Starinsky}, \citenamefont {Steiger}, \citenamefont {Stokstad},
  \citenamefont {Storey}, \citenamefont {Sur}, \citenamefont {Tafirout},
  \citenamefont {Tagg}, \citenamefont {Tanner}, \citenamefont {Taplin},
  \citenamefont {Thorman}, \citenamefont {Thornewell}, \citenamefont {Trent},
  \citenamefont {Tserkovnyak}, \citenamefont {Van~Berg}, \citenamefont {Van~de
  Water}, \citenamefont {Virtue}, \citenamefont {Waltham}, \citenamefont
  {Wang}, \citenamefont {Wark}, \citenamefont {West}, \citenamefont {Wilhelmy},
  \citenamefont {Wilkerson}, \citenamefont {Wilson}, \citenamefont {Wittich},
  \citenamefont {Wouters},\ and\ \citenamefont {Yeh}}]{ahmad_measurement_2001}%
  \BibitemOpen
  \bibfield  {author} {\bibinfo {author} {\bibfnamefont {Q.~R.}\ \bibnamefont
  {Ahmad}}, \bibinfo {author} {\bibfnamefont {R.~C.}\ \bibnamefont {Allen}},
  \bibinfo {author} {\bibfnamefont {T.~C.}\ \bibnamefont {Andersen}}, \bibinfo
  {author} {\bibfnamefont {J.~D.}\ \bibnamefont {Anglin}}, \bibinfo {author}
  {\bibfnamefont {G.}~\bibnamefont {Bühler}}, \bibinfo {author} {\bibfnamefont
  {J.~C.}\ \bibnamefont {Barton}}, \bibinfo {author} {\bibfnamefont {E.~W.}\
  \bibnamefont {Beier}}, \bibinfo {author} {\bibfnamefont {M.}~\bibnamefont
  {Bercovitch}}, \bibinfo {author} {\bibfnamefont {J.}~\bibnamefont {Bigu}},
  \bibinfo {author} {\bibfnamefont {S.}~\bibnamefont {Biller}}, \bibinfo
  {author} {\bibfnamefont {R.~A.}\ \bibnamefont {Black}}, \bibinfo {author}
  {\bibfnamefont {I.}~\bibnamefont {Blevis}}, \bibinfo {author} {\bibfnamefont
  {R.~J.}\ \bibnamefont {Boardman}}, \bibinfo {author} {\bibfnamefont
  {J.}~\bibnamefont {Boger}}, \bibinfo {author} {\bibfnamefont
  {E.}~\bibnamefont {Bonvin}}, \bibinfo {author} {\bibfnamefont {M.~G.}\
  \bibnamefont {Boulay}}, \bibinfo {author} {\bibfnamefont {M.~G.}\
  \bibnamefont {Bowler}}, \bibinfo {author} {\bibfnamefont {T.~J.}\
  \bibnamefont {Bowles}}, \bibinfo {author} {\bibfnamefont {S.~J.}\
  \bibnamefont {Brice}}, \bibinfo {author} {\bibfnamefont {M.~C.}\ \bibnamefont
  {Browne}}, \bibinfo {author} {\bibfnamefont {T.~V.}\ \bibnamefont {Bullard}},
  \bibinfo {author} {\bibfnamefont {T.~H.}\ \bibnamefont {Burritt}}, \bibinfo
  {author} {\bibfnamefont {K.}~\bibnamefont {Cameron}}, \bibinfo {author}
  {\bibfnamefont {J.}~\bibnamefont {Cameron}}, \bibinfo {author} {\bibfnamefont
  {Y.~D.}\ \bibnamefont {Chan}}, \bibinfo {author} {\bibfnamefont
  {M.}~\bibnamefont {Chen}}, \bibinfo {author} {\bibfnamefont {H.~H.}\
  \bibnamefont {Chen}}, \bibinfo {author} {\bibfnamefont {X.}~\bibnamefont
  {Chen}}, \bibinfo {author} {\bibfnamefont {M.~C.}\ \bibnamefont {Chon}},
  \bibinfo {author} {\bibfnamefont {B.~T.}\ \bibnamefont {Cleveland}}, \bibinfo
  {author} {\bibfnamefont {E.~T.~H.}\ \bibnamefont {Clifford}}, \bibinfo
  {author} {\bibfnamefont {J.~H.~M.}\ \bibnamefont {Cowan}}, \bibinfo {author}
  {\bibfnamefont {D.~F.}\ \bibnamefont {Cowen}}, \bibinfo {author}
  {\bibfnamefont {G.~A.}\ \bibnamefont {Cox}}, \bibinfo {author} {\bibfnamefont
  {Y.}~\bibnamefont {Dai}}, \bibinfo {author} {\bibfnamefont {X.}~\bibnamefont
  {Dai}}, \bibinfo {author} {\bibfnamefont {F.}~\bibnamefont {Dalnoki-Veress}},
  \bibinfo {author} {\bibfnamefont {W.~F.}\ \bibnamefont {Davidson}}, \bibinfo
  {author} {\bibfnamefont {P.~J.}\ \bibnamefont {Doe}}, \bibinfo {author}
  {\bibfnamefont {G.}~\bibnamefont {Doucas}}, \bibinfo {author} {\bibfnamefont
  {M.~R.}\ \bibnamefont {Dragowsky}}, \bibinfo {author} {\bibfnamefont {C.~A.}\
  \bibnamefont {Duba}}, \bibinfo {author} {\bibfnamefont {F.~A.}\ \bibnamefont
  {Duncan}}, \bibinfo {author} {\bibfnamefont {J.}~\bibnamefont {Dunmore}},
  \bibinfo {author} {\bibfnamefont {E.~D.}\ \bibnamefont {Earle}}, \bibinfo
  {author} {\bibfnamefont {S.~R.}\ \bibnamefont {Elliott}}, \bibinfo {author}
  {\bibfnamefont {H.~C.}\ \bibnamefont {Evans}}, \bibinfo {author}
  {\bibfnamefont {G.~T.}\ \bibnamefont {Ewan}}, \bibinfo {author}
  {\bibfnamefont {J.}~\bibnamefont {Farine}}, \bibinfo {author} {\bibfnamefont
  {H.}~\bibnamefont {Fergani}}, \bibinfo {author} {\bibfnamefont {A.~P.}\
  \bibnamefont {Ferraris}}, \bibinfo {author} {\bibfnamefont {R.~J.}\
  \bibnamefont {Ford}}, \bibinfo {author} {\bibfnamefont {M.~M.}\ \bibnamefont
  {Fowler}}, \bibinfo {author} {\bibfnamefont {K.}~\bibnamefont {Frame}},
  \bibinfo {author} {\bibfnamefont {E.~D.}\ \bibnamefont {Frank}}, \bibinfo
  {author} {\bibfnamefont {W.}~\bibnamefont {Frati}}, \bibinfo {author}
  {\bibfnamefont {J.~V.}\ \bibnamefont {Germani}}, \bibinfo {author}
  {\bibfnamefont {S.}~\bibnamefont {Gil}}, \bibinfo {author} {\bibfnamefont
  {A.}~\bibnamefont {Goldschmidt}}, \bibinfo {author} {\bibfnamefont {D.~R.}\
  \bibnamefont {Grant}}, \bibinfo {author} {\bibfnamefont {R.~L.}\ \bibnamefont
  {Hahn}}, \bibinfo {author} {\bibfnamefont {A.~L.}\ \bibnamefont {Hallin}},
  \bibinfo {author} {\bibfnamefont {E.~D.}\ \bibnamefont {Hallman}}, \bibinfo
  {author} {\bibfnamefont {A.}~\bibnamefont {Hamer}}, \bibinfo {author}
  {\bibfnamefont {A.~A.}\ \bibnamefont {Hamian}}, \bibinfo {author}
  {\bibfnamefont {R.~U.}\ \bibnamefont {Haq}}, \bibinfo {author} {\bibfnamefont
  {C.~K.}\ \bibnamefont {Hargrove}}, \bibinfo {author} {\bibfnamefont {P.~J.}\
  \bibnamefont {Harvey}}, \bibinfo {author} {\bibfnamefont {R.}~\bibnamefont
  {Hazama}}, \bibinfo {author} {\bibfnamefont {R.}~\bibnamefont {Heaton}},
  \bibinfo {author} {\bibfnamefont {K.~M.}\ \bibnamefont {Heeger}}, \bibinfo
  {author} {\bibfnamefont {W.~J.}\ \bibnamefont {Heintzelman}}, \bibinfo
  {author} {\bibfnamefont {J.}~\bibnamefont {Heise}}, \bibinfo {author}
  {\bibfnamefont {R.~L.}\ \bibnamefont {Helmer}}, \bibinfo {author}
  {\bibfnamefont {J.~D.}\ \bibnamefont {Hepburn}}, \bibinfo {author}
  {\bibfnamefont {H.}~\bibnamefont {Heron}}, \bibinfo {author} {\bibfnamefont
  {J.}~\bibnamefont {Hewett}}, \bibinfo {author} {\bibfnamefont
  {A.}~\bibnamefont {Hime}}, \bibinfo {author} {\bibfnamefont {M.}~\bibnamefont
  {Howe}}, \bibinfo {author} {\bibfnamefont {J.~G.}\ \bibnamefont {Hykawy}},
  \bibinfo {author} {\bibfnamefont {M.~C.~P.}\ \bibnamefont {Isaac}}, \bibinfo
  {author} {\bibfnamefont {P.}~\bibnamefont {Jagam}}, \bibinfo {author}
  {\bibfnamefont {N.~A.}\ \bibnamefont {Jelley}}, \bibinfo {author}
  {\bibfnamefont {C.}~\bibnamefont {Jillings}}, \bibinfo {author}
  {\bibfnamefont {G.}~\bibnamefont {Jonkmans}}, \bibinfo {author}
  {\bibfnamefont {J.}~\bibnamefont {Karn}}, \bibinfo {author} {\bibfnamefont
  {P.~T.}\ \bibnamefont {Keener}}, \bibinfo {author} {\bibfnamefont
  {K.}~\bibnamefont {Kirch}}, \bibinfo {author} {\bibfnamefont {J.~R.}\
  \bibnamefont {Klein}}, \bibinfo {author} {\bibfnamefont {A.~B.}\ \bibnamefont
  {Knox}}, \bibinfo {author} {\bibfnamefont {R.~J.}\ \bibnamefont {Komar}},
  \bibinfo {author} {\bibfnamefont {R.}~\bibnamefont {Kouzes}}, \bibinfo
  {author} {\bibfnamefont {T.}~\bibnamefont {Kutter}}, \bibinfo {author}
  {\bibfnamefont {C.~C.~M.}\ \bibnamefont {Kyba}}, \bibinfo {author}
  {\bibfnamefont {J.}~\bibnamefont {Law}}, \bibinfo {author} {\bibfnamefont
  {I.~T.}\ \bibnamefont {Lawson}}, \bibinfo {author} {\bibfnamefont
  {M.}~\bibnamefont {Lay}}, \bibinfo {author} {\bibfnamefont {H.~W.}\
  \bibnamefont {Lee}}, \bibinfo {author} {\bibfnamefont {K.~T.}\ \bibnamefont
  {Lesko}}, \bibinfo {author} {\bibfnamefont {J.~R.}\ \bibnamefont {Leslie}},
  \bibinfo {author} {\bibfnamefont {I.}~\bibnamefont {Levine}}, \bibinfo
  {author} {\bibfnamefont {W.}~\bibnamefont {Locke}}, \bibinfo {author}
  {\bibfnamefont {M.~M.}\ \bibnamefont {Lowry}}, \bibinfo {author}
  {\bibfnamefont {S.}~\bibnamefont {Luoma}}, \bibinfo {author} {\bibfnamefont
  {J.}~\bibnamefont {Lyon}}, \bibinfo {author} {\bibfnamefont {S.}~\bibnamefont
  {Majerus}}, \bibinfo {author} {\bibfnamefont {H.~B.}\ \bibnamefont {Mak}},
  \bibinfo {author} {\bibfnamefont {A.~D.}\ \bibnamefont {Marino}}, \bibinfo
  {author} {\bibfnamefont {N.}~\bibnamefont {McCauley}}, \bibinfo {author}
  {\bibfnamefont {A.~B.}\ \bibnamefont {McDonald}}, \bibinfo {author}
  {\bibfnamefont {D.~S.}\ \bibnamefont {McDonald}}, \bibinfo {author}
  {\bibfnamefont {K.}~\bibnamefont {McFarlane}}, \bibinfo {author}
  {\bibfnamefont {G.}~\bibnamefont {McGregor}}, \bibinfo {author}
  {\bibfnamefont {W.}~\bibnamefont {McLatchie}}, \bibinfo {author}
  {\bibfnamefont {R.~Meijer}\ \bibnamefont {Drees}}, \bibinfo {author}
  {\bibfnamefont {H.}~\bibnamefont {Mes}}, \bibinfo {author} {\bibfnamefont
  {C.}~\bibnamefont {Mifflin}}, \bibinfo {author} {\bibfnamefont {G.~G.}\
  \bibnamefont {Miller}}, \bibinfo {author} {\bibfnamefont {G.}~\bibnamefont
  {Milton}}, \bibinfo {author} {\bibfnamefont {B.~A.}\ \bibnamefont {Moffat}},
  \bibinfo {author} {\bibfnamefont {M.}~\bibnamefont {Moorhead}}, \bibinfo
  {author} {\bibfnamefont {C.~W.}\ \bibnamefont {Nally}}, \bibinfo {author}
  {\bibfnamefont {M.~S.}\ \bibnamefont {Neubauer}}, \bibinfo {author}
  {\bibfnamefont {F.~M.}\ \bibnamefont {Newcomer}}, \bibinfo {author}
  {\bibfnamefont {H.~S.}\ \bibnamefont {Ng}}, \bibinfo {author} {\bibfnamefont
  {A.~J.}\ \bibnamefont {Noble}}, \bibinfo {author} {\bibfnamefont {E.~B.}\
  \bibnamefont {Norman}}, \bibinfo {author} {\bibfnamefont {V.~M.}\
  \bibnamefont {Novikov}}, \bibinfo {author} {\bibfnamefont {M.}~\bibnamefont
  {O'Neill}}, \bibinfo {author} {\bibfnamefont {C.~E.}\ \bibnamefont {Okada}},
  \bibinfo {author} {\bibfnamefont {R.~W.}\ \bibnamefont {Ollerhead}}, \bibinfo
  {author} {\bibfnamefont {M.}~\bibnamefont {Omori}}, \bibinfo {author}
  {\bibfnamefont {J.~L.}\ \bibnamefont {Orrell}}, \bibinfo {author}
  {\bibfnamefont {S.~M.}\ \bibnamefont {Oser}}, \bibinfo {author}
  {\bibfnamefont {A.~W.~P.}\ \bibnamefont {Poon}}, \bibinfo {author}
  {\bibfnamefont {T.~J.}\ \bibnamefont {Radcliffe}}, \bibinfo {author}
  {\bibfnamefont {A.}~\bibnamefont {Roberge}}, \bibinfo {author} {\bibfnamefont
  {B.~C.}\ \bibnamefont {Robertson}}, \bibinfo {author} {\bibfnamefont
  {R.~G.~H.}\ \bibnamefont {Robertson}}, \bibinfo {author} {\bibfnamefont
  {J.~K.}\ \bibnamefont {Rowley}}, \bibinfo {author} {\bibfnamefont {V.~L.}\
  \bibnamefont {Rusu}}, \bibinfo {author} {\bibfnamefont {E.}~\bibnamefont
  {Saettler}}, \bibinfo {author} {\bibfnamefont {K.~K.}\ \bibnamefont
  {Schaffer}}, \bibinfo {author} {\bibfnamefont {A.}~\bibnamefont {Schuelke}},
  \bibinfo {author} {\bibfnamefont {M.~H.}\ \bibnamefont {Schwendener}},
  \bibinfo {author} {\bibfnamefont {H.}~\bibnamefont {Seifert}}, \bibinfo
  {author} {\bibfnamefont {M.}~\bibnamefont {Shatkay}}, \bibinfo {author}
  {\bibfnamefont {J.~J.}\ \bibnamefont {Simpson}}, \bibinfo {author}
  {\bibfnamefont {D.}~\bibnamefont {Sinclair}}, \bibinfo {author}
  {\bibfnamefont {P.}~\bibnamefont {Skensved}}, \bibinfo {author}
  {\bibfnamefont {A.~R.}\ \bibnamefont {Smith}}, \bibinfo {author}
  {\bibfnamefont {M.~W.~E.}\ \bibnamefont {Smith}}, \bibinfo {author}
  {\bibfnamefont {N.}~\bibnamefont {Starinsky}}, \bibinfo {author}
  {\bibfnamefont {T.~D.}\ \bibnamefont {Steiger}}, \bibinfo {author}
  {\bibfnamefont {R.~G.}\ \bibnamefont {Stokstad}}, \bibinfo {author}
  {\bibfnamefont {R.~S.}\ \bibnamefont {Storey}}, \bibinfo {author}
  {\bibfnamefont {B.}~\bibnamefont {Sur}}, \bibinfo {author} {\bibfnamefont
  {R.}~\bibnamefont {Tafirout}}, \bibinfo {author} {\bibfnamefont
  {N.}~\bibnamefont {Tagg}}, \bibinfo {author} {\bibfnamefont {N.~W.}\
  \bibnamefont {Tanner}}, \bibinfo {author} {\bibfnamefont {R.~K.}\
  \bibnamefont {Taplin}}, \bibinfo {author} {\bibfnamefont {M.}~\bibnamefont
  {Thorman}}, \bibinfo {author} {\bibfnamefont {P.}~\bibnamefont {Thornewell}},
  \bibinfo {author} {\bibfnamefont {P.~T.}\ \bibnamefont {Trent}}, \bibinfo
  {author} {\bibfnamefont {Y.~I.}\ \bibnamefont {Tserkovnyak}}, \bibinfo
  {author} {\bibfnamefont {R.}~\bibnamefont {Van~Berg}}, \bibinfo {author}
  {\bibfnamefont {R.~G.}\ \bibnamefont {Van~de Water}}, \bibinfo {author}
  {\bibfnamefont {C.~J.}\ \bibnamefont {Virtue}}, \bibinfo {author}
  {\bibfnamefont {C.~E.}\ \bibnamefont {Waltham}}, \bibinfo {author}
  {\bibfnamefont {J.-X.}\ \bibnamefont {Wang}}, \bibinfo {author}
  {\bibfnamefont {D.~L.}\ \bibnamefont {Wark}}, \bibinfo {author}
  {\bibfnamefont {N.}~\bibnamefont {West}}, \bibinfo {author} {\bibfnamefont
  {J.~B.}\ \bibnamefont {Wilhelmy}}, \bibinfo {author} {\bibfnamefont {J.~F.}\
  \bibnamefont {Wilkerson}}, \bibinfo {author} {\bibfnamefont {J.}~\bibnamefont
  {Wilson}}, \bibinfo {author} {\bibfnamefont {P.}~\bibnamefont {Wittich}},
  \bibinfo {author} {\bibfnamefont {J.~M.}\ \bibnamefont {Wouters}}, \ and\
  \bibinfo {author} {\bibfnamefont {M.}~\bibnamefont {Yeh}},\ }\bibfield
  {title} {\enquote {\bibinfo {title} {Measurement of the {Rate} of nu\_e + d
  -{\textgreater} p + p + e- {Interactions} {Produced} by {B} 8 {Solar}
  {Neutrinos} at the {Sudbury} {Neutrino} {Observatory}},}\ }\href {\doibase
  10.1103/PhysRevLett.87.071301} {\bibfield  {journal} {\bibinfo  {journal}
  {Physical Review Letters}\ }\textbf {\bibinfo {volume} {87}},\ \bibinfo
  {pages} {071301} (\bibinfo {year} {2001})}\BibitemShut {NoStop}%
\bibitem [{\citenamefont {Duan}\ \emph {et~al.}(2010)\citenamefont {Duan},
  \citenamefont {Fuller},\ and\ \citenamefont {Qian}}]{duan_collective_2010}%
  \BibitemOpen
  \bibfield  {author} {\bibinfo {author} {\bibfnamefont {Huaiyu}\ \bibnamefont
  {Duan}}, \bibinfo {author} {\bibfnamefont {George~M.}\ \bibnamefont
  {Fuller}}, \ and\ \bibinfo {author} {\bibfnamefont {Yong-Zhong}\ \bibnamefont
  {Qian}},\ }\bibfield  {title} {\enquote {\bibinfo {title} {Collective
  {Neutrino} {Oscillations}},}\ }\href {\doibase
  10.1146/annurev.nucl.012809.104524} {\bibfield  {journal} {\bibinfo
  {journal} {Annual Review of Nuclear and Particle Science}\ }\textbf {\bibinfo
  {volume} {60}},\ \bibinfo {pages} {569--594} (\bibinfo {year}
  {2010})}\BibitemShut {NoStop}%
\bibitem [{\citenamefont {Malkus}\ \emph {et~al.}(2016)\citenamefont {Malkus},
  \citenamefont {McLaughlin},\ and\ \citenamefont
  {Surman}}]{malkus_symmetric_2016}%
  \BibitemOpen
  \bibfield  {author} {\bibinfo {author} {\bibfnamefont {A.}~\bibnamefont
  {Malkus}}, \bibinfo {author} {\bibfnamefont {G.~C.}\ \bibnamefont
  {McLaughlin}}, \ and\ \bibinfo {author} {\bibfnamefont {R.}~\bibnamefont
  {Surman}},\ }\bibfield  {title} {\enquote {\bibinfo {title} {Symmetric and
  standard matter neutrino resonances above merging compact objects},}\ }\href
  {\doibase 10.1103/PhysRevD.93.045021} {\bibfield  {journal} {\bibinfo
  {journal} {Physical Review D}\ }\textbf {\bibinfo {volume} {93}},\ \bibinfo
  {pages} {045021} (\bibinfo {year} {2016})}\BibitemShut {NoStop}%
\bibitem [{\citenamefont {Wu}\ \emph {et~al.}(2017)\citenamefont {Wu},
  \citenamefont {Tamborra}, \citenamefont {Just},\ and\ \citenamefont
  {Janka}}]{wu_imprints_2017}%
  \BibitemOpen
  \bibfield  {author} {\bibinfo {author} {\bibfnamefont {Meng-Ru}\ \bibnamefont
  {Wu}}, \bibinfo {author} {\bibfnamefont {Irene}\ \bibnamefont {Tamborra}},
  \bibinfo {author} {\bibfnamefont {Oliver}\ \bibnamefont {Just}}, \ and\
  \bibinfo {author} {\bibfnamefont {Hans-Thomas}\ \bibnamefont {Janka}},\
  }\bibfield  {title} {\enquote {\bibinfo {title} {Imprints of neutrino-pair
  flavor conversions on nucleosynthesis in ejecta from neutron-star merger
  remnants},}\ }\href {\doibase 10.1103/PhysRevD.96.123015} {\bibfield
  {journal} {\bibinfo  {journal} {Physical Review D}\ }\textbf {\bibinfo
  {volume} {96}},\ \bibinfo {pages} {123015} (\bibinfo {year}
  {2017})}\BibitemShut {NoStop}%
\bibitem [{\citenamefont {George}\ \emph {et~al.}(2020)\citenamefont {George},
  \citenamefont {Wu}, \citenamefont {Tamborra}, \citenamefont
  {Ardevol-Pulpillo},\ and\ \citenamefont {Janka}}]{george_fast_2020}%
  \BibitemOpen
  \bibfield  {author} {\bibinfo {author} {\bibfnamefont {Manu}\ \bibnamefont
  {George}}, \bibinfo {author} {\bibfnamefont {Meng-Ru}\ \bibnamefont {Wu}},
  \bibinfo {author} {\bibfnamefont {Irene}\ \bibnamefont {Tamborra}}, \bibinfo
  {author} {\bibfnamefont {Ricard}\ \bibnamefont {Ardevol-Pulpillo}}, \ and\
  \bibinfo {author} {\bibfnamefont {Hans-Thomas}\ \bibnamefont {Janka}},\
  }\bibfield  {title} {\enquote {\bibinfo {title} {Fast neutrino flavor
  conversion, ejecta properties, and nucleosynthesis in newly-formed
  hypermassive remnants of neutron-star mergers},}\ }\href {\doibase
  10.1103/PhysRevD.102.103015} {\bibfield  {journal} {\bibinfo  {journal}
  {Physical Review D}\ }\textbf {\bibinfo {volume} {102}},\ \bibinfo {pages}
  {103015} (\bibinfo {year} {2020})}\BibitemShut {NoStop}%
\bibitem [{\citenamefont {Xiong}\ \emph {et~al.}(2020)\citenamefont {Xiong},
  \citenamefont {Sieverding}, \citenamefont {Sen},\ and\ \citenamefont
  {Qian}}]{xiong_potential_2020}%
  \BibitemOpen
  \bibfield  {author} {\bibinfo {author} {\bibfnamefont {Zewei}\ \bibnamefont
  {Xiong}}, \bibinfo {author} {\bibfnamefont {Andre}\ \bibnamefont
  {Sieverding}}, \bibinfo {author} {\bibfnamefont {Manibrata}\ \bibnamefont
  {Sen}}, \ and\ \bibinfo {author} {\bibfnamefont {Yong-Zhong}\ \bibnamefont
  {Qian}},\ }\bibfield  {title} {\enquote {\bibinfo {title} {Potential {Impact}
  of {Fast} {Flavor} {Oscillations} on {Neutrino}-driven {Winds} and {Their}
  {Nucleosynthesis}},}\ }\href {\doibase 10.3847/1538-4357/abac5e} {\bibfield
  {journal} {\bibinfo  {journal} {The Astrophysical Journal}\ }\textbf
  {\bibinfo {volume} {900}},\ \bibinfo {pages} {144} (\bibinfo {year}
  {2020})}\BibitemShut {NoStop}%
\bibitem [{\citenamefont {Sawyer}(2005)}]{sawyer_speed-up_2005}%
  \BibitemOpen
  \bibfield  {author} {\bibinfo {author} {\bibfnamefont {R.~F.}\ \bibnamefont
  {Sawyer}},\ }\bibfield  {title} {\enquote {\bibinfo {title} {Speed-up of
  neutrino transformations in a supernova environment},}\ }\href {\doibase
  10.1103/PhysRevD.72.045003} {\bibfield  {journal} {\bibinfo  {journal}
  {Physical Review D}\ }\textbf {\bibinfo {volume} {72}},\ \bibinfo {pages}
  {045003} (\bibinfo {year} {2005})}\BibitemShut {NoStop}%
\bibitem [{\citenamefont {Sawyer}(2016)}]{sawyer_neutrino_2016}%
  \BibitemOpen
  \bibfield  {author} {\bibinfo {author} {\bibfnamefont {R.~F.}\ \bibnamefont
  {Sawyer}},\ }\bibfield  {title} {\enquote {\bibinfo {title} {Neutrino cloud
  instabilities just above the neutrino sphere of a supernova},}\ }\href
  {\doibase 10.1103/PhysRevLett.116.081101} {\bibfield  {journal} {\bibinfo
  {journal} {Physical Review Letters}\ }\textbf {\bibinfo {volume} {116}},\
  \bibinfo {pages} {081101} (\bibinfo {year} {2016})}\BibitemShut {NoStop}%
\bibitem [{\citenamefont {Wu}\ and\ \citenamefont
  {Tamborra}(2017)}]{wu_fast_2017}%
  \BibitemOpen
  \bibfield  {author} {\bibinfo {author} {\bibfnamefont {Meng-Ru}\ \bibnamefont
  {Wu}}\ and\ \bibinfo {author} {\bibfnamefont {Irene}\ \bibnamefont
  {Tamborra}},\ }\bibfield  {title} {\enquote {\bibinfo {title} {Fast neutrino
  conversions: {Ubiquitous} in compact binary merger remnants},}\ }\href
  {\doibase 10.1103/PhysRevD.95.103007} {\bibfield  {journal} {\bibinfo
  {journal} {Physical Review D}\ }\textbf {\bibinfo {volume} {95}},\ \bibinfo
  {pages} {103007} (\bibinfo {year} {2017})}\BibitemShut {NoStop}%
\bibitem [{\citenamefont {Morinaga}\ \emph {et~al.}(2020)\citenamefont
  {Morinaga}, \citenamefont {Nagakura}, \citenamefont {Kato},\ and\
  \citenamefont {Yamada}}]{morinaga_fast_2020}%
  \BibitemOpen
  \bibfield  {author} {\bibinfo {author} {\bibfnamefont {Taiki}\ \bibnamefont
  {Morinaga}}, \bibinfo {author} {\bibfnamefont {Hiroki}\ \bibnamefont
  {Nagakura}}, \bibinfo {author} {\bibfnamefont {Chinami}\ \bibnamefont
  {Kato}}, \ and\ \bibinfo {author} {\bibfnamefont {Shoichi}\ \bibnamefont
  {Yamada}},\ }\bibfield  {title} {\enquote {\bibinfo {title} {Fast
  neutrino-flavor conversion in the preshock region of core-collapse
  supernovae},}\ }\href {\doibase 10.1103/PhysRevResearch.2.012046} {\bibfield
  {journal} {\bibinfo  {journal} {Physical Review Research}\ }\textbf {\bibinfo
  {volume} {2}},\ \bibinfo {pages} {012046(R)} (\bibinfo {year}
  {2020})}\BibitemShut {NoStop}%
\bibitem [{\citenamefont {Morinaga}(2021)}]{morinaga_fast_2021}%
  \BibitemOpen
  \bibfield  {author} {\bibinfo {author} {\bibfnamefont {Taiki}\ \bibnamefont
  {Morinaga}},\ }\bibfield  {title} {\enquote {\bibinfo {title} {Fast neutrino
  flavor instability and neutrino flavor lepton number crossings},}\ }\href
  {http://arxiv.org/abs/2103.15267} {\bibfield  {journal} {\bibinfo  {journal}
  {arXiv:2103.15267 [astro-ph, physics:hep-ph]}\ } (\bibinfo {year}
  {2021})}\BibitemShut {NoStop}%
\bibitem [{\citenamefont {Tamborra}\ \emph {et~al.}(2017)\citenamefont
  {Tamborra}, \citenamefont {Hüdepohl}, \citenamefont {Raffelt},\ and\
  \citenamefont {Janka}}]{tamborra_flavor-dependent_2017}%
  \BibitemOpen
  \bibfield  {author} {\bibinfo {author} {\bibfnamefont {Irene}\ \bibnamefont
  {Tamborra}}, \bibinfo {author} {\bibfnamefont {Lorenz}\ \bibnamefont
  {Hüdepohl}}, \bibinfo {author} {\bibfnamefont {Georg~G.}\ \bibnamefont
  {Raffelt}}, \ and\ \bibinfo {author} {\bibfnamefont {Hans-Thomas}\
  \bibnamefont {Janka}},\ }\bibfield  {title} {\enquote {\bibinfo {title}
  {Flavor-dependent {Neutrino} {Angular} {Distribution} in {Core}-collapse
  {Supernovae}},}\ }\href {\doibase 10.3847/1538-4357/aa6a18} {\bibfield
  {journal} {\bibinfo  {journal} {The Astrophysical Journal}\ }\textbf
  {\bibinfo {volume} {839}},\ \bibinfo {pages} {132} (\bibinfo {year}
  {2017})}\BibitemShut {NoStop}%
\bibitem [{\citenamefont {Abbar}\ \emph {et~al.}(2019)\citenamefont {Abbar},
  \citenamefont {Duan}, \citenamefont {Sumiyoshi}, \citenamefont {Takiwaki},\
  and\ \citenamefont {Volpe}}]{abbar_occurrence_2019}%
  \BibitemOpen
  \bibfield  {author} {\bibinfo {author} {\bibfnamefont {Sajad}\ \bibnamefont
  {Abbar}}, \bibinfo {author} {\bibfnamefont {Huaiyu}\ \bibnamefont {Duan}},
  \bibinfo {author} {\bibfnamefont {Kohsuke}\ \bibnamefont {Sumiyoshi}},
  \bibinfo {author} {\bibfnamefont {Tomoya}\ \bibnamefont {Takiwaki}}, \ and\
  \bibinfo {author} {\bibfnamefont {Maria~Cristina}\ \bibnamefont {Volpe}},\
  }\bibfield  {title} {\enquote {\bibinfo {title} {On the occurrence of fast
  neutrino flavor conversions in multidimensional supernova models},}\ }\href
  {\doibase 10.1103/PhysRevD.100.043004} {\bibfield  {journal} {\bibinfo
  {journal} {Physical Review D}\ }\textbf {\bibinfo {volume} {100}},\ \bibinfo
  {pages} {043004} (\bibinfo {year} {2019})}\BibitemShut {NoStop}%
\bibitem [{\citenamefont {{M. D. Azari}}\ \emph {et~al.}(2019)\citenamefont
  {{M. D. Azari}}, \citenamefont {Yamada}, \citenamefont {Morinaga},
  \citenamefont {Iwakami}, \citenamefont {Okawa}, \citenamefont {Nagakura},\
  and\ \citenamefont {Sumiyoshi}}]{m_d_azari_linear_2019}%
  \BibitemOpen
  \bibfield  {author} {\bibinfo {author} {\bibnamefont {{M. D. Azari}}},
  \bibinfo {author} {\bibfnamefont {Shoichi}\ \bibnamefont {Yamada}}, \bibinfo
  {author} {\bibfnamefont {Taiki}\ \bibnamefont {Morinaga}}, \bibinfo {author}
  {\bibfnamefont {Wakana}\ \bibnamefont {Iwakami}}, \bibinfo {author}
  {\bibfnamefont {Hirotada}\ \bibnamefont {Okawa}}, \bibinfo {author}
  {\bibfnamefont {Hiroki}\ \bibnamefont {Nagakura}}, \ and\ \bibinfo {author}
  {\bibfnamefont {Kohsuke}\ \bibnamefont {Sumiyoshi}},\ }\bibfield  {title}
  {\enquote {\bibinfo {title} {Linear analysis of fast-pairwise collective
  neutrino oscillations in core-collapse supernovae based on the results of
  {Boltzmann} simulations},}\ }\href {\doibase 10.1103/PhysRevD.99.103011}
  {\bibfield  {journal} {\bibinfo  {journal} {Physical Review D}\ }\textbf
  {\bibinfo {volume} {99}},\ \bibinfo {pages} {103011} (\bibinfo {year}
  {2019})}\BibitemShut {NoStop}%
\bibitem [{\citenamefont {{M. Delfan Azari}}\ \emph {et~al.}(2020)\citenamefont
  {{M. Delfan Azari}}, \citenamefont {Yamada}, \citenamefont {Morinaga},
  \citenamefont {Nagakura}, \citenamefont {Furusawa}, \citenamefont {Harada},
  \citenamefont {Okawa}, \citenamefont {Iwakami},\ and\ \citenamefont
  {Sumiyoshi}}]{m_delfan_azari_fast_2020}%
  \BibitemOpen
  \bibfield  {author} {\bibinfo {author} {\bibnamefont {{M. Delfan Azari}}},
  \bibinfo {author} {\bibfnamefont {S.}~\bibnamefont {Yamada}}, \bibinfo
  {author} {\bibfnamefont {T.}~\bibnamefont {Morinaga}}, \bibinfo {author}
  {\bibfnamefont {H.}~\bibnamefont {Nagakura}}, \bibinfo {author}
  {\bibfnamefont {S.}~\bibnamefont {Furusawa}}, \bibinfo {author}
  {\bibfnamefont {A.}~\bibnamefont {Harada}}, \bibinfo {author} {\bibfnamefont
  {H.}~\bibnamefont {Okawa}}, \bibinfo {author} {\bibfnamefont
  {W.}~\bibnamefont {Iwakami}}, \ and\ \bibinfo {author} {\bibfnamefont
  {K}~\bibnamefont {Sumiyoshi}},\ }\bibfield  {title} {\enquote {\bibinfo
  {title} {Fast collective neutrino oscillations inside the neutrino sphere in
  core-collapse supernovae},}\ }\href {\doibase 10.1103/PhysRevD.101.023018}
  {\bibfield  {journal} {\bibinfo  {journal} {Physical Review D}\ }\textbf
  {\bibinfo {volume} {101}},\ \bibinfo {pages} {023018} (\bibinfo {year}
  {2020})}\BibitemShut {NoStop}%
\bibitem [{\citenamefont {Abbar}\ \emph {et~al.}(2020)\citenamefont {Abbar},
  \citenamefont {Duan}, \citenamefont {Sumiyoshi}, \citenamefont {Takiwaki},\
  and\ \citenamefont {Volpe}}]{abbar_fast_2020}%
  \BibitemOpen
  \bibfield  {author} {\bibinfo {author} {\bibfnamefont {S.}~\bibnamefont
  {Abbar}}, \bibinfo {author} {\bibfnamefont {H.}~\bibnamefont {Duan}},
  \bibinfo {author} {\bibfnamefont {K.}~\bibnamefont {Sumiyoshi}}, \bibinfo
  {author} {\bibfnamefont {T.}~\bibnamefont {Takiwaki}}, \ and\ \bibinfo
  {author} {\bibfnamefont {M.~C.}\ \bibnamefont {Volpe}},\ }\bibfield  {title}
  {\enquote {\bibinfo {title} {Fast neutrino flavor conversion modes in
  multidimensional core-collapse supernova models: {The} role of the asymmetric
  neutrino distributions},}\ }\href {\doibase 10.1103/PhysRevD.101.043016}
  {\bibfield  {journal} {\bibinfo  {journal} {Physical Review D}\ }\textbf
  {\bibinfo {volume} {101}},\ \bibinfo {pages} {043016} (\bibinfo {year}
  {2020})}\BibitemShut {NoStop}%
\bibitem [{\citenamefont {Nagakura}\ \emph {et~al.}(2019)\citenamefont
  {Nagakura}, \citenamefont {Morinaga}, \citenamefont {Kato},\ and\
  \citenamefont {Yamada}}]{nagakura_fast-pairwise_2019}%
  \BibitemOpen
  \bibfield  {author} {\bibinfo {author} {\bibfnamefont {Hiroki}\ \bibnamefont
  {Nagakura}}, \bibinfo {author} {\bibfnamefont {Taiki}\ \bibnamefont
  {Morinaga}}, \bibinfo {author} {\bibfnamefont {Chinami}\ \bibnamefont
  {Kato}}, \ and\ \bibinfo {author} {\bibfnamefont {Shoichi}\ \bibnamefont
  {Yamada}},\ }\bibfield  {title} {\enquote {\bibinfo {title} {Fast-pairwise
  {Collective} {Neutrino} {Oscillations} {Associated} with {Asymmetric}
  {Neutrino} {Emissions} in {Core}-collapse {Supernovae}},}\ }\href {\doibase
  10.3847/1538-4357/ab4cf2} {\bibfield  {journal} {\bibinfo  {journal} {The
  Astrophysical Journal}\ }\textbf {\bibinfo {volume} {886}},\ \bibinfo {pages}
  {139} (\bibinfo {year} {2019})}\BibitemShut {NoStop}%
\bibitem [{\citenamefont {Glas}\ \emph {et~al.}(2020)\citenamefont {Glas},
  \citenamefont {{H. Thomas Janka}}, \citenamefont {Capozzi}, \citenamefont
  {Sen}, \citenamefont {Dasgupta}, \citenamefont {Mirizzi},\ and\ \citenamefont
  {Sigl}}]{glas_fast_2020}%
  \BibitemOpen
  \bibfield  {author} {\bibinfo {author} {\bibfnamefont {Robert}\ \bibnamefont
  {Glas}}, \bibinfo {author} {\bibnamefont {{H. Thomas Janka}}}, \bibinfo
  {author} {\bibfnamefont {F.}~\bibnamefont {Capozzi}}, \bibinfo {author}
  {\bibfnamefont {M.}~\bibnamefont {Sen}}, \bibinfo {author} {\bibfnamefont
  {B.}~\bibnamefont {Dasgupta}}, \bibinfo {author} {\bibfnamefont
  {A.}~\bibnamefont {Mirizzi}}, \ and\ \bibinfo {author} {\bibfnamefont
  {G.}~\bibnamefont {Sigl}},\ }\bibfield  {title} {\enquote {\bibinfo {title}
  {Fast neutrino flavor instability in the neutron-star convection layer of
  three-dimensional supernova models},}\ }\href {\doibase
  10.1103/PhysRevD.101.063001} {\bibfield  {journal} {\bibinfo  {journal}
  {Physical Review D}\ }\textbf {\bibinfo {volume} {101}},\ \bibinfo {pages}
  {063001} (\bibinfo {year} {2020})}\BibitemShut {NoStop}%
\bibitem [{\citenamefont {Capozzi}\ \emph {et~al.}(2021)\citenamefont
  {Capozzi}, \citenamefont {Abbar}, \citenamefont {Bollig},\ and\ \citenamefont
  {Janka}}]{capozzi_fast_2021}%
  \BibitemOpen
  \bibfield  {author} {\bibinfo {author} {\bibfnamefont {Francesco}\
  \bibnamefont {Capozzi}}, \bibinfo {author} {\bibfnamefont {Sajad}\
  \bibnamefont {Abbar}}, \bibinfo {author} {\bibfnamefont {Robert}\
  \bibnamefont {Bollig}}, \ and\ \bibinfo {author} {\bibfnamefont {H.-Thomas}\
  \bibnamefont {Janka}},\ }\bibfield  {title} {\enquote {\bibinfo {title} {Fast
  neutrino flavor conversions in one-dimensional core-collapse supernova models
  with and without muon creation},}\ }\href {\doibase
  10.1103/PhysRevD.103.063013} {\bibfield  {journal} {\bibinfo  {journal}
  {Phys. Rev. D}\ }\textbf {\bibinfo {volume} {103}},\ \bibinfo {pages}
  {063013} (\bibinfo {year} {2021})}\BibitemShut {NoStop}%
\bibitem [{\citenamefont {Abbar}\ \emph {et~al.}(2021)\citenamefont {Abbar},
  \citenamefont {Capozzi}, \citenamefont {Glas}, \citenamefont {Janka},\ and\
  \citenamefont {Tamborra}}]{abbar_characteristics_2021}%
  \BibitemOpen
  \bibfield  {author} {\bibinfo {author} {\bibfnamefont {Sajad}\ \bibnamefont
  {Abbar}}, \bibinfo {author} {\bibfnamefont {Francesco}\ \bibnamefont
  {Capozzi}}, \bibinfo {author} {\bibfnamefont {Robert}\ \bibnamefont {Glas}},
  \bibinfo {author} {\bibfnamefont {H.-Thomas}\ \bibnamefont {Janka}}, \ and\
  \bibinfo {author} {\bibfnamefont {Irene}\ \bibnamefont {Tamborra}},\
  }\bibfield  {title} {\enquote {\bibinfo {title} {On the characteristics of
  fast neutrino flavor instabilities in three-dimensional core-collapse
  supernova models},}\ }\href {\doibase 10/gmt59m} {\bibfield  {journal}
  {\bibinfo  {journal} {Physical Review D}\ }\textbf {\bibinfo {volume}
  {103}},\ \bibinfo {pages} {063033} (\bibinfo {year} {2021})}\BibitemShut
  {NoStop}%
\bibitem [{\citenamefont {Dasgupta}\ and\ \citenamefont
  {Mirizzi}(2015)}]{dasgupta_temporal_2015}%
  \BibitemOpen
  \bibfield  {author} {\bibinfo {author} {\bibfnamefont {Basudeb}\ \bibnamefont
  {Dasgupta}}\ and\ \bibinfo {author} {\bibfnamefont {Alessandro}\ \bibnamefont
  {Mirizzi}},\ }\bibfield  {title} {\enquote {\bibinfo {title} {Temporal
  instability enables neutrino flavor conversions deep inside supernovae},}\
  }\href {\doibase 10/gkp7d3} {\bibfield  {journal} {\bibinfo  {journal}
  {Physical Review D}\ }\textbf {\bibinfo {volume} {92}},\ \bibinfo {pages}
  {125030} (\bibinfo {year} {2015})}\BibitemShut {NoStop}%
\bibitem [{\citenamefont {Capozzi}\ \emph {et~al.}(2019)\citenamefont
  {Capozzi}, \citenamefont {Dasgupta}, \citenamefont {Mirizzi}, \citenamefont
  {Sen},\ and\ \citenamefont {Sigl}}]{capozzi_collisional_2019}%
  \BibitemOpen
  \bibfield  {author} {\bibinfo {author} {\bibfnamefont {Francesco}\
  \bibnamefont {Capozzi}}, \bibinfo {author} {\bibfnamefont {Basudeb}\
  \bibnamefont {Dasgupta}}, \bibinfo {author} {\bibfnamefont {Alessandro}\
  \bibnamefont {Mirizzi}}, \bibinfo {author} {\bibfnamefont {Manibrata}\
  \bibnamefont {Sen}}, \ and\ \bibinfo {author} {\bibfnamefont {Günter}\
  \bibnamefont {Sigl}},\ }\bibfield  {title} {\enquote {\bibinfo {title}
  {Collisional {Triggering} of {Fast} {Flavor} {Conversions} of {Supernova}
  {Neutrinos}},}\ }\href {\doibase 10.1103/PhysRevLett.122.091101} {\bibfield
  {journal} {\bibinfo  {journal} {Physical Review Letters}\ }\textbf {\bibinfo
  {volume} {122}},\ \bibinfo {pages} {091101} (\bibinfo {year}
  {2019})}\BibitemShut {NoStop}%
\bibitem [{\citenamefont {Abbar}\ \emph {et~al.}(2015)\citenamefont {Abbar},
  \citenamefont {Duan},\ and\ \citenamefont {Shalgar}}]{abbar_flavor_2015}%
  \BibitemOpen
  \bibfield  {author} {\bibinfo {author} {\bibfnamefont {Sajad}\ \bibnamefont
  {Abbar}}, \bibinfo {author} {\bibfnamefont {Huaiyu}\ \bibnamefont {Duan}}, \
  and\ \bibinfo {author} {\bibfnamefont {Shashank}\ \bibnamefont {Shalgar}},\
  }\bibfield  {title} {\enquote {\bibinfo {title} {Flavor instabilities in the
  multiangle neutrino line model},}\ }\href {\doibase 10/gmpwhc} {\bibfield
  {journal} {\bibinfo  {journal} {Physical Review D}\ }\textbf {\bibinfo
  {volume} {92}},\ \bibinfo {pages} {065019} (\bibinfo {year} {2015})},\
  \bibinfo {note} {publisher: American Physical Society}\BibitemShut {NoStop}%
\bibitem [{\citenamefont {Abbar}\ and\ \citenamefont
  {Volpe}(2019)}]{abbar_fast_2019}%
  \BibitemOpen
  \bibfield  {author} {\bibinfo {author} {\bibfnamefont {Sajad}\ \bibnamefont
  {Abbar}}\ and\ \bibinfo {author} {\bibfnamefont {Maria~Cristina}\
  \bibnamefont {Volpe}},\ }\bibfield  {title} {\enquote {\bibinfo {title} {On
  fast neutrino flavor conversion modes in the nonlinear regime},}\ }\href
  {\doibase 10.1016/j.physletb.2019.02.002} {\bibfield  {journal} {\bibinfo
  {journal} {Physics Letters B}\ }\textbf {\bibinfo {volume} {790}},\ \bibinfo
  {pages} {545--550} (\bibinfo {year} {2019})}\BibitemShut {NoStop}%
\bibitem [{\citenamefont {Padilla-Gay}\ \emph {et~al.}(2021)\citenamefont
  {Padilla-Gay}, \citenamefont {Shalgar},\ and\ \citenamefont
  {Tamborra}}]{padilla-gay_multi-dimensional_2021}%
  \BibitemOpen
  \bibfield  {author} {\bibinfo {author} {\bibfnamefont {Ian}\ \bibnamefont
  {Padilla-Gay}}, \bibinfo {author} {\bibfnamefont {Shashank}\ \bibnamefont
  {Shalgar}}, \ and\ \bibinfo {author} {\bibfnamefont {Irene}\ \bibnamefont
  {Tamborra}},\ }\bibfield  {title} {\enquote {\bibinfo {title}
  {Multi-{Dimensional} {Solution} of {Fast} {Neutrino} {Conversions} in
  {Binary} {Neutron} {Star} {Merger} {Remnants}},}\ }\href {\doibase
  10.1088/1475-7516/2021/01/017} {\bibfield  {journal} {\bibinfo  {journal} {J.
  Cosmol. Astropart. Phys.}\ }\textbf {\bibinfo {volume} {01}},\ \bibinfo
  {pages} {017} (\bibinfo {year} {2021})}\BibitemShut {NoStop}%
\bibitem [{\citenamefont {Padilla-Gay}\ and\ \citenamefont
  {Shalgar}(2021)}]{padilla-gay_fast_2021}%
  \BibitemOpen
  \bibfield  {author} {\bibinfo {author} {\bibfnamefont {Ian}\ \bibnamefont
  {Padilla-Gay}}\ and\ \bibinfo {author} {\bibfnamefont {Shashank}\
  \bibnamefont {Shalgar}},\ }\bibfield  {title} {\enquote {\bibinfo {title}
  {Fast flavor conversion of neutrinos in presence of matter bulk velocity},}\
  }\href {https://arxiv.org/abs/2108.00012v1} {\bibfield  {journal} {\bibinfo
  {journal} {arxiv:2108:00012 [astro-ph]}\ } (\bibinfo {year}
  {2021})}\BibitemShut {NoStop}%
\bibitem [{\citenamefont {Shalgar}\ and\ \citenamefont
  {Tamborra}(2021{\natexlab{a}})}]{shalgar_change_2021}%
  \BibitemOpen
  \bibfield  {author} {\bibinfo {author} {\bibfnamefont {Shashank}\
  \bibnamefont {Shalgar}}\ and\ \bibinfo {author} {\bibfnamefont {Irene}\
  \bibnamefont {Tamborra}},\ }\bibfield  {title} {\enquote {\bibinfo {title} {A
  change of direction in pairwise neutrino conversion physics: {The} effect of
  collisions},}\ }\href@noop {} {\bibfield  {journal} {\bibinfo  {journal}
  {Phys. Rev. D}\ }\textbf {\bibinfo {volume} {103}},\ \bibinfo {pages}
  {063002} (\bibinfo {year} {2021}{\natexlab{a}})}\BibitemShut {NoStop}%
\bibitem [{\citenamefont {Shalgar}\ and\ \citenamefont
  {Tamborra}(2021{\natexlab{b}})}]{shalgar_three_2021}%
  \BibitemOpen
  \bibfield  {author} {\bibinfo {author} {\bibfnamefont {Shashank}\
  \bibnamefont {Shalgar}}\ and\ \bibinfo {author} {\bibfnamefont {Irene}\
  \bibnamefont {Tamborra}},\ }\bibfield  {title} {\enquote {\bibinfo {title}
  {Three flavor revolution in fast pairwise neutrino conversion},}\ }\href
  {\doibase 10/gmt59n} {\bibfield  {journal} {\bibinfo  {journal} {Physical
  Review D}\ }\textbf {\bibinfo {volume} {104}},\ \bibinfo {pages} {023011}
  (\bibinfo {year} {2021}{\natexlab{b}})}\BibitemShut {NoStop}%
\bibitem [{\citenamefont {Xiong}\ and\ \citenamefont
  {Qian}(2021)}]{xiong_stationary_2021}%
  \BibitemOpen
  \bibfield  {author} {\bibinfo {author} {\bibfnamefont {Zewei}\ \bibnamefont
  {Xiong}}\ and\ \bibinfo {author} {\bibfnamefont {Yong-Zhong}\ \bibnamefont
  {Qian}},\ }\bibfield  {title} {\enquote {\bibinfo {title} {Stationary
  solutions for fast flavor oscillations of a homogeneous dense neutrino
  gas},}\ }\href {\doibase 10/gmt59q} {\bibfield  {journal} {\bibinfo
  {journal} {Physics Letters B}\ }\textbf {\bibinfo {volume} {820}},\ \bibinfo
  {pages} {136550} (\bibinfo {year} {2021})}\BibitemShut {NoStop}%
\bibitem [{\citenamefont {Shalgar}\ and\ \citenamefont
  {Tamborra}(2021{\natexlab{c}})}]{shalgar_symmetry_2021}%
  \BibitemOpen
  \bibfield  {author} {\bibinfo {author} {\bibfnamefont {Shashank}\
  \bibnamefont {Shalgar}}\ and\ \bibinfo {author} {\bibfnamefont {Irene}\
  \bibnamefont {Tamborra}},\ }\bibfield  {title} {\enquote {\bibinfo {title}
  {Symmetry breaking induced by pairwise conversion of neutrinos in compact
  sources},}\ }\href {http://arxiv.org/abs/2106.15622} {\bibfield  {journal}
  {\bibinfo  {journal} {arXiv:2106.15622 [astro-ph, physics:hep-ph]}\ }
  (\bibinfo {year} {2021}{\natexlab{c}})}\BibitemShut {NoStop}%
\bibitem [{\citenamefont {Mirizzi}\ \emph {et~al.}(2015)\citenamefont
  {Mirizzi}, \citenamefont {Mangano},\ and\ \citenamefont
  {Saviano}}]{mirizzi_self-induced_2015}%
  \BibitemOpen
  \bibfield  {author} {\bibinfo {author} {\bibfnamefont {Alessandro}\
  \bibnamefont {Mirizzi}}, \bibinfo {author} {\bibfnamefont {Gianpiero}\
  \bibnamefont {Mangano}}, \ and\ \bibinfo {author} {\bibfnamefont {Ninetta}\
  \bibnamefont {Saviano}},\ }\bibfield  {title} {\enquote {\bibinfo {title}
  {Self-induced flavor instabilities of a dense neutrino stream in a
  two-dimensional model},}\ }\href {\doibase 10.1103/PhysRevD.92.021702}
  {\bibfield  {journal} {\bibinfo  {journal} {Physical Review D}\ }\textbf
  {\bibinfo {volume} {92}},\ \bibinfo {pages} {021702} (\bibinfo {year}
  {2015})}\BibitemShut {NoStop}%
\bibitem [{\citenamefont {Bhattacharyya}\ and\ \citenamefont
  {Dasgupta}(2020)}]{bhattacharyya_late-time_2020}%
  \BibitemOpen
  \bibfield  {author} {\bibinfo {author} {\bibfnamefont {Soumya}\ \bibnamefont
  {Bhattacharyya}}\ and\ \bibinfo {author} {\bibfnamefont {Basudeb}\
  \bibnamefont {Dasgupta}},\ }\bibfield  {title} {\enquote {\bibinfo {title}
  {Late-time behavior of fast neutrino oscillations},}\ }\href {\doibase
  10.1103/PhysRevD.102.063018} {\bibfield  {journal} {\bibinfo  {journal}
  {Physical Review D}\ }\textbf {\bibinfo {volume} {102}},\ \bibinfo {pages}
  {063018} (\bibinfo {year} {2020})}\BibitemShut {NoStop}%
\bibitem [{\citenamefont {Wu}\ \emph {et~al.}(2021)\citenamefont {Wu},
  \citenamefont {George}, \citenamefont {Lin},\ and\ \citenamefont
  {Xiong}}]{wu_collective_2021}%
  \BibitemOpen
  \bibfield  {author} {\bibinfo {author} {\bibfnamefont {Meng-Ru}\ \bibnamefont
  {Wu}}, \bibinfo {author} {\bibfnamefont {Manu}\ \bibnamefont {George}},
  \bibinfo {author} {\bibfnamefont {Chun-Yu}\ \bibnamefont {Lin}}, \ and\
  \bibinfo {author} {\bibfnamefont {Zewei}\ \bibnamefont {Xiong}},\ }\bibfield
  {title} {\enquote {\bibinfo {title} {Collective fast neutrino flavor
  conversions in an {1D} box: ({I}) initial condition and long-term
  evolution},}\ }\href {http://arxiv.org/abs/2108.09886} {\bibfield  {journal}
  {\bibinfo  {journal} {arXiv:2108.09886 [astro-ph, physics:hep-ph]}\ }
  (\bibinfo {year} {2021})}\BibitemShut {NoStop}%
\bibitem [{\citenamefont {Bhattacharyya}\ and\ \citenamefont
  {Dasgupta}(2021)}]{bhattacharyya_fast_2021}%
  \BibitemOpen
  \bibfield  {author} {\bibinfo {author} {\bibfnamefont {Soumya}\ \bibnamefont
  {Bhattacharyya}}\ and\ \bibinfo {author} {\bibfnamefont {Basudeb}\
  \bibnamefont {Dasgupta}},\ }\bibfield  {title} {\enquote {\bibinfo {title}
  {Fast {Flavor} {Depolarization} of {Supernova} {Neutrinos}},}\ }\href
  {\doibase 10.1103/PhysRevLett.126.061302} {\bibfield  {journal} {\bibinfo
  {journal} {Phys. Rev. Lett.}\ }\textbf {\bibinfo {volume} {126}},\ \bibinfo
  {pages} {061302} (\bibinfo {year} {2021})}\BibitemShut {NoStop}%
\bibitem [{\citenamefont {Richers}\ \emph {et~al.}(2021)\citenamefont
  {Richers}, \citenamefont {Willcox}, \citenamefont {Ford},\ and\ \citenamefont
  {Myers}}]{richers_particle--cell_2021}%
  \BibitemOpen
  \bibfield  {author} {\bibinfo {author} {\bibfnamefont {Sherwood}\
  \bibnamefont {Richers}}, \bibinfo {author} {\bibfnamefont {Don~E.}\
  \bibnamefont {Willcox}}, \bibinfo {author} {\bibfnamefont {Nicole~M.}\
  \bibnamefont {Ford}}, \ and\ \bibinfo {author} {\bibfnamefont {Andrew}\
  \bibnamefont {Myers}},\ }\bibfield  {title} {\enquote {\bibinfo {title}
  {Particle-in-cell simulation of the neutrino fast flavor instability},}\
  }\href {\doibase 10.1103/PhysRevD.103.083013} {\bibfield  {journal} {\bibinfo
   {journal} {Physical Review D}\ }\textbf {\bibinfo {volume} {103}},\ \bibinfo
  {pages} {083013} (\bibinfo {year} {2021})}\BibitemShut {NoStop}%
\bibitem [{noa(2021)}]{noauthor_emu_2021}%
  \BibitemOpen
  \href {\doibase 10.6084/m9.figshare.14158547.v2} {\enquote {\bibinfo {title}
  {Emu {Illustrations}},}\ } (\bibinfo {year} {2021}),\ \bibinfo {note}
  {publisher: figshare}\BibitemShut {NoStop}%
\bibitem [{\citenamefont {Virtanen}\ \emph {et~al.}(2020)\citenamefont
  {Virtanen}, \citenamefont {Gommers}, \citenamefont {Oliphant}, \citenamefont
  {Haberland}, \citenamefont {Reddy}, \citenamefont {Cournapeau}, \citenamefont
  {Burovski}, \citenamefont {Peterson}, \citenamefont {Weckesser},
  \citenamefont {Bright}, \citenamefont {van~der Walt}, \citenamefont {Brett},
  \citenamefont {Wilson}, \citenamefont {Millman}, \citenamefont {Mayorov},
  \citenamefont {Nelson}, \citenamefont {Jones}, \citenamefont {Kern},
  \citenamefont {Larson}, \citenamefont {Carey}, \citenamefont {Polat},
  \citenamefont {Feng}, \citenamefont {Moore}, \citenamefont {VanderPlas},
  \citenamefont {Laxalde}, \citenamefont {Perktold}, \citenamefont {Cimrman},
  \citenamefont {Henriksen}, \citenamefont {Quintero}, \citenamefont {Harris},
  \citenamefont {Archibald}, \citenamefont {Ribeiro}, \citenamefont
  {Pedregosa},\ and\ \citenamefont {van Mulbregt}}]{virtanen_scipy_2020}%
  \BibitemOpen
  \bibfield  {author} {\bibinfo {author} {\bibfnamefont {Pauli}\ \bibnamefont
  {Virtanen}}, \bibinfo {author} {\bibfnamefont {Ralf}\ \bibnamefont
  {Gommers}}, \bibinfo {author} {\bibfnamefont {Travis~E.}\ \bibnamefont
  {Oliphant}}, \bibinfo {author} {\bibfnamefont {Matt}\ \bibnamefont
  {Haberland}}, \bibinfo {author} {\bibfnamefont {Tyler}\ \bibnamefont
  {Reddy}}, \bibinfo {author} {\bibfnamefont {David}\ \bibnamefont
  {Cournapeau}}, \bibinfo {author} {\bibfnamefont {Evgeni}\ \bibnamefont
  {Burovski}}, \bibinfo {author} {\bibfnamefont {Pearu}\ \bibnamefont
  {Peterson}}, \bibinfo {author} {\bibfnamefont {Warren}\ \bibnamefont
  {Weckesser}}, \bibinfo {author} {\bibfnamefont {Jonathan}\ \bibnamefont
  {Bright}}, \bibinfo {author} {\bibfnamefont {Stéfan~J.}\ \bibnamefont
  {van~der Walt}}, \bibinfo {author} {\bibfnamefont {Matthew}\ \bibnamefont
  {Brett}}, \bibinfo {author} {\bibfnamefont {Joshua}\ \bibnamefont {Wilson}},
  \bibinfo {author} {\bibfnamefont {K.~Jarrod}\ \bibnamefont {Millman}},
  \bibinfo {author} {\bibfnamefont {Nikolay}\ \bibnamefont {Mayorov}}, \bibinfo
  {author} {\bibfnamefont {Andrew R.~J.}\ \bibnamefont {Nelson}}, \bibinfo
  {author} {\bibfnamefont {Eric}\ \bibnamefont {Jones}}, \bibinfo {author}
  {\bibfnamefont {Robert}\ \bibnamefont {Kern}}, \bibinfo {author}
  {\bibfnamefont {Eric}\ \bibnamefont {Larson}}, \bibinfo {author}
  {\bibfnamefont {C.~J.}\ \bibnamefont {Carey}}, \bibinfo {author}
  {\bibfnamefont {İlhan}\ \bibnamefont {Polat}}, \bibinfo {author}
  {\bibfnamefont {Yu}~\bibnamefont {Feng}}, \bibinfo {author} {\bibfnamefont
  {Eric~W.}\ \bibnamefont {Moore}}, \bibinfo {author} {\bibfnamefont {Jake}\
  \bibnamefont {VanderPlas}}, \bibinfo {author} {\bibfnamefont {Denis}\
  \bibnamefont {Laxalde}}, \bibinfo {author} {\bibfnamefont {Josef}\
  \bibnamefont {Perktold}}, \bibinfo {author} {\bibfnamefont {Robert}\
  \bibnamefont {Cimrman}}, \bibinfo {author} {\bibfnamefont {Ian}\ \bibnamefont
  {Henriksen}}, \bibinfo {author} {\bibfnamefont {E.~A.}\ \bibnamefont
  {Quintero}}, \bibinfo {author} {\bibfnamefont {Charles~R.}\ \bibnamefont
  {Harris}}, \bibinfo {author} {\bibfnamefont {Anne~M.}\ \bibnamefont
  {Archibald}}, \bibinfo {author} {\bibfnamefont {Antônio~H.}\ \bibnamefont
  {Ribeiro}}, \bibinfo {author} {\bibfnamefont {Fabian}\ \bibnamefont
  {Pedregosa}}, \ and\ \bibinfo {author} {\bibfnamefont {Paul}\ \bibnamefont
  {van Mulbregt}},\ }\bibfield  {title} {\enquote {\bibinfo {title} {{SciPy}
  1.0: fundamental algorithms for scientific computing in {Python}},}\ }\href
  {\doibase 10.1038/s41592-019-0686-2} {\bibfield  {journal} {\bibinfo
  {journal} {Nature Methods}\ }\textbf {\bibinfo {volume} {17}},\ \bibinfo
  {pages} {261--272} (\bibinfo {year} {2020})}\BibitemShut {NoStop}%
\bibitem [{\citenamefont {Fornberg}\ and\ \citenamefont
  {Martel}(2014)}]{fornberg_spherical_2014}%
  \BibitemOpen
  \bibfield  {author} {\bibinfo {author} {\bibfnamefont {Bengt}\ \bibnamefont
  {Fornberg}}\ and\ \bibinfo {author} {\bibfnamefont {Jordan~M.}\ \bibnamefont
  {Martel}},\ }\bibfield  {title} {\enquote {\bibinfo {title} {On spherical
  harmonics based numerical quadrature over the surface of a sphere},}\ }\href
  {\doibase 10/gmq8xt} {\bibfield  {journal} {\bibinfo  {journal} {Advances in
  Computational Mathematics}\ }\textbf {\bibinfo {volume} {40}},\ \bibinfo
  {pages} {1169--1184} (\bibinfo {year} {2014})}\BibitemShut {NoStop}%
\bibitem [{\citenamefont {Johns}\ \emph {et~al.}(2020)\citenamefont {Johns},
  \citenamefont {Nagakura}, \citenamefont {Fuller},\ and\ \citenamefont
  {Burrows}}]{johns_fast_2020}%
  \BibitemOpen
  \bibfield  {author} {\bibinfo {author} {\bibfnamefont {Lucas}\ \bibnamefont
  {Johns}}, \bibinfo {author} {\bibfnamefont {Hiroki}\ \bibnamefont
  {Nagakura}}, \bibinfo {author} {\bibfnamefont {George~M.}\ \bibnamefont
  {Fuller}}, \ and\ \bibinfo {author} {\bibfnamefont {Adam}\ \bibnamefont
  {Burrows}},\ }\bibfield  {title} {\enquote {\bibinfo {title} {Fast
  oscillations, collisionless relaxation, and spurious evolution of supernova
  neutrino flavor},}\ }\href {\doibase 10.1103/PhysRevD.102.103017} {\bibfield
  {journal} {\bibinfo  {journal} {Physical Review D}\ }\textbf {\bibinfo
  {volume} {102}},\ \bibinfo {pages} {103017} (\bibinfo {year}
  {2020})}\BibitemShut {NoStop}%
\bibitem [{\citenamefont {Rrapaj}(2020)}]{rrapaj_exact_2020}%
  \BibitemOpen
  \bibfield  {author} {\bibinfo {author} {\bibfnamefont {Ermal}\ \bibnamefont
  {Rrapaj}},\ }\bibfield  {title} {\enquote {\bibinfo {title} {Exact solution
  of multiangle quantum many-body collective neutrino-flavor oscillations},}\
  }\href {\doibase 10.1103/PhysRevC.101.065805} {\bibfield  {journal} {\bibinfo
   {journal} {Physical Review C}\ }\textbf {\bibinfo {volume} {101}},\ \bibinfo
  {pages} {065805} (\bibinfo {year} {2020})}\BibitemShut {NoStop}%
\bibitem [{\citenamefont
  {Roggero}(2021{\natexlab{a}})}]{roggero_dynamical_2021}%
  \BibitemOpen
  \bibfield  {author} {\bibinfo {author} {\bibfnamefont {Alessandro}\
  \bibnamefont {Roggero}},\ }\bibfield  {title} {\enquote {\bibinfo {title}
  {Dynamical {Phase} {Transitions} in models of {Collective} {Neutrino}
  {Oscillations}},}\ }\href {http://arxiv.org/abs/2103.11497} {\bibfield
  {journal} {\bibinfo  {journal} {arXiv:2103.11497 [astro-ph, physics:hep-ph,
  physics:nucl-th]}\ } (\bibinfo {year} {2021}{\natexlab{a}})}\BibitemShut
  {NoStop}%
\bibitem [{\citenamefont
  {Roggero}(2021{\natexlab{b}})}]{roggero_entanglement_2021}%
  \BibitemOpen
  \bibfield  {author} {\bibinfo {author} {\bibfnamefont {Alessandro}\
  \bibnamefont {Roggero}},\ }\bibfield  {title} {\enquote {\bibinfo {title}
  {Entanglement and {Many}-{Body} effects in {Collective} {Neutrino}
  {Oscillations}},}\ }\href {http://arxiv.org/abs/2102.10188} {\bibfield
  {journal} {\bibinfo  {journal} {arXiv:2102.10188 [astro-ph, physics:hep-ph,
  physics:nucl-th]}\ } (\bibinfo {year} {2021}{\natexlab{b}})}\BibitemShut
  {NoStop}%
\bibitem [{\citenamefont {Hall}\ \emph {et~al.}(2021)\citenamefont {Hall},
  \citenamefont {Roggero}, \citenamefont {Baroni},\ and\ \citenamefont
  {Carlson}}]{hall_simulation_2021}%
  \BibitemOpen
  \bibfield  {author} {\bibinfo {author} {\bibfnamefont {Benjamin}\
  \bibnamefont {Hall}}, \bibinfo {author} {\bibfnamefont {Alessandro}\
  \bibnamefont {Roggero}}, \bibinfo {author} {\bibfnamefont {Alessandro}\
  \bibnamefont {Baroni}}, \ and\ \bibinfo {author} {\bibfnamefont {Joseph}\
  \bibnamefont {Carlson}},\ }\bibfield  {title} {\enquote {\bibinfo {title}
  {Simulation of collective neutrino oscillations on a quantum computer},}\
  }\href {\doibase 10/gmt59x} {\bibfield  {journal} {\bibinfo  {journal}
  {Physical Review D}\ }\textbf {\bibinfo {volume} {104}},\ \bibinfo {pages}
  {063009} (\bibinfo {year} {2021})}\BibitemShut {NoStop}%
\bibitem [{\citenamefont {Meurer}\ \emph {et~al.}(2017)\citenamefont {Meurer},
  \citenamefont {Smith}, \citenamefont {Paprocki}, \citenamefont {Čertík},
  \citenamefont {Kirpichev}, \citenamefont {Rocklin}, \citenamefont {Kumar},
  \citenamefont {Ivanov}, \citenamefont {Moore}, \citenamefont {Singh},
  \citenamefont {Rathnayake}, \citenamefont {Vig}, \citenamefont {Granger},
  \citenamefont {Muller}, \citenamefont {Bonazzi}, \citenamefont {Gupta},
  \citenamefont {Vats}, \citenamefont {Johansson}, \citenamefont {Pedregosa},
  \citenamefont {Curry}, \citenamefont {Terrel}, \citenamefont {Roučka},
  \citenamefont {Saboo}, \citenamefont {Fernando}, \citenamefont {Kulal},
  \citenamefont {Cimrman},\ and\ \citenamefont {Scopatz}}]{meurer_sympy_2017}%
  \BibitemOpen
  \bibfield  {author} {\bibinfo {author} {\bibfnamefont {Aaron}\ \bibnamefont
  {Meurer}}, \bibinfo {author} {\bibfnamefont {Christopher~P.}\ \bibnamefont
  {Smith}}, \bibinfo {author} {\bibfnamefont {Mateusz}\ \bibnamefont
  {Paprocki}}, \bibinfo {author} {\bibfnamefont {Ondřej}\ \bibnamefont
  {Čertík}}, \bibinfo {author} {\bibfnamefont {Sergey~B.}\ \bibnamefont
  {Kirpichev}}, \bibinfo {author} {\bibfnamefont {Matthew}\ \bibnamefont
  {Rocklin}}, \bibinfo {author} {\bibfnamefont {AMiT}\ \bibnamefont {Kumar}},
  \bibinfo {author} {\bibfnamefont {Sergiu}\ \bibnamefont {Ivanov}}, \bibinfo
  {author} {\bibfnamefont {Jason~K.}\ \bibnamefont {Moore}}, \bibinfo {author}
  {\bibfnamefont {Sartaj}\ \bibnamefont {Singh}}, \bibinfo {author}
  {\bibfnamefont {Thilina}\ \bibnamefont {Rathnayake}}, \bibinfo {author}
  {\bibfnamefont {Sean}\ \bibnamefont {Vig}}, \bibinfo {author} {\bibfnamefont
  {Brian~E.}\ \bibnamefont {Granger}}, \bibinfo {author} {\bibfnamefont
  {Richard~P.}\ \bibnamefont {Muller}}, \bibinfo {author} {\bibfnamefont
  {Francesco}\ \bibnamefont {Bonazzi}}, \bibinfo {author} {\bibfnamefont
  {Harsh}\ \bibnamefont {Gupta}}, \bibinfo {author} {\bibfnamefont {Shivam}\
  \bibnamefont {Vats}}, \bibinfo {author} {\bibfnamefont {Fredrik}\
  \bibnamefont {Johansson}}, \bibinfo {author} {\bibfnamefont {Fabian}\
  \bibnamefont {Pedregosa}}, \bibinfo {author} {\bibfnamefont {Matthew~J.}\
  \bibnamefont {Curry}}, \bibinfo {author} {\bibfnamefont {Andy~R.}\
  \bibnamefont {Terrel}}, \bibinfo {author} {\bibfnamefont {Štěpán}\
  \bibnamefont {Roučka}}, \bibinfo {author} {\bibfnamefont {Ashutosh}\
  \bibnamefont {Saboo}}, \bibinfo {author} {\bibfnamefont {Isuru}\ \bibnamefont
  {Fernando}}, \bibinfo {author} {\bibfnamefont {Sumith}\ \bibnamefont
  {Kulal}}, \bibinfo {author} {\bibfnamefont {Robert}\ \bibnamefont {Cimrman}},
  \ and\ \bibinfo {author} {\bibfnamefont {Anthony}\ \bibnamefont {Scopatz}},\
  }\bibfield  {title} {\enquote {\bibinfo {title} {{SymPy}: symbolic computing
  in {Python}},}\ }\href {\doibase 10.7717/peerj-cs.103} {\bibfield  {journal}
  {\bibinfo  {journal} {PeerJ Computer Science}\ }\textbf {\bibinfo {volume}
  {3}},\ \bibinfo {pages} {e103} (\bibinfo {year} {2017})},\ \bibinfo {note}
  {publisher: PeerJ Inc.}\BibitemShut {Stop}%
\bibitem [{\citenamefont {Walt}\ \emph {et~al.}(2011)\citenamefont {Walt},
  \citenamefont {Colbert},\ and\ \citenamefont {Varoquaux}}]{walt_numpy_2011}%
  \BibitemOpen
  \bibfield  {author} {\bibinfo {author} {\bibfnamefont {S.~van~der}\
  \bibnamefont {Walt}}, \bibinfo {author} {\bibfnamefont {S.~C.}\ \bibnamefont
  {Colbert}}, \ and\ \bibinfo {author} {\bibfnamefont {G.}~\bibnamefont
  {Varoquaux}},\ }\bibfield  {title} {\enquote {\bibinfo {title} {The {NumPy}
  {Array}: {A} {Structure} for {Efficient} {Numerical} {Computation}},}\ }\href
  {\doibase 10.1109/MCSE.2011.37} {\bibfield  {journal} {\bibinfo  {journal}
  {Computing in Science Engineering}\ }\textbf {\bibinfo {volume} {13}},\
  \bibinfo {pages} {22--30} (\bibinfo {year} {2011})},\ \bibinfo {note}
  {conference Name: Computing in Science Engineering}\BibitemShut {NoStop}%
\bibitem [{\citenamefont {Hunter}(2007)}]{hunter_matplotlib_2007}%
  \BibitemOpen
  \bibfield  {author} {\bibinfo {author} {\bibfnamefont {J.~D.}\ \bibnamefont
  {Hunter}},\ }\bibfield  {title} {\enquote {\bibinfo {title} {Matplotlib: {A}
  {2D} {Graphics} {Environment}},}\ }\href {\doibase 10.1109/MCSE.2007.55}
  {\bibfield  {journal} {\bibinfo  {journal} {Computing in Science
  Engineering}\ }\textbf {\bibinfo {volume} {9}},\ \bibinfo {pages} {90--95}
  (\bibinfo {year} {2007})},\ \bibinfo {note} {conference Name: Computing in
  Science Engineering}\BibitemShut {NoStop}%
\bibitem [{\citenamefont {Turk}\ \emph {et~al.}(2011)\citenamefont {Turk},
  \citenamefont {Smith}, \citenamefont {Oishi}, \citenamefont {Skory},
  \citenamefont {Skillman}, \citenamefont {Abel},\ and\ \citenamefont
  {Norman}}]{turk_yt_2011}%
  \BibitemOpen
  \bibfield  {author} {\bibinfo {author} {\bibfnamefont {Matthew~J.}\
  \bibnamefont {Turk}}, \bibinfo {author} {\bibfnamefont {Britton~D.}\
  \bibnamefont {Smith}}, \bibinfo {author} {\bibfnamefont {Jeffrey~S.}\
  \bibnamefont {Oishi}}, \bibinfo {author} {\bibfnamefont {Stephen}\
  \bibnamefont {Skory}}, \bibinfo {author} {\bibfnamefont {Samuel~W.}\
  \bibnamefont {Skillman}}, \bibinfo {author} {\bibfnamefont {Tom}\
  \bibnamefont {Abel}}, \ and\ \bibinfo {author} {\bibfnamefont {Michael~L.}\
  \bibnamefont {Norman}},\ }\bibfield  {title} {\enquote {\bibinfo {title} {yt:
  a multi-code analysis toolkit for astrophysical simulation data},}\ }\href
  {\doibase 10.1088/0067-0049/192/1/9} {\bibfield  {journal} {\bibinfo
  {journal} {The Astrophysical Journal Supplement Series}\ }\textbf {\bibinfo
  {volume} {192}},\ \bibinfo {pages} {9} (\bibinfo {year} {2011})}\BibitemShut
  {NoStop}%
\end{thebibliography}%

\appendix
\section{Convergence Study}
\label{app:convergence}
\begin{figure}
    \centering
    \includegraphics[width=\linewidth]{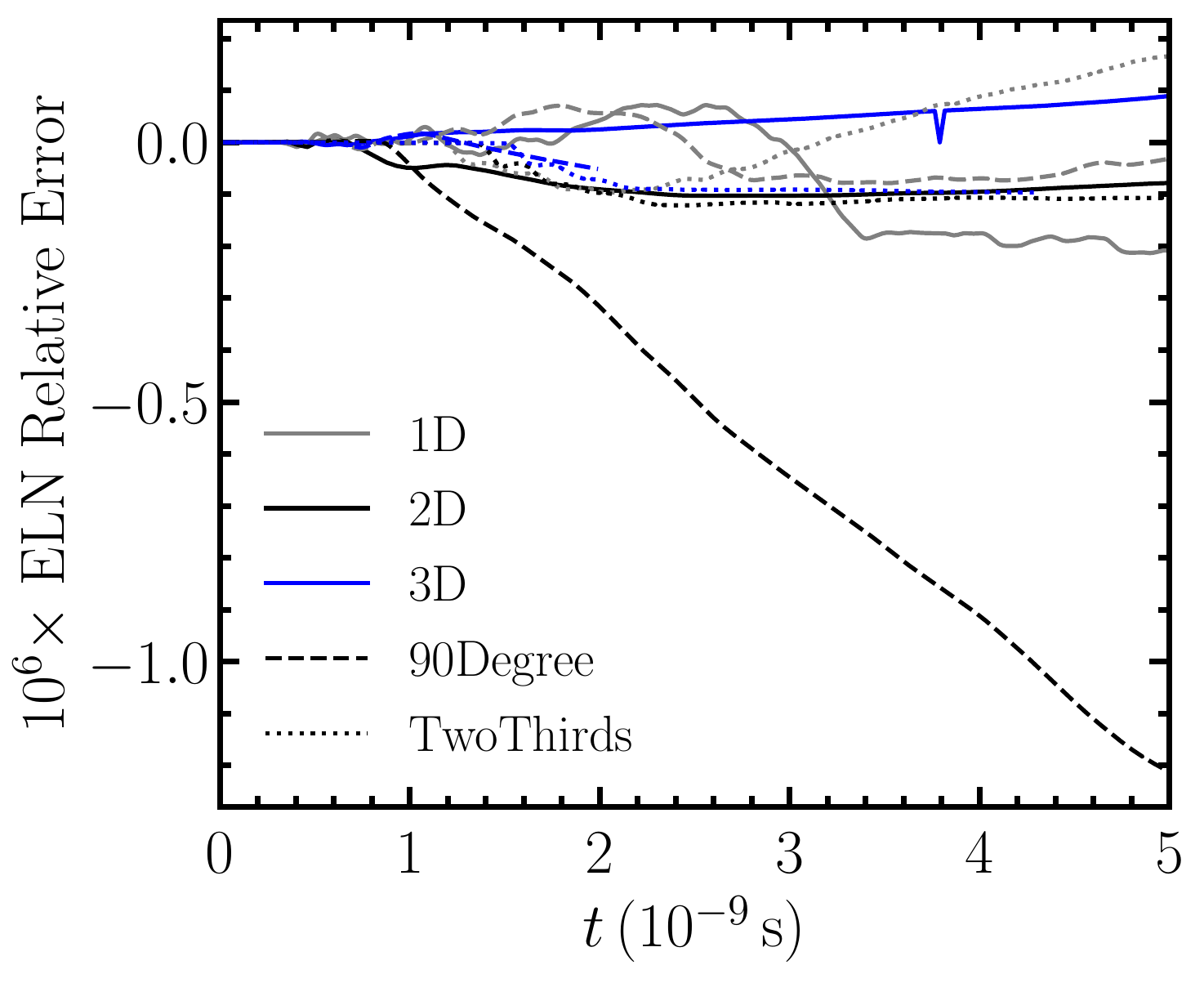}
    \caption{Violation of the conservation of integrated electron flavor content, computed as $(\delta n_{ee}-\delta \bar{n}_{ee})/\mathrm{Tr}(n+\bar{n})$ and $\delta n = n(t)-n(0)$. The symmetry of the neutrino self-interaction part of the Hamiltonian requires this to be zero, but this is not enforced by the code.}
    \label{fig:lepton_number}
\end{figure}
The symmetries of the self-interaction Hamiltonian guarantee that, if the vacuum and matter parts of the Hamiltonian are neglected, the difference between the number of neutrinos and antineutrinos remains a constant. This is not explicitly enforced in our code, and in fact we do include a vacuum contribution to the Hamiltonian, but it is so weak to be of little consequence. Figure~\ref{fig:lepton_number} shows the violation of this conservation of electron lepton number, computed as $(\delta n_{ee}-\delta \bar{n}_{ee})/\mathrm{Tr}(n+\bar{n})$ and $\delta n = n(t)-n(0)$. In all simulations, the integrated electron lepton number is preserved to approximately one part in $10^6$.

\begin{figure*}
    \centering
    \includegraphics[width=\linewidth]{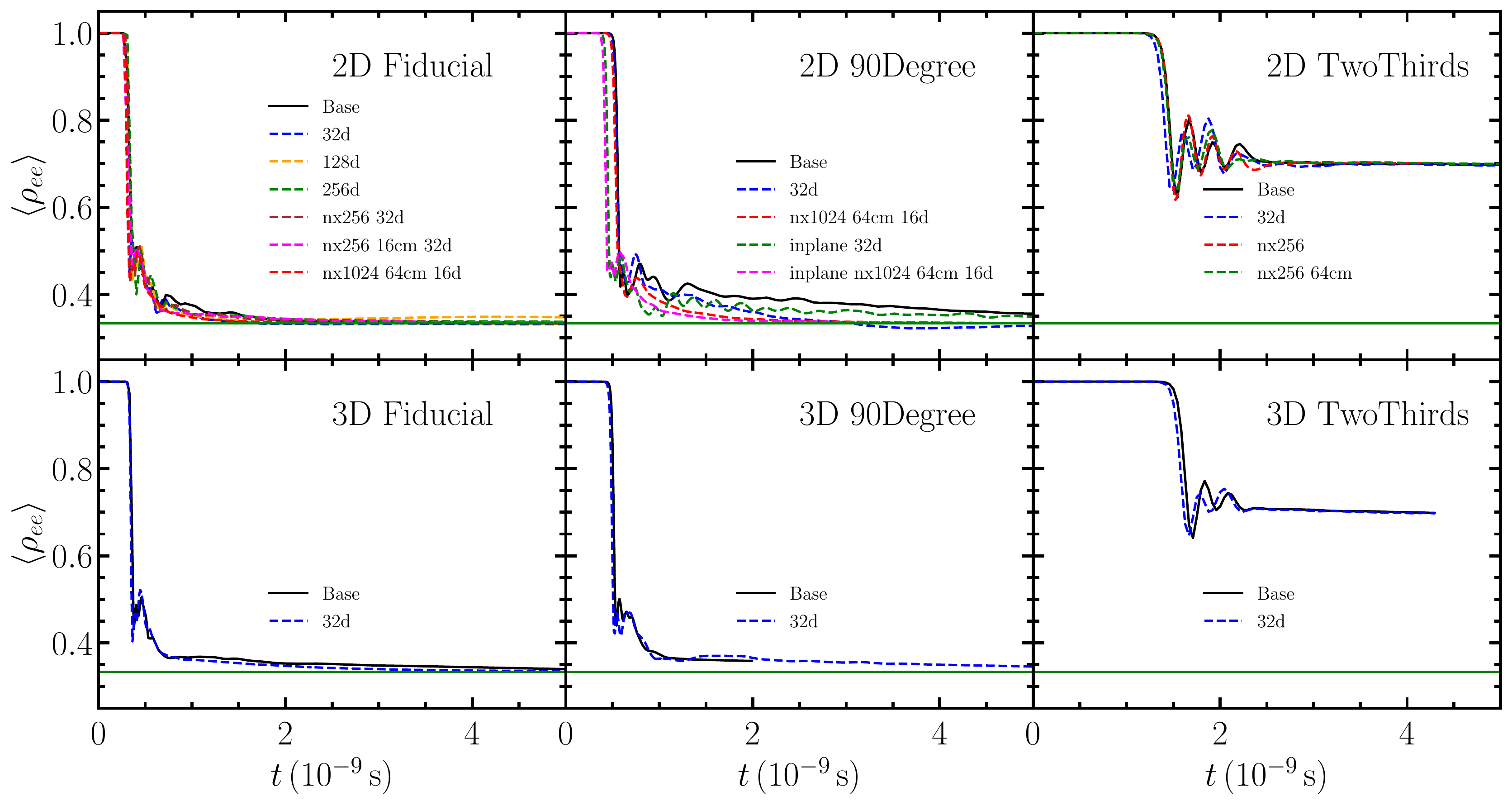}
    \caption{Evolution of the domain-averaged electron neutrino abundance considering various numerical choices. Solid curves reproduce data shown in Fig.~\ref{fig:avgfee}. The top row shows results from 2D simulations while the bottom shows those from 3D simulations. Each column shows results from each of our three initial conditions. Labels with a "d" refer to equatorial directions to be compared to the standard resolution of 64; 32d means 32 equatorial directions or 378 particles per cell, 128 means 6022 particles per cell, an d256d means 24088 particles per cell. "nx" labels the number of grid cells on each extended dimension of the simulation, keeping the domain size constant, where "nx128" is the standard resolution for the multidimensional simulations. "cm" labels the domain size of each dimension of the cube in centimeters. The horizontal green line marks complete flavor mixing. "inplane" for the 2D 90Degree simulations means the computational domain extends over the $\hat{x}-\hat{z}$ plane instead of $\hat{y}-\hat{z}$, slightly increasing the instability growth rate. Overall, our numerical choices have little effect on the instability growth rates and the late-time electron neutrino abundances.}
    \label{fig:fee_convergence}
\end{figure*}
The most obvious relevant prediction from these simulations is the expectation value of the number of electron neutrinos present. In Figure~\ref{fig:fee_convergence} we show the evolution of the abundance of electron neutrinos for 2D (top) and 3D (bottom) simulations of all three initial conditions, similar to Fig.~\ref{fig:avgfee}. We performed a much larger set of numerical studies for 2D simulations because of their relatively lower computational cost. The growth rates of the instability and the equilibrium abundances are quite robust to changes in the number of particles per cell (comparing to simulations labeled "32d", "128d", and "256d" after the number of directions in the $\hat{x}-\hat{y}$ plane in momentum space), the size of the domain (comparing to simulations labeled "16cm" and "64cm" indicating the cube size), and the number of grid cells in each direction  (comparing to simulations labeled "nx256" and "nx1024" indicating the number of grid cells in each non-homogeneous direction). The results seem to be most strongly dependent on the number of particles per cell. For the 2D 90Degree simulations, the growth rate of the instability can be enhanced by making the computational plane in the $\hat{x}-\hat{z}$ direction to match the principal directions of the neutrino and antineutrino distributions (labeled by "inplane"). This permits the growth of the true fastest growing mode, which has a wavevector that is between those two directions.

\begin{figure}
    \centering
    \includegraphics[width=\linewidth]{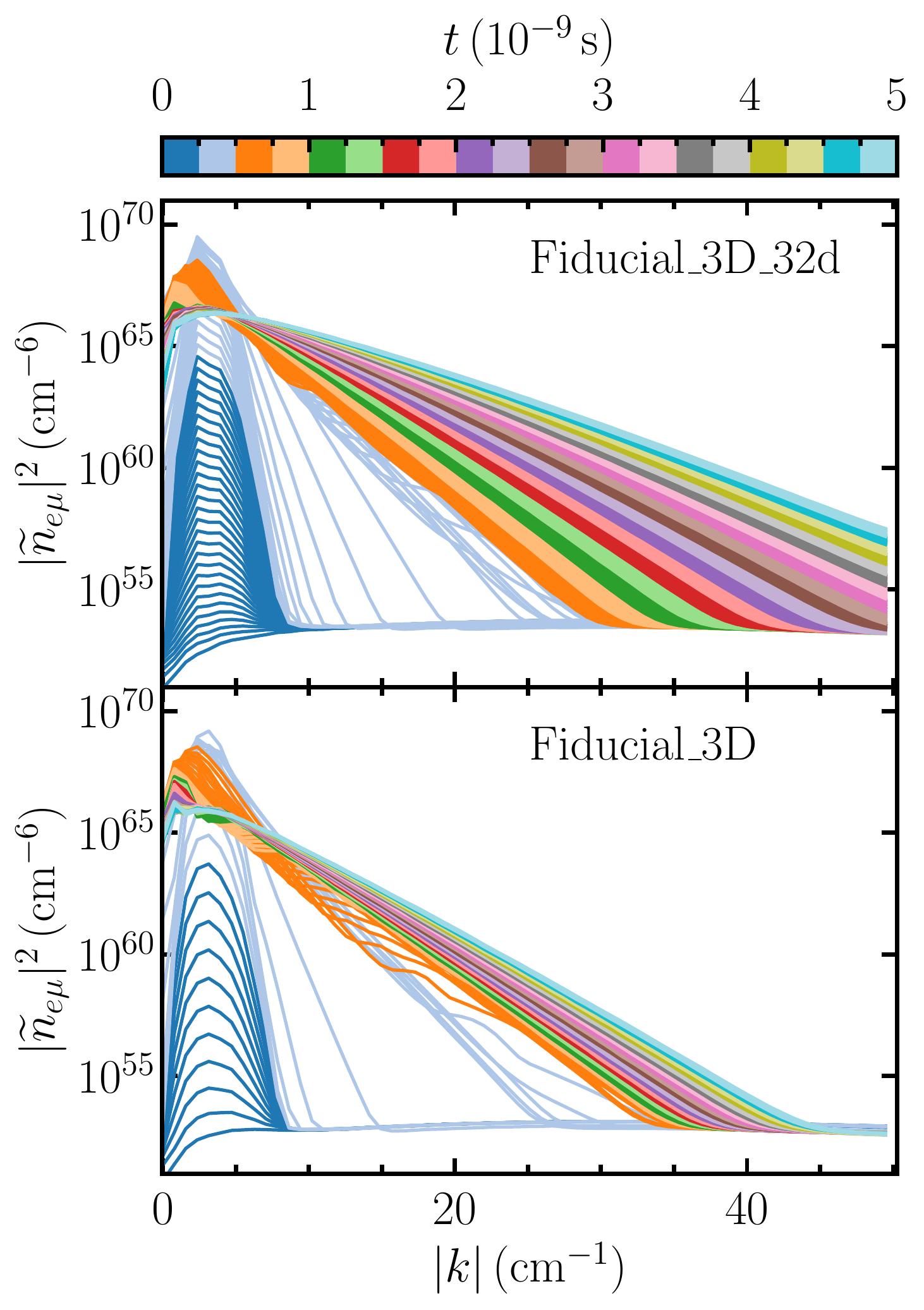}
    \caption{Time evolution of the power spectrum of $n_{e\mu}$ for the Fiducial\_3D simulation (bottom panel) employing 64 equatorial directions (i.e. 1506 particles per cell) and then an otherwise identical simulation employing only 32 equatorial directions (top panel, 378 particles per cell). The spread of the exponential tail toward higher values of $k$ is more strongly affected by this angular resolution than any other numerical choice.}
    \label{fig:fft_3d_convergence}
\end{figure}
The evolution of the power spectrum is a telling indicator of numerical effects, so we investigate it in some detail. Fig.~\ref{fig:fft_3d_convergence} shows the evolution of the power spectrum of $n_{e\mu}$ for the standard-resolution Fiducial\_3D simulation (bottom panel), along with an otherwise identical simulation where we set the number of equatorial directions to 32 (top panel, corresponding to 378 particles per cell and the "32d" dashed blue curve in the bottom left panel of Fig.~\ref{fig:fee_convergence}). The top panel has a higher line density simply as a result of more frequent data output. The most obvious difference is that with more particles per cell, the exponential tail spreads more slowly. In addition, looking at the lowest orange curve in each panel, it is clear that the "tail-kick" feature described in Section~\ref{sec:power_spectrum} occurs at a later time in the higher-resolution simulations (e.g., the bump in the lowest orange curve is farther up the slope in the top panel than in the bottom panel). We argue in Sec.~\ref{sec:angular_structure} that this timing difference is because more particles per cell means that the low-$l$ part of the angular power spectrum has a smaller amplitude. It then takes slightly longer for the angular power spectrum to grow to the point that it interacts significantly with the high-$k$ part of the Fourier spectrum. In both cases, the peak of the distribution drifts toward $k=0$ at the same rate for the first $1\,\mathrm{ns}$, but beyond that the Fiducial\_3D simulation develops a peak at a single $k$ not evident in the Fiducial\_3D\_32d simulation. We do not believe that feature is physical. The details of the spectrum this close to $k=0$ show limitations from the domain size - a larger domain would afford finer resolution in $k$.

\begin{figure}
    \centering
    \includegraphics[width=0.9\linewidth]{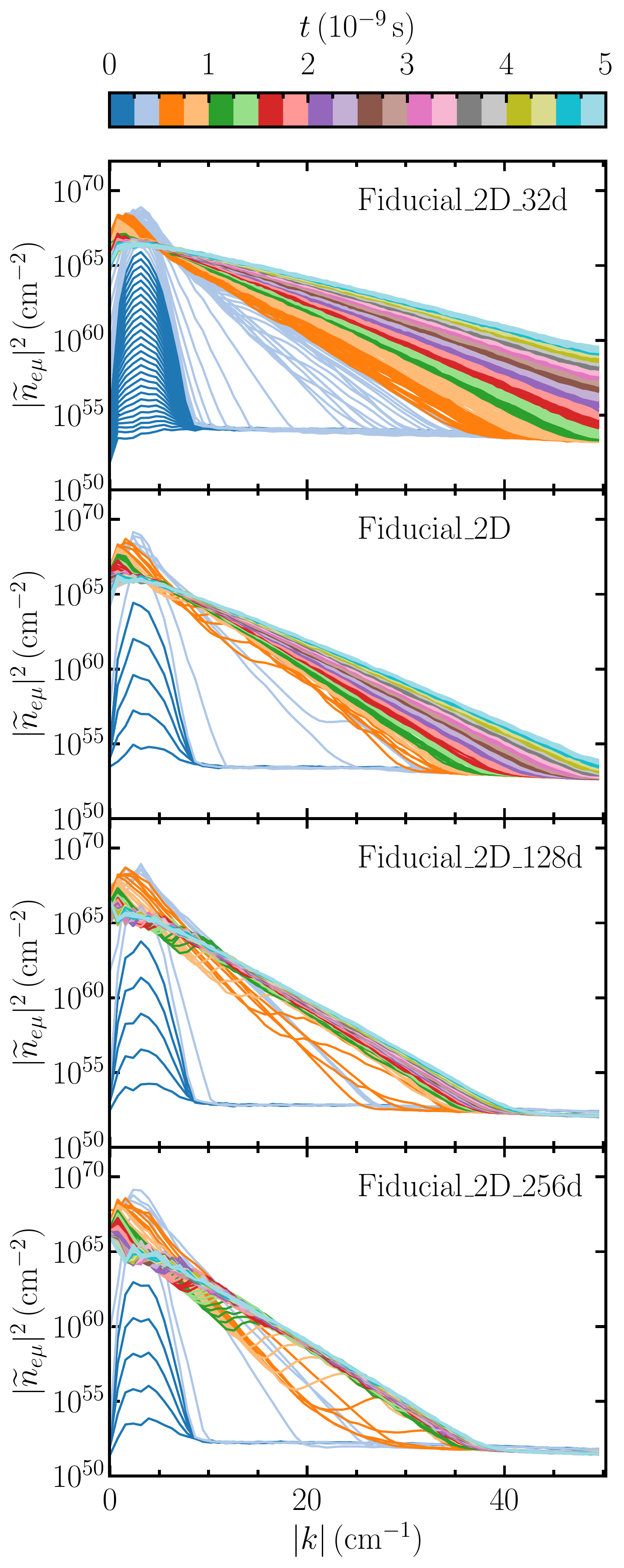}
    \caption{Time evolution of the power spectrum of $n_{e\mu}$ for the Fiducial\_2D simulation (second panel) employing 64 equatorial directions (i.e. 1506 particles per cell). The other panels show the equivalent results from simulations employing only 32, 128, and 256 equatorial directions (378, 6022, and 24088 particles per cell, respectively). The spread of the exponential tail toward higher values of $k$ is more strongly affected by this angular resolution than any other numerical choice.}
    \label{fig:fft_2d_convergence}
\end{figure}
The same can be said about the 2D simulations shown in Fig.~\ref{fig:fft_2d_convergence}. From top to bottom we show simulations with 32, 64, 128, and 256 equatorial directions (378, 1506, 6022, and 24088 particles per cell, respectively). As before, higher angular resolution causes the exponential tail to drift more slowly. In addition, higher angular resolution also causes more choppy features in the spectrum near $k=0$ at the scale of the $k$ resolution. The limiting numerical factor for the $0 \leq k \leq 10$ part of the spectrum for the Fiducial\_2D\_256d simulation at $t=5\,\mathrm{ns}$ is likely the domain size.

We will focus on the drift of the exponential tail as a probe of numerical artifacts. Interestingly, for the same angular resolution the 3D results in Fig.~\ref{fig:fft_3d_convergence} perform better than the 2D results in Fig.~\ref{fig:fft_2d_convergence} in that the exponential tail drifts more slowly. The spread of the exponential tail after $t=1\,\mathrm{ns}$ is significantly smaller for the Fiducial\_3D simulation than for the Fiducial\_2D simulation. In addition, the difference between the Fiducial\_3D\_32d and Fiducial\_3D exponential tails is much greater than the difference between the Fiducial\_2D\_32d and Fiducial\_2D tails. This pattern extends down to 1D simulations - there is a small difference between simulations of different angular resolution, which led us to expect that the exponential drift was not a numerical artifact in \cite{richers_particle--cell_2021}. The numerical origin of this drift is not clear.

\begin{figure}
    \centering
    \includegraphics[width=\linewidth]{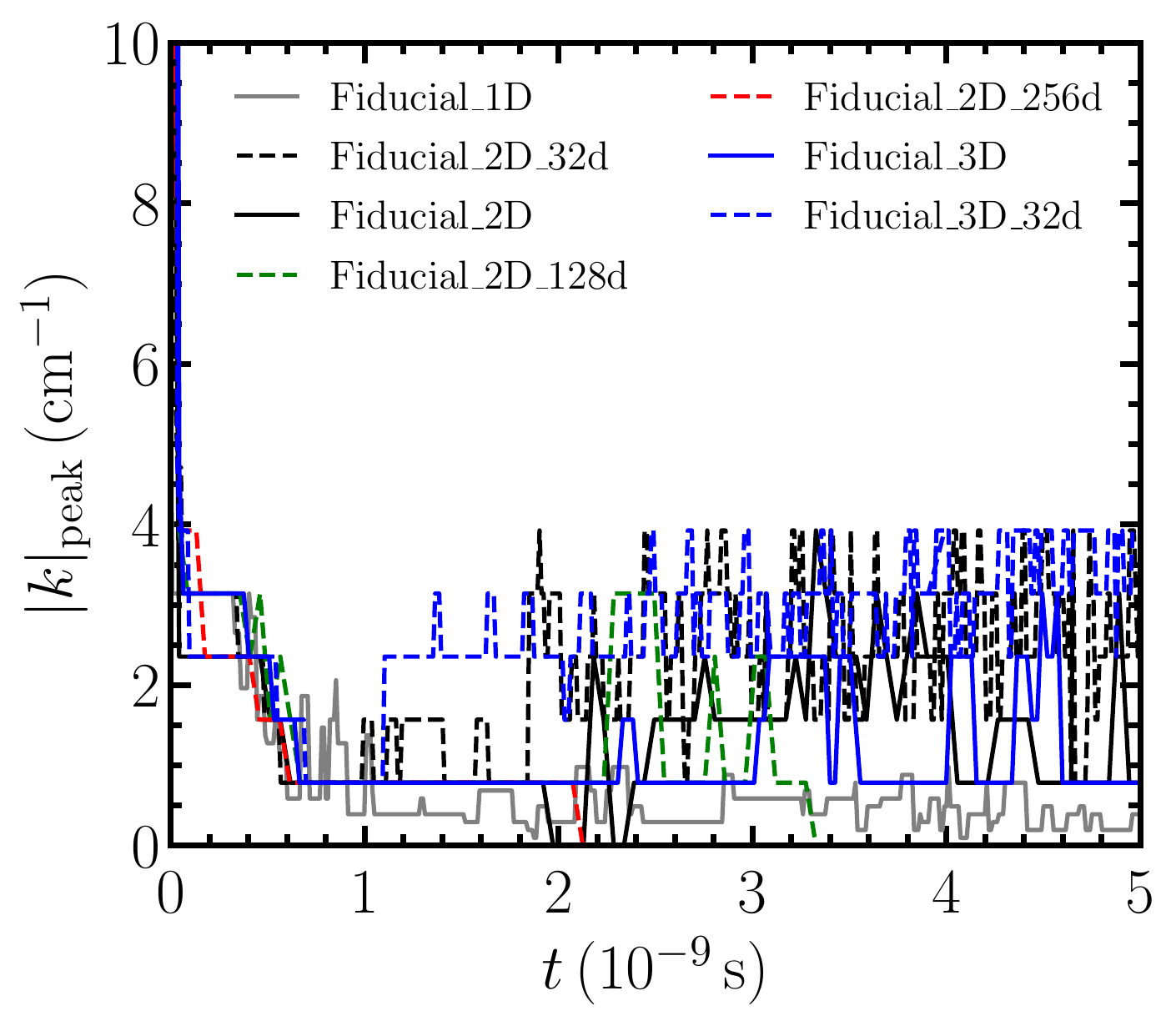}
    \caption{Evolution of the value of $k$ at the peak of the power spectrum of $n_{e\mu}$ for 1D, 2D, and 3D simulations of varying angular resolution. The low-resolution (32d) simulations stay tightly packed until about $1\,\mathrm{ns}$, while the standard resolution results are very similar until about $2\,\mathrm{ns}$. The highest-resolution 2D simulation predicts a peak very close to $k=0$ throughout the post-saturation phase.}
    \label{fig:fft_peak_convergence}
\end{figure}
Finally, we look at the evolution of the peak of the power spectrum with time in Fig.~\ref{fig:fft_peak_convergence}. Initially, the randomized initial conditions result in a peak at high wavenumber, but almost immediately shoots down to the value of $k\approx3\,\mathrm{cm}^{-1}$ corresponding to the fastest growing wavelength of $\lambda=2.2\,\mathrm{cm}$. After the instability saturates at $t\approx0.3\,\mathrm{ns}$, the peak begins to drift toward $k=0$ over the next few tenths of a nanosecond, and should remain there for the rest of the simulation. Looking first at the standard resolutions (solid curves), all three dimensionalities agree well until $t\approx2\,\mathrm{ns}$, at which point all three dimensionalities show peaks drifting away from $k=0$. The higher-resolution 2D results (red and green dashed curves) tend to stay much closer to $k=0$ at late times, while the lower-resolution 2D and 3D results (dashed black and blue, respectively) drift upward as early as $t\approx1\,\mathrm{ns}$.

\end{document}